\documentclass[]{aastex701}
\newcommand{\kms}{km s$^{-1}$\xspace}
\newcommand{\HI}{{\rm H\,{\scriptsize I }}\xspace}

\usepackage{xcolor} 
\newcommand{\zenodosite}{%
\href{https://zenodo.org/records/19904867?preview=1&token=eyJhbGciOiJIUzUxMiJ9.eyJpZCI6IjBhZTM5ZDUwLWNhN2YtNGViNi04MmI4LTk4OGNhMWFlMGRhNyIsImRhdGEiOnt9LCJyYW5kb20iOiI4ODUwNzJhYjFiMDM5NGQ2MjJjNTUwZWY1ZWUwZWMxYSJ9.FFsKntsIcKJIy_M0DZn9fK-TsYQu01t_1k0E2MCS0L2CipmkTn7J6uyPwyupNxlDFSZPYmG3QjWeVUAq5xNpDw}{(doi:10.5281/zenodo.19904867)}%
}

\usepackage{etoolbox}
\usepackage{booktabs}    
\usepackage{xcolor}      
\usepackage{amsmath}     
\usepackage{amsmath} 
\usepackage{caption}
\usepackage{subcaption}
\usepackage{siunitx}
\usepackage{rotating} 
\usepackage{xspace}
\usepackage{graphicx}
\setkeys{Gin}{draft=false}

\begin{document}

\title{OH Line Detections in Southern Galaxies of the IRAS Revised Bright Galaxy Sample}

\author{han zhao}
\affiliation{College of Physics, Guizhou University, 550025 Guiyang, PR China}
\email{}

\author[orcid=0000-0000-0000-0001,sname='North America']{zhongzu Wu}
\altaffiliation{}
\affiliation{College of Physics, Guizhou University, 550025 Guiyang, PR China}
\email[show]{zzwu08@google.com}

\author{Bo Zhang}
\affiliation{Shanghai Astronomical Observatory, Chinese Academy of Sciences, 80 Nandan Road, Shanghai 200030, PR China}
\email{}

\author{Timur Mufakharov}
\affiliation{Special Astrophysical Observatory of RAS, Nizhny Arkhyz 369167, Russia}
\email{}

\author{Yongjun Chen}
\affiliation{Shanghai Astronomical Observatory, Chinese Academy of Sciences, 80 Nandan Road, Shanghai 200030, PR China}
\email{}

\author{Zhiqiang Shen}
\affiliation{Shanghai Astronomical Observatory, Chinese Academy of Sciences, 80 Nandan Road, Shanghai 200030, PR China}
\email{}

\author{Yulia Sotnikova}
\affiliation{Special Astrophysical Observatory of RAS, Nizhny Arkhyz 369167, Russia}
\email{}


\begin{abstract}
We present a systematic study of OH main-line emission and absorption in 186 southern galaxies from the IRAS Revised Bright Galaxy Sample, using archival MeerKAT snapshot data. OH features are detected in 38 galaxies, including eight with OH maser emission (three new) and 30 showing OH absorption, mostly unreported previously. Four absorption systems exhibit weak OH emission superposed on strong absorption. OH-emitting regions are generally more compact than the associated radio continuum. Most absorption profiles are well fit by two Gaussian components (1667 and 1665 MHz), with an average integrated line ratio of $\sim$1.5. LIRGs show an OH emission detection rate of $\sim$13\%, versus significantly lower rates in non-LIRGs. For sources with radio continuum flux densities $\gtrsim 20$ mJy, OH absorption detection rates reach $\sim$36\% (LIRGs) and $\sim$27\% (non-LIRGs), while no OH absorption features were detected among sources with lower radio continuum flux densities. This suggests that sufficient background continuum is likely an important factor for the detection of OH absorption. Detected OH emitters follow the empirical $L_{\rm OH}$--$L_{\rm FIR}$ relation, consistent with far-infrared pumping, while non-detections show upper limits below the relation. No significant differences are found between OH absorbers and non-detections in infrared luminosity or radio continuum compactness. Stacked spectra of non-detections reveal no significant OH features, suggesting that sensitivity and orientation alone do not fully explain the absence of absorption. In contrast, mid-infrared colors (e.g., W2--W3) and $q_{\rm TIR}$ differ between the two populations. OH absorption galaxies occupy an intermediate regime in $L_{\rm HCN}/L_{\rm CO}$ between OH megamasers and non-detections, implying that OH absorption detectability is linked to dense molecular gas conditions, with extreme star formation potentially suppressing its occurrence.
\end{abstract}



\section{Introduction} 

The {\it Infrared Astronomical Satellite} (IRAS; \citealt{1984ApJ...278L...1N}) revealed a population of galaxies with exceptionally high infrared luminosities. Galaxies with $L_{\rm IR} \lesssim 10^{11}\,L_{\odot}$ are predominantly gas-rich spiral systems, whose infrared emission is largely attributed to star formation. In contrast, the most powerful infrared sources—luminous and ultraluminous infrared galaxies (LIRGs and ULIRGs), with $L_{\rm IR} = 10^{11}$--$10^{12}\,L_{\odot}$ and $>10^{12}\,L_{\odot}$, respectively—are primarily interacting or merging systems hosting enormous reservoirs of molecular gas \citep{1996ARA&A..34..749S}. 

The hydroxyl (OH) 18\,cm main lines at 1667 and 1665\,MHz provide a powerful probe of the molecular interstellar medium (ISM) in infrared-luminous galaxies. OH can be detected either in maser emission, tracing compact and dense nuclear regions pumped by strong far-infrared (FIR) radiation fields \citep{2008ApJ...677..985L,2018JApA...39...34H}, or in absorption against bright radio continuum sources, revealing the distribution and kinematics of cooler molecular gas along the line of sight. In general, OH megamasers (OHMs) are preferentially found in (U)LIRGs, whereas OH absorption is more commonly detected in infrared-bright galaxies with lower FIR luminosities (e.g., \citealt{1986AJ.....92.1291S,1989ApJ...338..804B,2018JApA...39...34H}). The occurrence of OH megamasers correlates with infrared luminosity, merger stage, and nuclear gas concentration \citep{2007ApJ...669L...9D}, while OH absorption is more sensitive to the geometry of the molecular gas and the structure of the background radio continuum. Despite this complementarity, the relationship between OH emission and absorption, and the physical conditions that influence their detectability, is still not well constrained.

Although OH emission and absorption provide valuable probes of the molecular ISM, the number of nearby IRAS galaxies with detected OH lines remains limited, and the overall detection rate is relatively low \citep[e.g.,][]{2010AJ....139.2066F,1986AJ.....92.1291S,2024ApJ...971..131Z}. Most detections to date originate from surveys conducted in the 1980s and 1990s using single-dish facilities such as Arecibo, the Green Bank Telescope, Parkes, and the Nançay Radio Telescope \citep[e.g.,][]{1985ApJ...293..394B,1986AJ.....92.1291S,1992AJ....103..728B,1985A&A...152L...9K,2010AJ....139.2066F}. These surveys typically targeted relatively small samples and were often biased toward northern hemisphere sources, leaving southern infrared-bright galaxies comparatively underexplored. Since the 2000s, OH observations have continued mainly with Arecibo, the Green Bank Telescope, and FAST, with increasing focus on intermediate-redshift sources ($z \gtrsim 0.1$) \citep[e.g.,][]{2000AJ....119.3003D,2001AJ....121.1278D,2002AJ....124..100D,2012IAUS..287..345W,2024ApJ...971..131Z,2025ApJ...986...70R}, partly driven by instrumental capabilities and survey strategies.

Recently, a new generation of high-sensitivity H\,{\scriptsize I} surveys using facilities such as MeerKAT, ASKAP, and the Westerbork Synthesis Radio Telescope (WSRT) has significantly improved this situation. These surveys employ wide-band receivers capable of simultaneously covering both H\,{\scriptsize I} and OH line frequencies, enabling efficient blind searches for OH lines over large samples. Examples include the MeerKAT Absorption Line Survey (MALS; \citealt{2021ApJ...907...11G}), Apertif on WSRT \citep{2021A&A...647A.193H}, and the ASKAP First Large Absorption Survey in H\,{\scriptsize I} (FLASH; \citealt{2022MNRAS.516.2947S}), as well as deep surveys such as LADUMA with MeerKAT \citep{2022ApJ...931L...7G}. These high-sensitivity wide-band \HI\ surveys, together with targeted observations, provide great opportunities to significantly increase the number of known OH line systems and extend the redshift range of OH studies to much higher redshifts \citep{2021ApJ...911...38R,2024MNRAS.528.3486B,2021A&A...647A.193H}. For example, OH megamasers have recently been detected at $z\simeq0.52$ by the LADUMA survey, at $z=0.7092$ by the MIGHTEE survey \citep{2024MNRAS.529.3484J}, and at $z=1.027$ \citep{2026arXiv260213396M}. Meanwhile, these wide-band facilities also enable systematic high-sensitivity surveys of low-redshift OH systems, providing new opportunities to discover previously unknown OH emitters and absorbers and to investigate the physical conditions of the interstellar medium in infrared-luminous galaxies. In particular, \citet{2021ApJS..257...35C} presented a MeerKAT snapshot survey of southern galaxies in the IRAS Revised Bright Galaxy Sample (RBGS; \citealt{2003AJ....126.1607S}), one of the most complete samples of star-forming galaxies in the local universe. Owing to the large collecting area, wide bandwidth, and interferometric imaging capability of MeerKAT, these archival data provide both high sensitivity and improved spatial information, mitigating the limitations of earlier single-dish surveys. This makes it possible to systematically explore the incidence of both weak OH maser emission and absorption, and to place them in a unified statistical framework.

The main aim of this paper is to carry out a systematic and homogeneous search for OH maser emission and absorption in southern IRAS RBGS galaxies using archival MeerKAT observations. By significantly expanding the sample of OH detections, especially in absorption, this work enables a more robust statistical comparison between OH-emitting, absorption, and non-detected systems. Details of the sample selection, observations, data reduction, and analysis are presented in Section~2. The results and discussion are given in Sections~3 and~4, respectively, and the main conclusions are summarized in Section~5.

\section{Meerkat archive data collection and reduction}

\subsection{The Sample and Data Collection}

The IRAS Revised Bright Galaxy Sample (RBGS), based on observations from IRAS, comprises 629 extragalactic sources with 60\,$\mu$m flux densities greater than 5.24\,Jy and Galactic latitudes $|b| > 5^\circ$ \citep{2003AJ....126.1607S}. The RBGS is one of the most complete and least biased samples of star-forming galaxies in the local universe \citep{2021ApJS..257...35C}. MeerKAT L-band observations are available for a subsample of 298 southern RBGS sources with declinations $\delta_{\rm J2000} < 0^\circ$, as presented by \citet{2021ApJS..257...35C}. We retrieved and inspected the archived MeerKAT data for all 298 sources in this work. During the data quality assessment, we found that 112 sources are unsuitable for reliable OH main line analysis. This classification is based on a combination of project-level metadata from the MeerKAT archive and direct inspection of the visibility data, including amplitude and phase as a function of frequency channel, as well as noise estimates from preliminary CLEAN imaging of channels near the expected OH line frequencies, determined from optical redshifts reported in the literature. The data quality issues are primarily caused by severe radio-frequency interference (RFI), unreliable spectral responses near the band edges for sources with OH lines falling in the 1660--1670 MHz frequency range and, in some cases, calibration problems. After excluding these sources, we retain a final sample of 186 galaxies with data of sufficient quality for OH emission and absorption searches. The RFI is mainly associated with satellite transmissions, including Inmarsat, GPS L1, GLONASS L1, and Iridium, affecting the frequency range 1526--1626 MHz (see the online MeerKAT RFI documentation\footnote{\url{https://skaafrica.atlassian.net/wiki/spaces/ESDKB/pages/305332225/Radio+Frequency+Interference+RFI}}). This frequency range corresponds to OH 1667 MHz line at redshifts z $\sim$ 0.025--0.093. In this interval, the contamination is often severe, leading to substantial data loss, significantly elevated noise levels, and, in many cases, a complete loss of sensitivity to OH features. As a result, a large fraction of sources within this redshift range cannot be reliably analyzed, introducing a pronounced redshift-dependent incompleteness in the sample.  Meanwhile, the severe RFI between 1526--1626\,MHz limits the accessibility of the OH 1612 and 1720\,MHz satellite lines. Although the observations cover 856--1712\,MHz, the effective frequency range is limited to 880--1670\,MHz \citep{2021ApJS..257...35C}. The OH 1720\,MHz line is accessible for sources with $0.03<z<0.058$, while the OH 1612\,MHz line is accessible for sources with $z>0.0565$. In our sample, 6 and 19 sources fall within these redshift ranges, respectively. After excluding one OH 1612\,MHz source and four OH 1720\,MHz sources affected by poor data quality, 5 and 15 sources were selected for the OH 1612 and 1720\,MHz satellite-line searches, respectively.

\subsection{Data Reduction}

The archived MeerKAT observations used in this work were originally presented by \citet{2021ApJS..257...35C}. The dataset consists of 14 observing runs, each approximately 8\,hr in duration, conducted between 2020 May 29 and 2021 March 3. The observations cover a frequency range of 856--1712\,MHz, divided into 4096 spectral channels, corresponding to a channel width of 208\,kHz (i.e., $\sim$38~km~s$^{-1}$ at the OH main-line frequencies). While the full bandwidth was used by \citet{2021ApJS..257...35C} to produce radio continuum images, in this work we extract $\sim$20\,MHz spectral windows centered on the expected frequencies of the OH 1667, 1612, and 1720\,MHz lines for each target, based on its optical redshift.

We initially downloaded the pipeline-calibrated datasets from the archive. However, for a subset of sources, these data were found to be unreliable, likely due to issues during data transfer, compression, or earlier processing. For these cases, we retrieved the raw datasets and performed calibration independently using the Common Astronomy Software Applications package (CASA; \citealt{2007ASPC..376..127M}). The calibration procedure included automated and manual flagging of corrupted data, followed by delay and bandpass calibration using primary flux calibrators. Phase and amplitude calibration were then carried out using standard phase calibrators, with the absolute flux scale tied to the flux calibrators. After an initial calibration, self-calibration was applied to the calibrators, followed by additional flagging and a second round of calibration. The final calibration solutions were then applied to both calibrators and target sources.

The calibrated target data were split from the measurement sets and imaged following the CASA VLA \HI\ tutorial\footnote{\url{https://casaguides.nrao.edu/index.php/HI_21cm_(1.4_GHz)_spectral_line_data_reduction:_LEDA_44055-CASA6.7.2}}. We first generated several single-channel images to estimate the rms noise level ($1\sigma$), and then produced full spectral cubes using \texttt{tclean} with an automated cleaning threshold of $3\sigma$. Line-free and RFI-free channels were identified and used for continuum subtraction in the $uv$-plane with \texttt{UVCONTSUB}, fitting a first-order polynomial. The continuum-subtracted data were subsequently re-imaged with \texttt{tclean} using the same $3\sigma$ threshold. Imaging was performed with Briggs weighting (robust = 0.5), and no self-calibration was applied to the target data. Although continuum images were also produced from line-free channels, we mainly adopt the higher-sensitivity full-band continuum images from \citet{2021ApJS..257...35C} for our analysis\footnote{\url{https://doi.org/10.48479/dnt7-6q05}}.

The resulting synthesized beam has a typical FWHM of $\sim$8\arcsec. The rms noise per channel (with a channel width of $\sim$208~kHz) is typically $\sim$0.5~mJy~beam$^{-1}$, with variations depending on data quality (See Tables~\ref{tab:ohemission} and \ref{tab:ohabsorption} for the detected OH emission and absorption sources, respectively, and the parameters of OH non-detections are available in the Zenodo data repository \zenodosite.) . Moment-0 maps were constructed by integrating over the velocity ranges of the OH 1667 and 1665 MHz lines. Based on these moment-0 maps, we classified the sources into detections, candidates, and non-detections using the peak signal-to-noise ratio, defined as $f_{\rm peak}/\sigma_{\rm mom0}$, where $f_{\rm peak}$ is the peak integrated flux density measured within one synthesized beam in the moment-0 map and $\sigma_{\rm mom0}$ is the rms noise measured from emission-free regions of the same map. Sources with S/N $> 4.5$ are considered detections, those with $3 < {\rm S/N} < 4.5$ are classified as candidates, and the remaining sources are treated as non-detections.

\section{Results}

\begin{figure}
\centering

\begin{subfigure}[t]{0.48\textwidth}
\centering
\includegraphics[width=\textwidth]{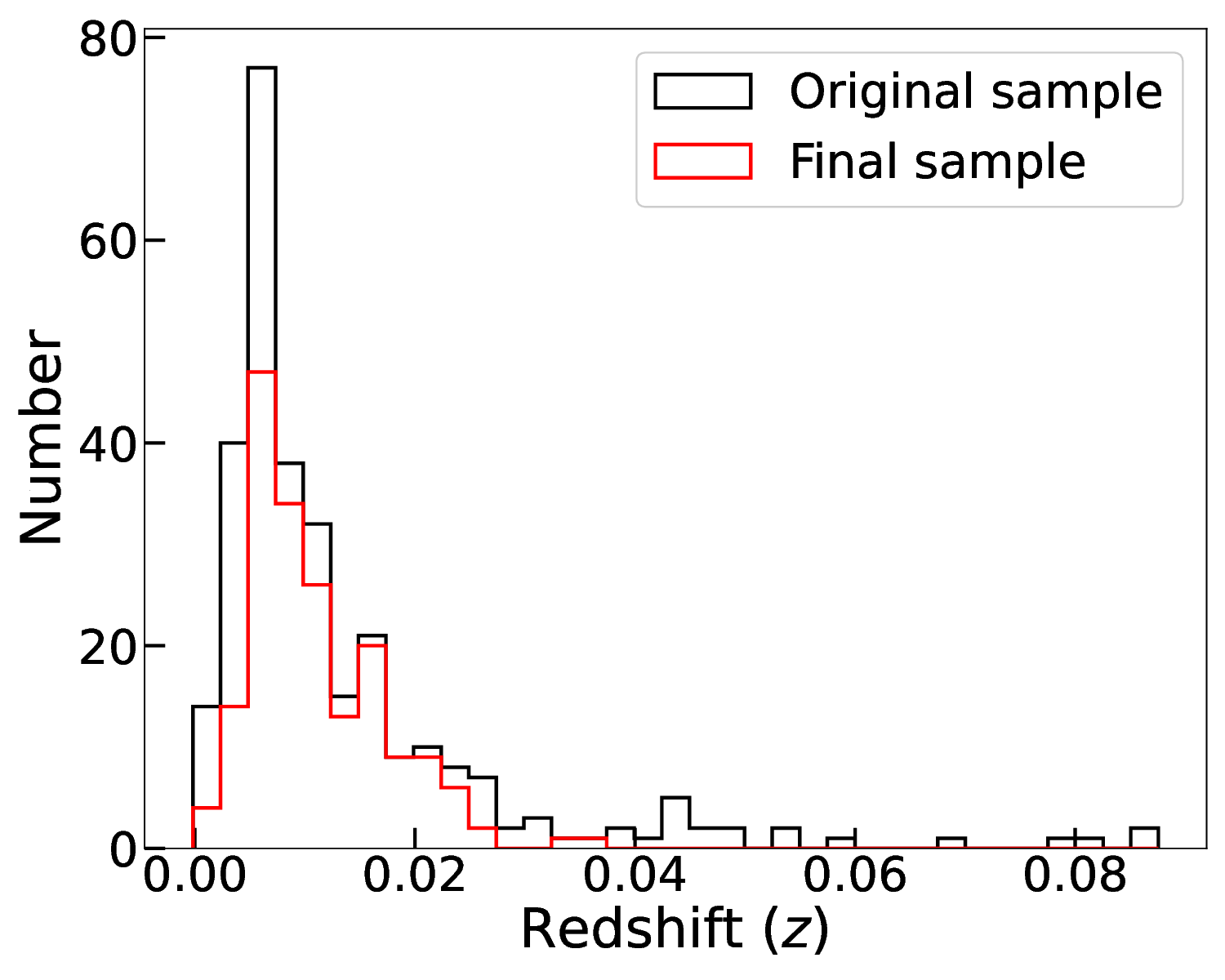}
\end{subfigure}
\hfill
\begin{subfigure}[t]{0.48\textwidth}
\centering
\includegraphics[width=\textwidth]{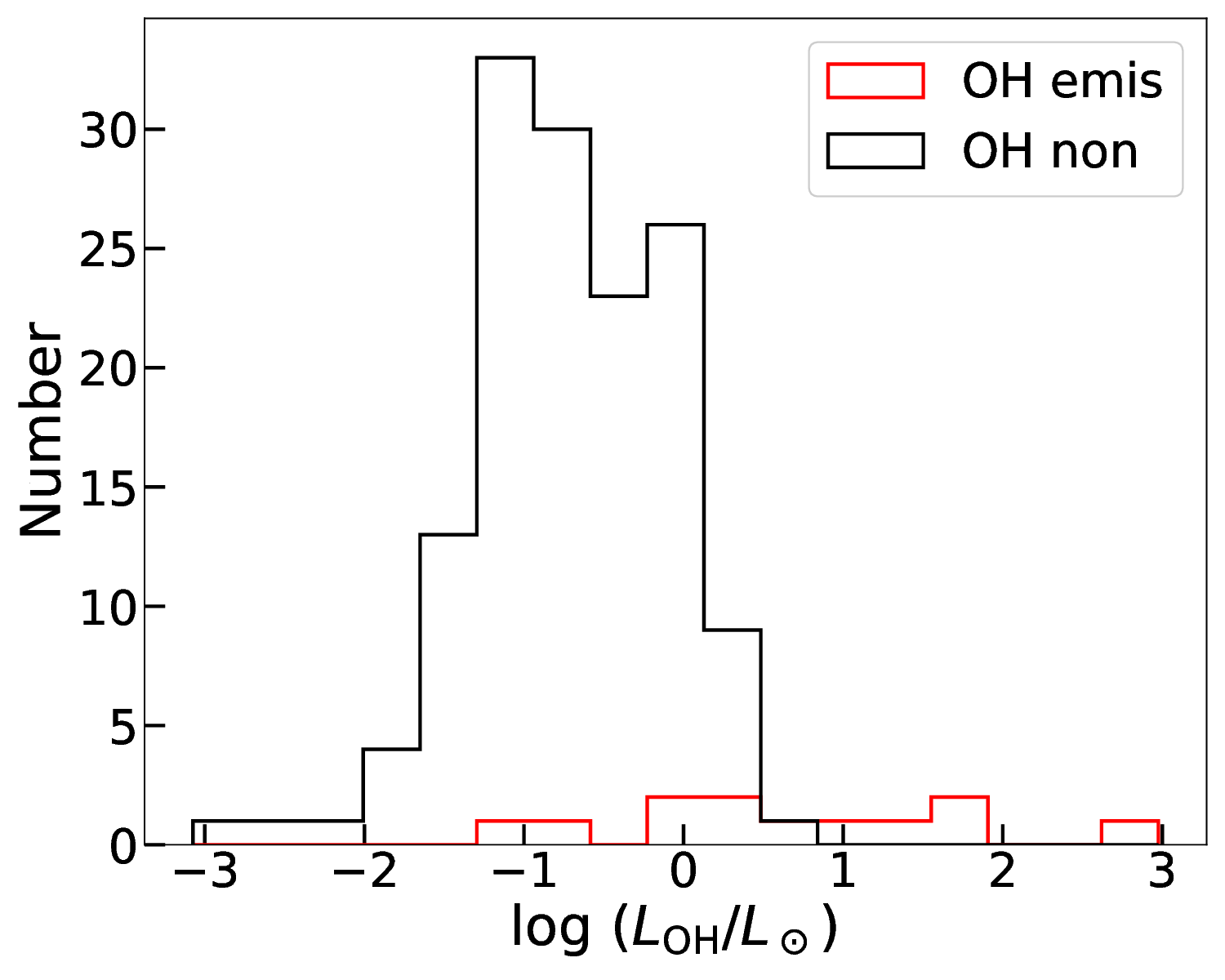}
\end{subfigure}

\caption{The redshift and OH luminosity ($L_{\rm OH}$) distributions of the sample. The left panel shows the redshift distribution of the original RBGS southern sample (298 sources; gray histogram) and the final sample (186 sources; red histogram). 
The redshifts are adopted from NED. In the right panel, the red and black histograms represent OH detections and non-detections, respectively. The $L_{\rm OH}$ values for detections are taken from Table \ref{tab:ohemission}, while the upper limits for non-detections are estimated from the rms noise levels of the spectra available in the Zenodo data repository (\href{https://doi.org/10.5281/zenodo.19904867}{doi:10.5281/zenodo.19904867}). }
\label{fig:LOH_redshift}
\end{figure}

We processed MeerKAT data for a sample of 186 southern galaxies selected from the IRAS RBGS to search for the OH 1665 and 1667\,MHz main lines (see Fig. \ref{fig:LOH_redshift}). We also searched for the OH 1612 and 1720\,MHz satellite lines in 5 and 15 sources, respectively, where the frequency coverage allowed. No significant satellite-line emission or absorption features were identified; therefore, the analysis presented in this paper focus primarily on the two OH main lines. OH line features are identified in 44 galaxies, including 38 robust detections at $\geq 4.5,\sigma$ and 6 absorption candidates at $\sim 3,\sigma$. Among the 38 secure detections, eight sources exhibit prominent OH emission, while 30 galaxies show OH absorption. No significant OH emission or absorption is detected in the remaining 142 galaxies. The properties of both the detected sources and the non-detections are presented and discussed in the following subsections.

\subsection{OH emission lines}

The OH line profiles of the eight OH maser galaxies detected in our sample are shown in Fig.~\ref{fig:oh_spectra_centered}, and the fitted parameters are listed in Table~\ref{tab:ohemission}. Five of these sources have previously been identified as OH maser galaxies in the literature; however, published OH line profiles are available for only four of them (see Fig.~\ref{fig:oh_spectra_centered} and Section~\ref{notesonind}). The OH spectra obtained with \textit{MeerKAT} are consistent with those reported in earlier studies. For one source, IRAS~16399$-$0937, no OH line profile has been published to date, but the measured peak OH flux density agrees well with the value reported in the literature (see Section~\ref{notesonind} for details). The remaining three sources represent new OH maser detections.

The rms noise levels of the OH spectra range from 0.6 to 1.0~mJy. The OH line profiles were fitted using Gaussian components, and seven of the eight galaxies require two or more distinct components to adequately reproduce the observed spectra. The resulting fit parameters are summarized in Table~\ref{tab:ohemission}. Based on these fits, we derived the OH line luminosities. For the five previously known OH maser galaxies, the derived OH luminosities ($L_{\mathrm{OH}}$) are consistent with values reported in the
literature \citep[see][and references therein]{2002AJ....124..100D}.  
Among the three newly detected sources, two galaxies (IRAS~16443$-$2915 and IRAS~12043$-$3140) clearly exceed the classical megamaser threshold of $L_{\mathrm{OH}} = 10~L_{\odot}$ (see Table~\ref{tab:ohemission}). The remaining source (IRAS~13242$-$5713) has $\log L_{\mathrm{OH}} > 0$, placing it in the so-called kilomaser regime, although this luminosity threshold has also been historically used to define OH megamasers \citep[see][]{2009A&A...502..529S,2011A&A...525A..91T,1990A&A...229..431H}. This source likely also exhibits two weak OH absorption features; however, the OH emission component dominates the spectrum, and we therefore classify it as an OH maser emitter.

Integrated OH line emission (moment--0) maps for the eight OH maser galaxies are shown in Fig.~\ref{OH8image}. In all cases, the OH emission is more compact than the corresponding radio continuum emission. Gaussian fitting of the OH emission maps in CASA resolves only one galaxy (F11506$-$3851), for which deconvolved major and minor axes and their associated
uncertainties are obtained (see Table \ref{tab:ohemission}). The remaining seven galaxies are unresolved, and reliable source sizes cannot be derived from Gaussian fitting. For these sources, we estimate upper limits on the sizes of the OH-emitting regions (see Table \ref{tab:ohemission}) using
\begin{equation}
    \mathrm{FWHM_{src}} \lesssim 
    \left( \frac{S_{\mathrm{int}}}{S_{\mathrm{peak}}} \right)
    \times \mathrm{FWHM_{beam}},
\end{equation}
where $S_{\mathrm{int}}$ and $S_{\mathrm{peak}}$ are the integrated and peak flux densities measured from the moment--0 maps, and $\mathrm{FWHM_{beam}}$ is the synthesized beam size.

\subsection{Galaxies with Weak OH Emission and Strong Absorption}

\begin{figure*}
\centering

\begin{subfigure}[t]{0.45\textwidth}
\centering
\includegraphics[width=\textwidth]{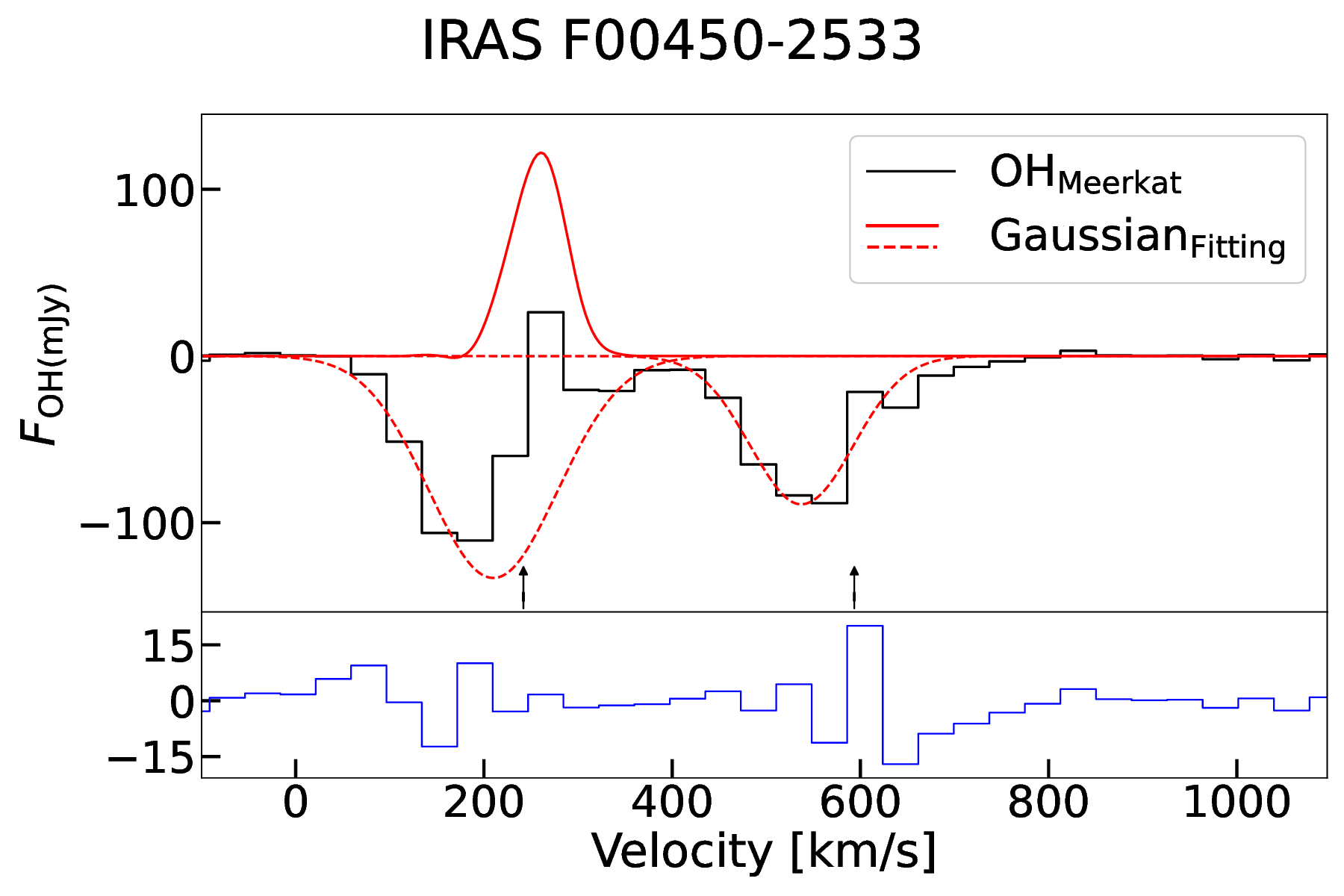}
\end{subfigure}
\hfill
\begin{subfigure}[t]{0.45\textwidth}
\centering
\includegraphics[width=\textwidth]{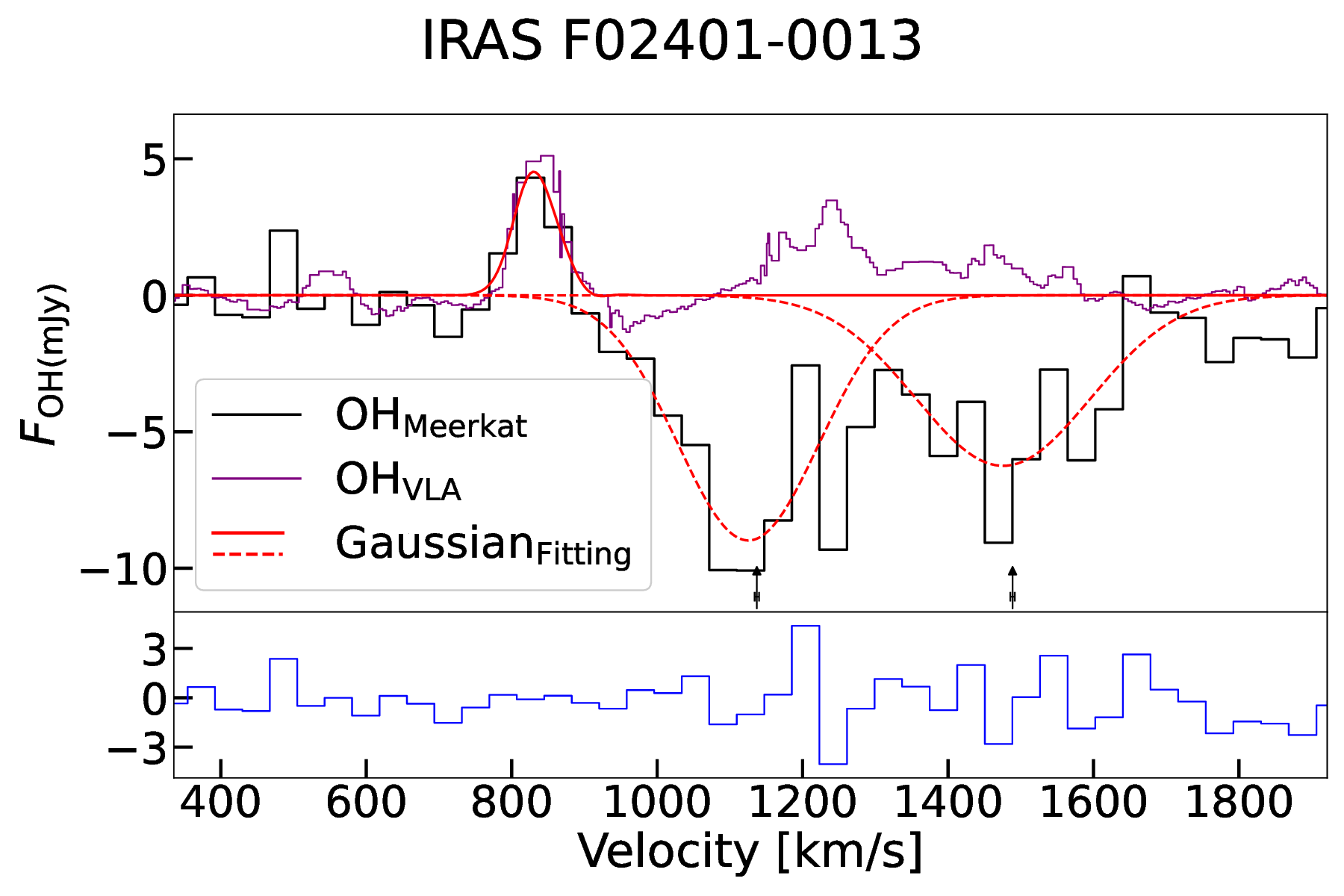}
\end{subfigure}

\vspace{0.6em}

\begin{subfigure}[t]{0.45\textwidth}
\centering
\includegraphics[width=\textwidth]{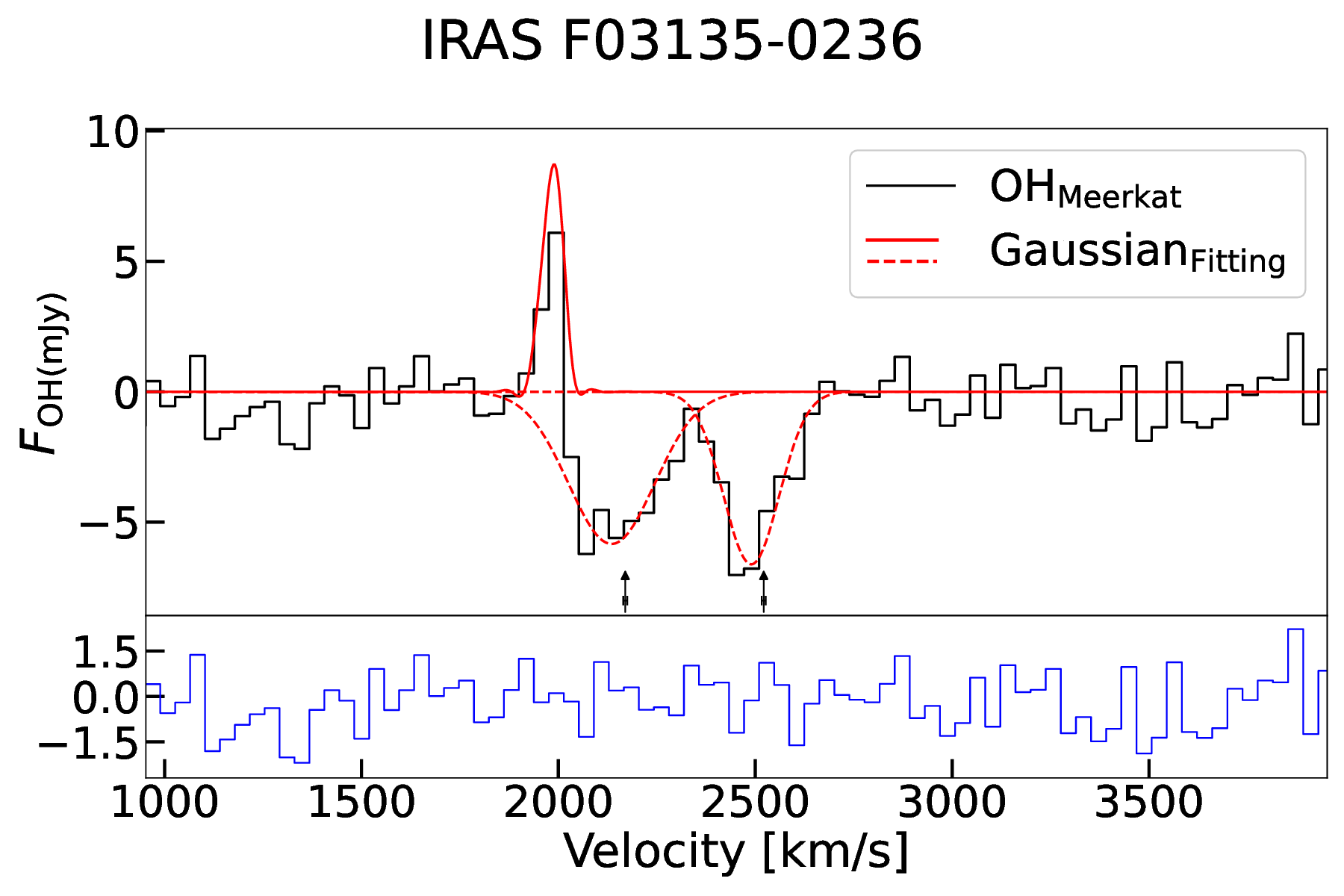}
\end{subfigure}
\hfill
\begin{subfigure}[t]{0.45\textwidth}
\centering
\includegraphics[width=\textwidth]{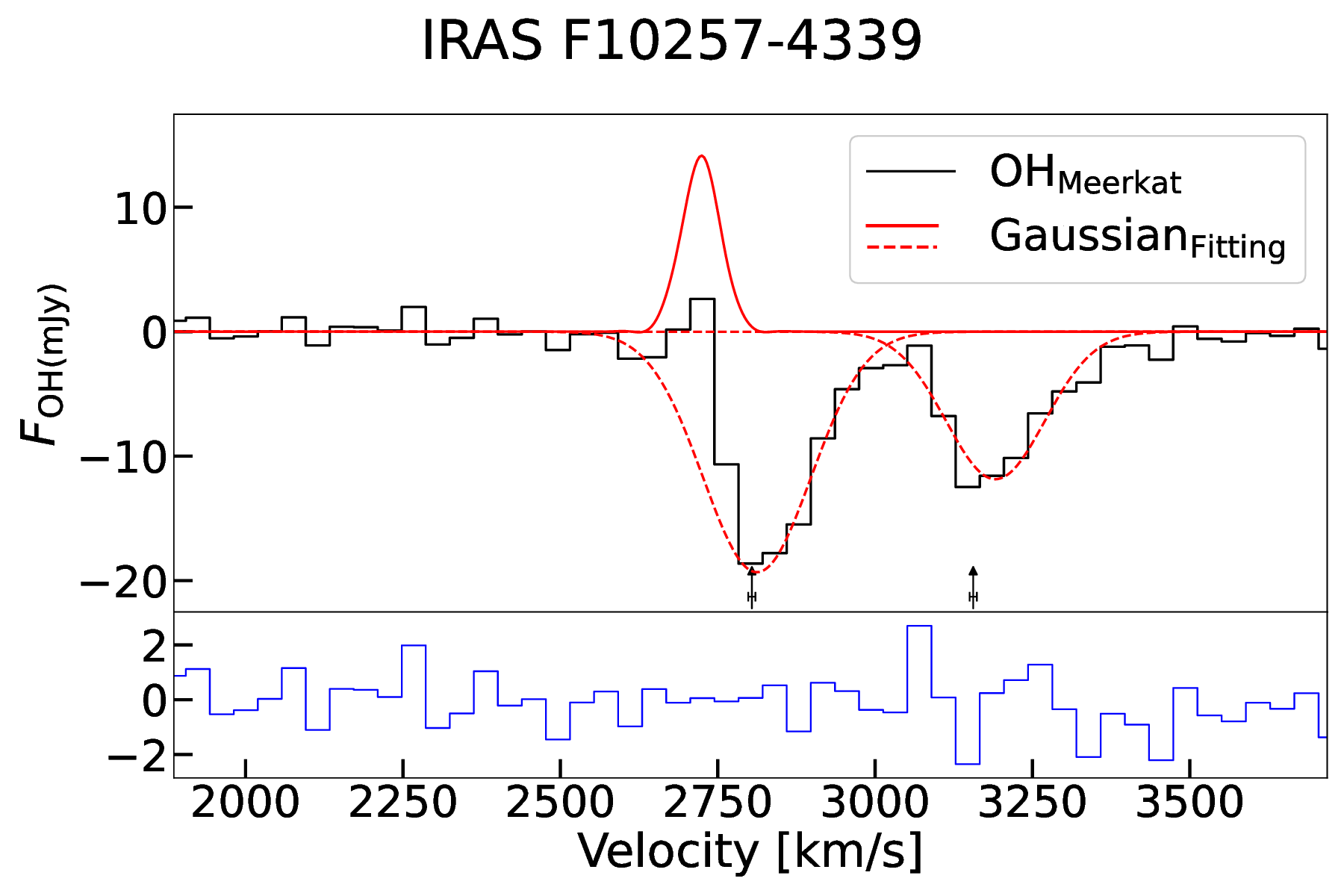}
\end{subfigure}

\caption{OH line profiles of the four galaxies showing weak OH emission accompanied by absorption. Velocities are given in the barycentric reference frame using the optical convention.
The black solid line shows the observed OH spectra.
The red solid and dotted lines indicate fitted Gaussian emission and absorption components.
The blue lines at the bottom represent residual spectra (data$-$model).
For IRAS~F02401$-$0013, the purple line shows literature OH spectrum \citep{1996ApJ...462..740G}.
Arrows mark the expected positions of the 1667~MHz (left) and 1665~MHz (right) OH main lines, with error bars indicating the uncertainty in the redshift.}

\label{oheands}

\end{figure*}

In addition to the eight galaxies with clear OH maser emission, we identify four sources that exhibit weak OH emission accompanied by strong absorption, where the absorption features appear partially filled by emission. Such systems may represent a transitional phase between pure OH absorption and OH maser emission \citep{1989ApJ...338..804B}. All four sources show high S/N OH absorption (Figs. \ref{oheands} and \ref{fig:radio-hst-1}), while the emission components are comparatively weak. Only IRAS~F00450$-$2533 exceeds our robust detection threshold ($>4.5\sigma$), whereas the remaining three sources reach only $\sim$3$\sigma$ and are therefore treated as candidates (see Fig. \ref{OH4image}). The OH spectra were extracted from regions of $\sim$20--30\arcsec\ centered on the brightest OH-emitting pixels (Table~\ref{tab:ohemission}). The spectra are presented in Fig.~\ref{oheands}, with Gaussian fit parameters listed in Table~\ref{tab:ohemission}. Although OH emission is visible in both the channel maps and spectra, the derived flux densities remain uncertain due to contamination from deep absorption features. 

IRAS~F00450$-$2533 (NGC~253) is a known OH emitter/absorber \citep{1974ApL....15..211W}. IRAS~F02401$-$0013 (NGC~1068) is a well-known OH maser galaxy identified in high-resolution VLA A-configuration observations, which reveal both a narrow, bright emission line at lower velocities and a broader, shallower emission component at higher velocities \citep[see Fig.~\ref{oheands}, and][]{1996ApJ...462..740G}. In our data, we detect a similar narrow, bright OH emission line centered at $\sim$830~\kms. However, the broad, shallow emission component reported in previous studies appears as a broad absorption feature in our spectrum, extending from the high-velocity end of the narrow line up to $\sim$2000~\kms\ (see Fig.~\ref{oheands}). We also find a small positional offset ($\sim$3\arcsec) between the narrow emission peak in our data and that reported by \cite{1996ApJ...462..740G}, likely due to differences in astrometry or angular resolution. Although the OH emission in this source does not exceed our formal $4.5\sigma$ threshold, the strong consistency with previous high-resolution observations supports its identification as an OH emission detection. Given the high S/N of the accompanying OH absorption, we classify this source as a confirmed OH emission+absorption system. IRAS~F03135$-$0236 (Mrk~1266) has previously been reported only in OH absorption \citep{1992AJ....103..728B}. IRAS~F10257$-$4339 (NGC~3256) is newly identified here as a candidate OH emitter/absorber. The remaining two sources require higher-resolution and higher-sensitivity observations to confirm the presence of OH emission.

\subsection{OH absorption lines}
\label{ohabssect}
We detect high-S/N OH absorption features in a total of 30 galaxies (see Fig.~\ref{fig:radio-hst-1}), along with six additional low-S/N candidates (S/N $\sim$ 3--4.5; OH absorption-line (moment–0) maps is available on Zenodo \zenodosite). The OH absorption is spatially associated with the radio continuum emission but is typically more compact than the continuum extent defined by the 3$\sigma$ contours. The fitted sizes of the absorption components are generally unresolved and consistent with the synthesized beam (see Table~\ref{tab:ohabsorption}). Among the detections, one source (IRAS~03316$-$3618) exhibits two spatially distinct OH absorption regions, separated by $\sim$18.3\arcsec\ ($\sim$1.9~kpc; see Table~\ref{tab:ohabsorption}). For this source, as well as for other galaxies with extended absorption, we re-extracted the spectra using regions that encompass the full absorption extent. The OH line profiles of the high-S/N sources are shown in Fig.~\ref{fig:absline1}, while those of the low-S/N candidates are presented on Zenodo \zenodosite. The derived spectral parameters are listed in Table~\ref{tab:ohabsorption}.

A literature search indicates that only six of these sources have been previously reported as OH absorption systems; thus, the majority of detections in our sample are new (see Table~\ref{tab:ohabsorption}).
Most galaxies in the sample are well described by two Gaussian components, likely corresponding to the OH main lines at 1667 and 1665~MHz. Five sources require three or more Gaussian components to adequately reproduce their spectra (see Fig. \ref{fig:absline1}). In these cases, we identify one component with the 1665~MHz line, while the remaining components are attributed to the 1667~MHz transition. The peak absorption depths span a wide range, from 0.2\% to 17\% (Table~\ref{tab:ohabsorption}). The distribution of the OH main-line ratios for the 30 OH absorption detections is presented on Zenodo \zenodosite. The mean and median integrated 1667/1665 ratios are similar, both $\sim$1.5, lower than the LTE optically thin thermal ratio of 1.8. Similar ratios have been found in OH absorbers \citep{1989ApJ...338..804B}, likely reflecting an intermediate optical-depth regime of the absorbing gas.

\begin{figure*}
\centering
\begin{subfigure}[t]{0.32\textwidth}
\centering
\includegraphics[width=\textwidth]{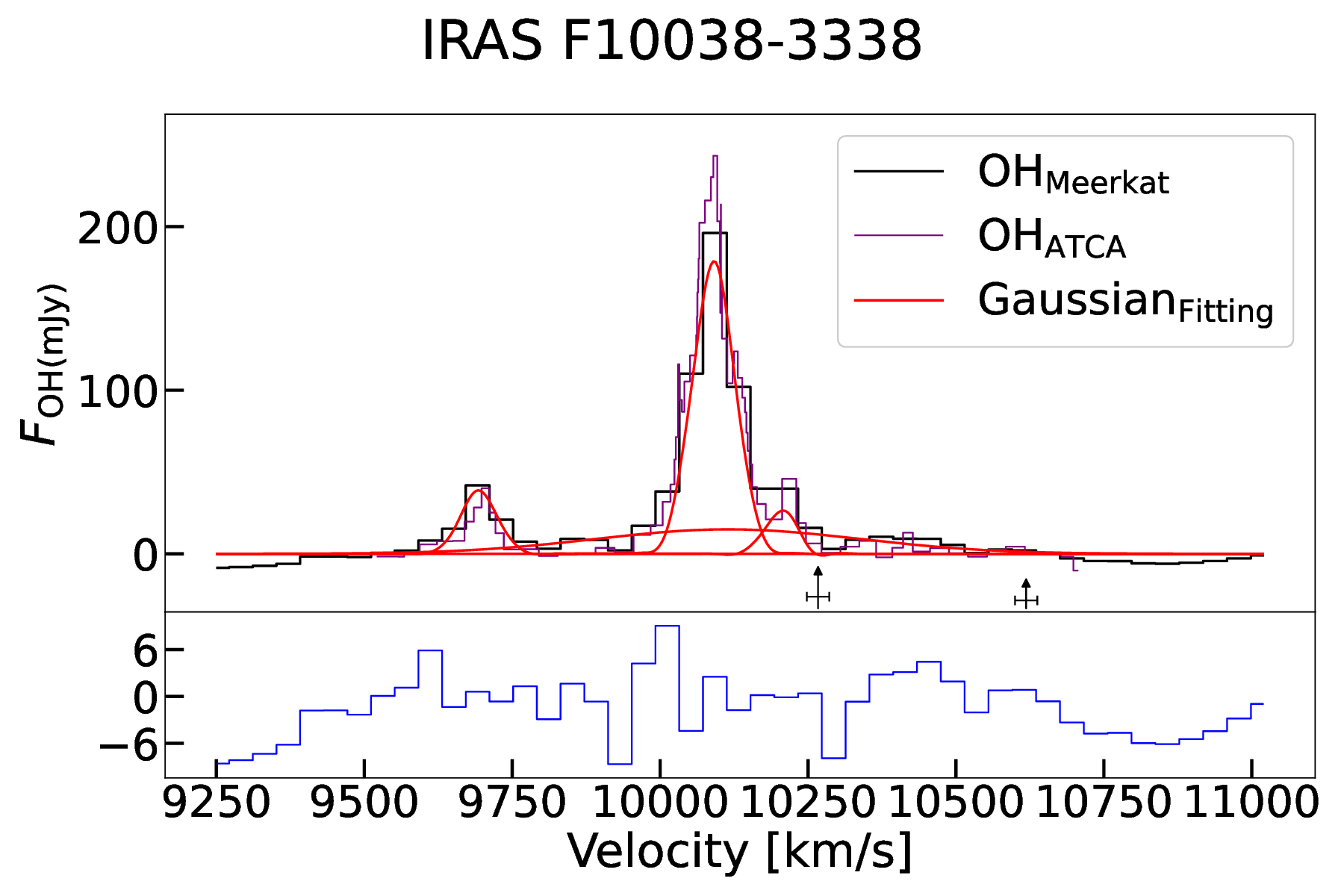}
\end{subfigure}
\hfill
\begin{subfigure}[t]{0.32\textwidth}
\centering
\includegraphics[width=\textwidth]{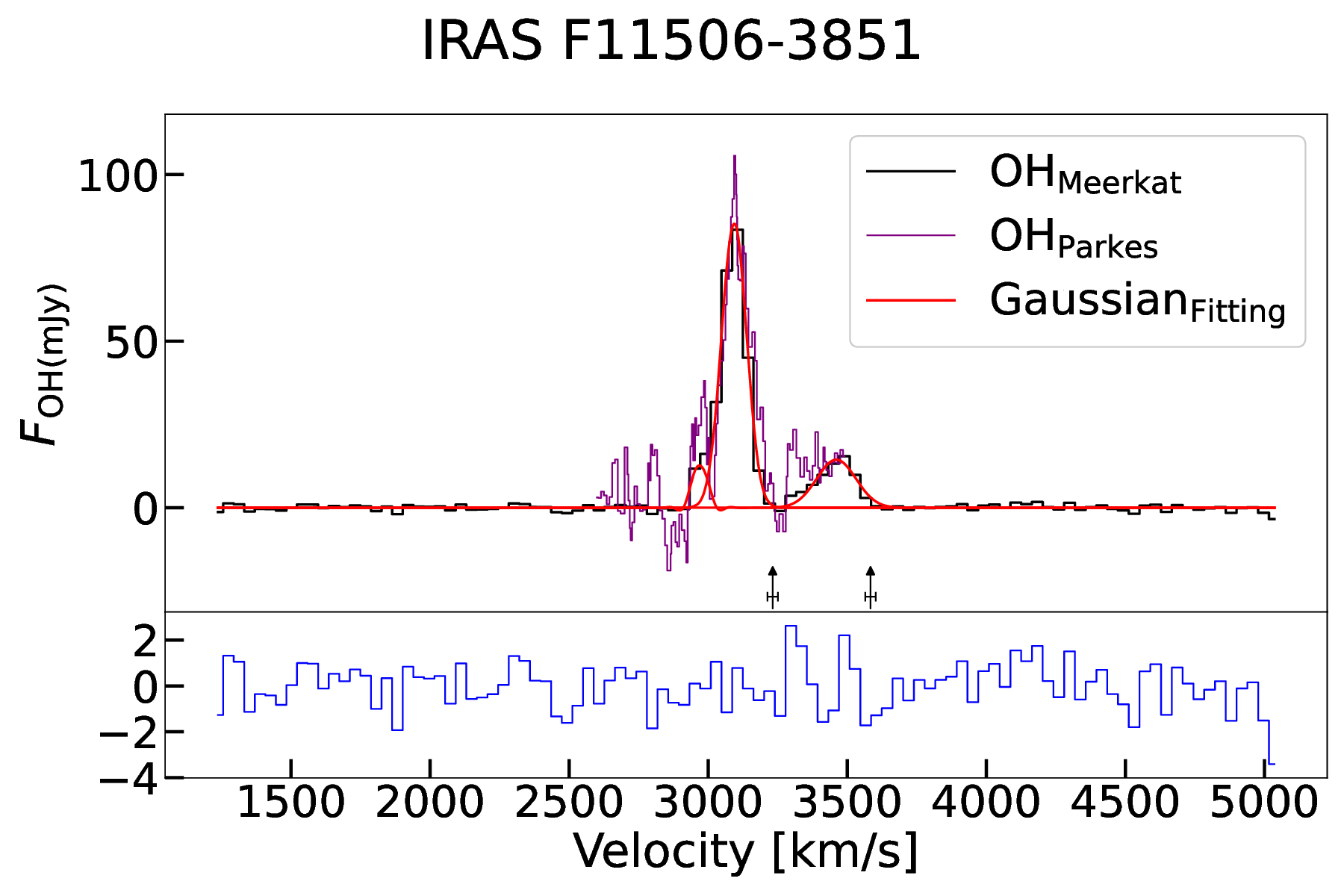}
\end{subfigure}
\hfill
\begin{subfigure}[t]{0.32\textwidth}
\centering
\includegraphics[width=\textwidth]{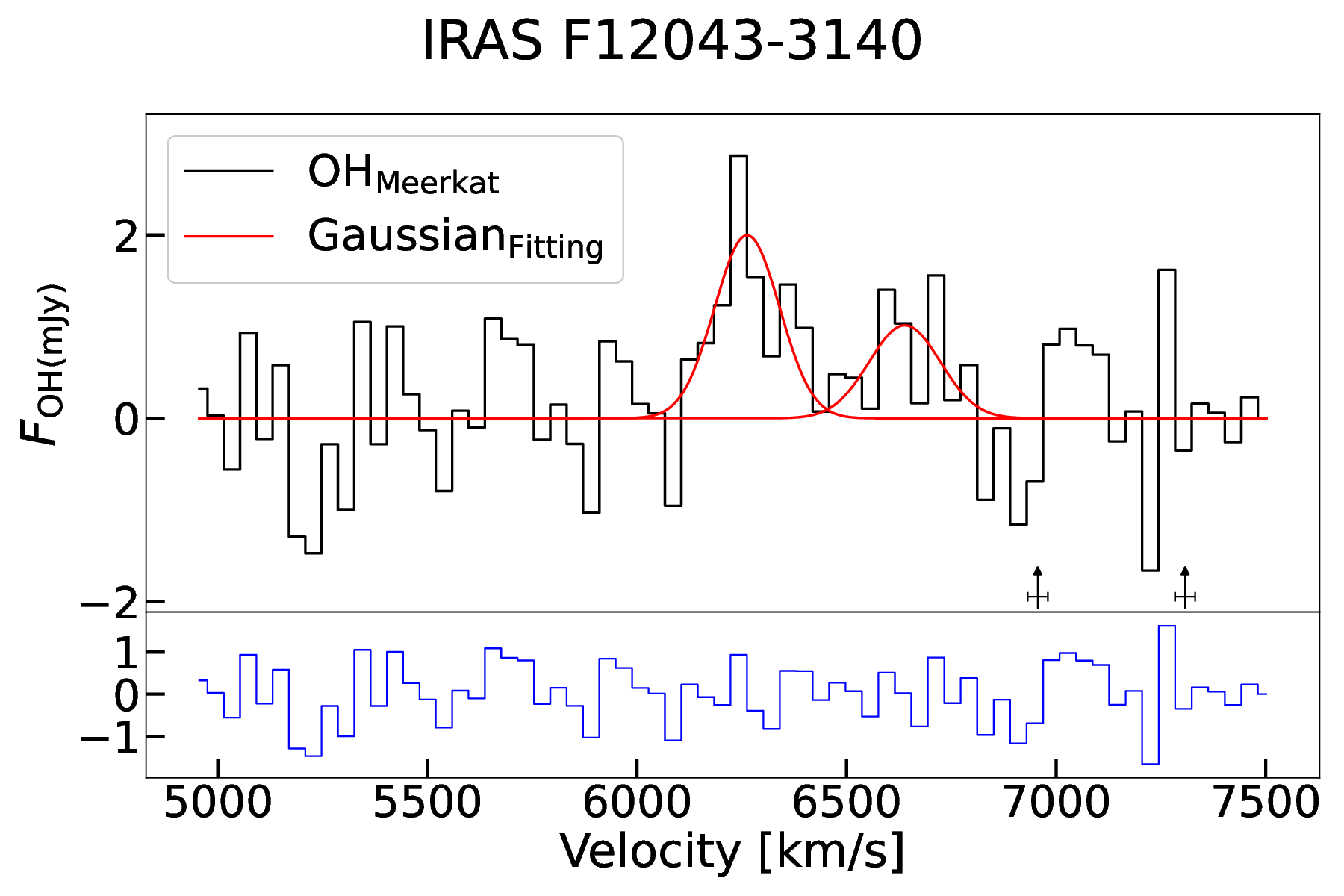}
\end{subfigure}

\vspace{0.6em}

\begin{subfigure}[t]{0.32\textwidth}
\centering
\includegraphics[width=\textwidth]{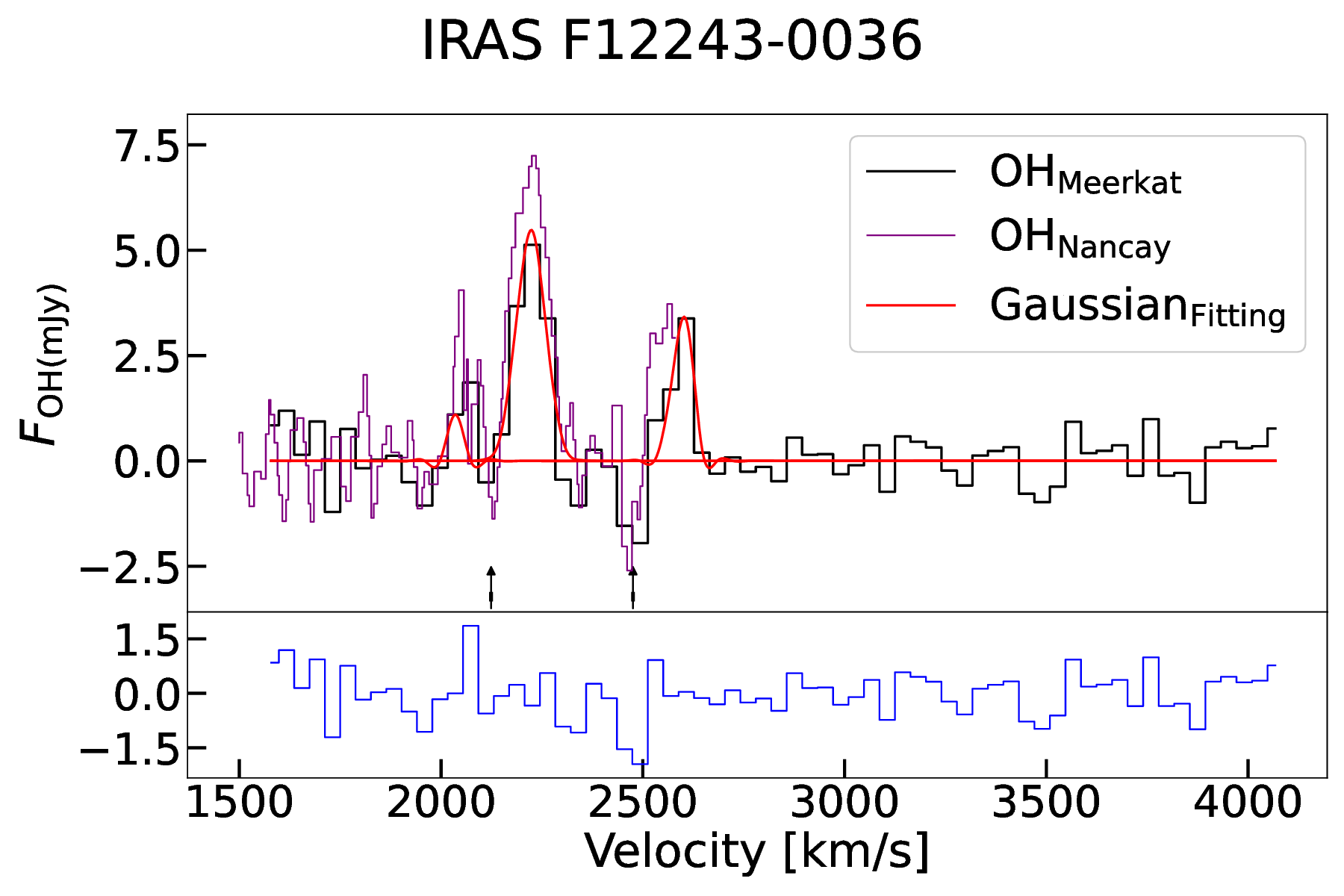}
\end{subfigure}
\hfill
\begin{subfigure}[t]{0.32\textwidth}
\centering
\includegraphics[width=\textwidth]{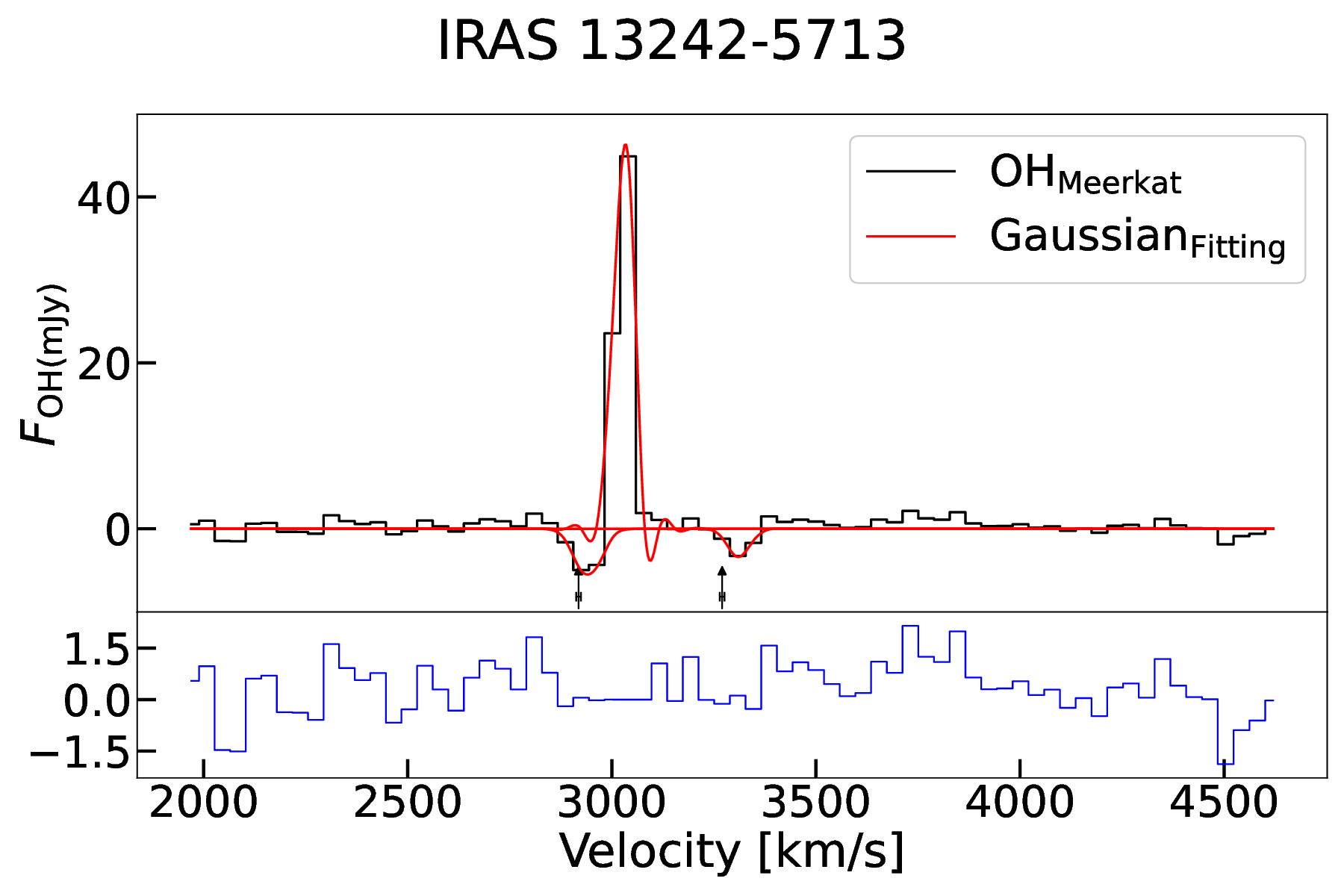}
\end{subfigure}
\hfill
\begin{subfigure}[t]{0.32\textwidth}
\centering
\includegraphics[width=\textwidth]{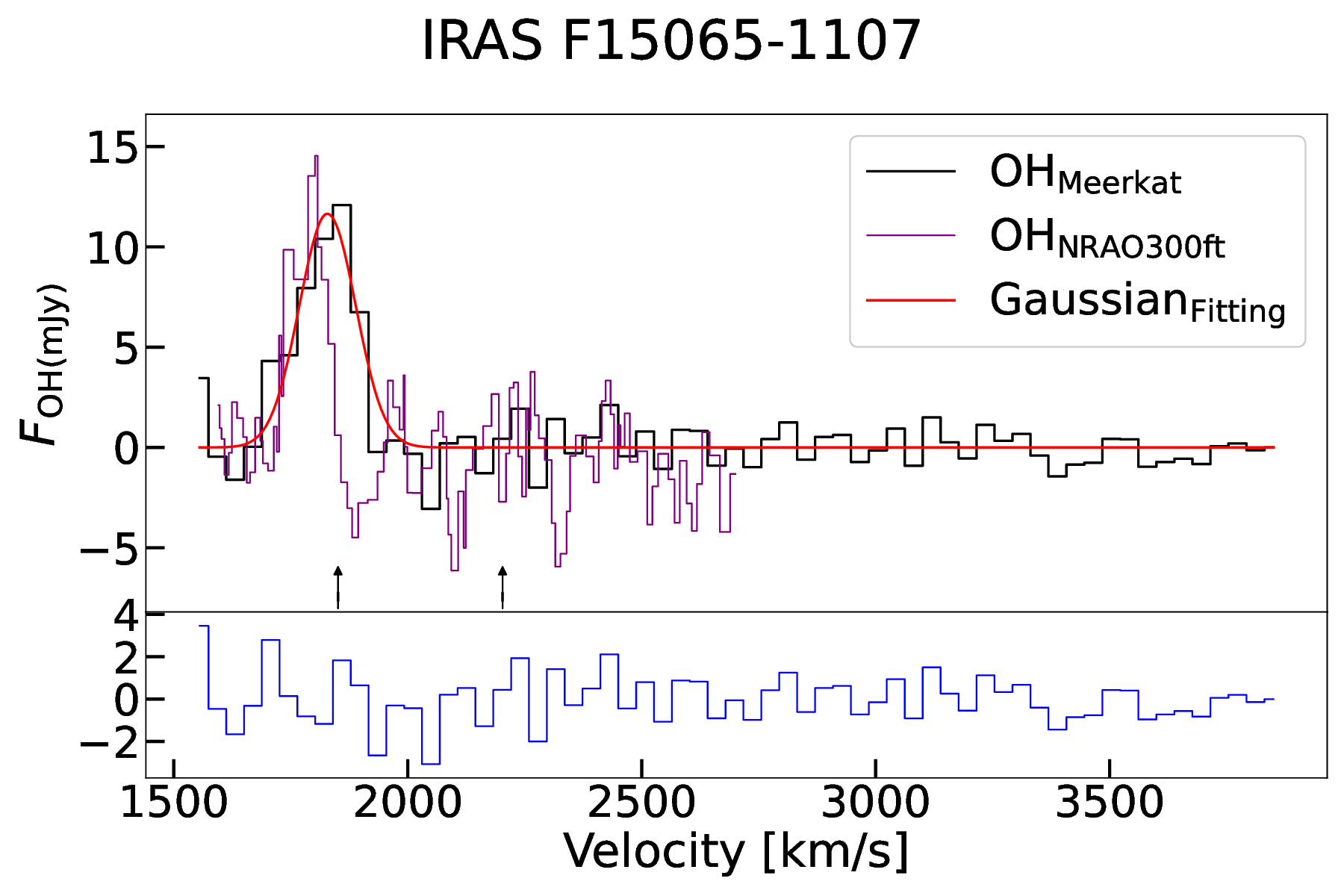}
\end{subfigure}

\vspace{0.6em}

\begin{center}
\begin{subfigure}[t]{0.32\textwidth}
\centering
\includegraphics[width=\textwidth]{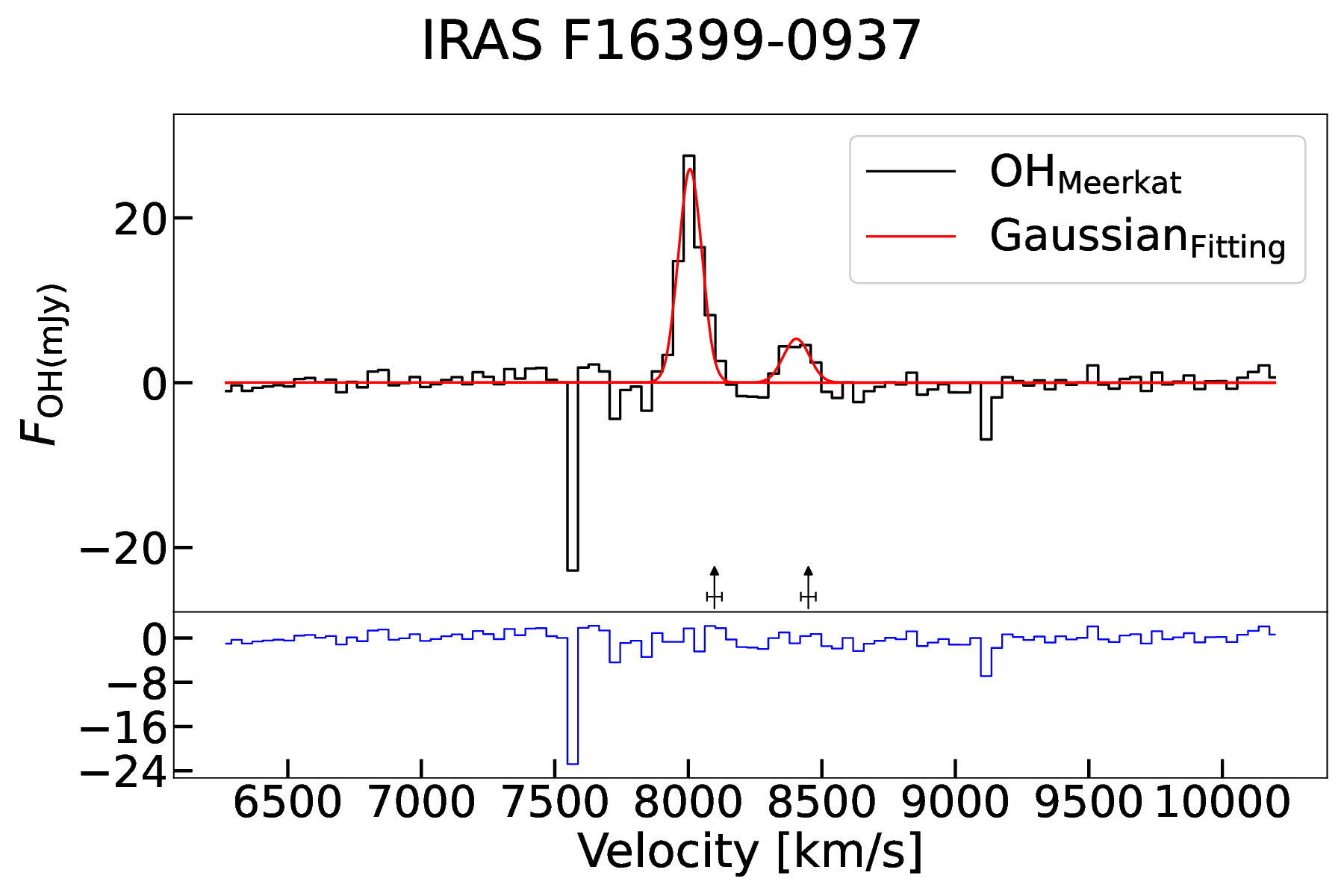}
\end{subfigure}
\hspace{0.05\textwidth}
\begin{subfigure}[t]{0.32\textwidth}
\centering
\includegraphics[width=\textwidth]{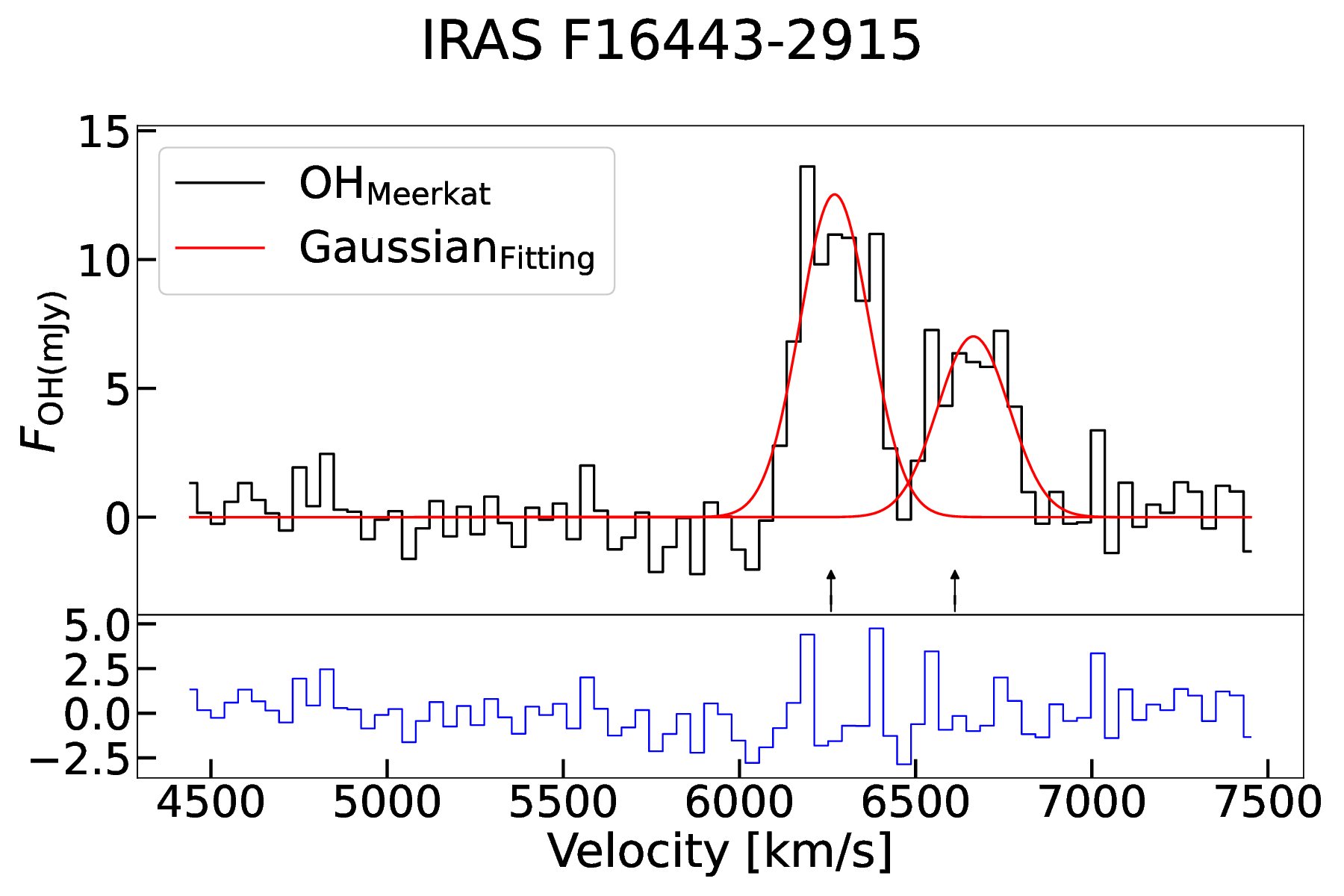}
\end{subfigure}
\end{center}

\caption{OH line profiles of the eight detected galaxies. Velocities are given in the barycentric reference frame using the optical convention. The solid black line shows the OH spectra detected in this work, while the red line indicates the Gaussian fits to the spectra. The blue line represents the residual spectra (data$-$model). The purple line shows OH spectra available in the literature for the following sources: IRAS~F10038$-$3338 \citep{1996MNRAS.280.1143K}, IRAS~F11506$-$3851 \citep{1992MNRAS.258..725S}, IRAS~F12243$-$0036 \citep{1988A&A...201L..13M}, and IRAS~F15065$-$1107 \citep{1992AJ....103..728B}. Arrows indicate the expected positions of the 1667~MHz (left) and 1665~MHz (right) OH main lines, with error bars indicating the uncertainty in the redshift.}

\label{fig:oh_spectra_centered}

\end{figure*}

\begin{figure*}
\centering

\begin{subfigure}[t]{0.32\textwidth}
\centering
\includegraphics[width=\textwidth]{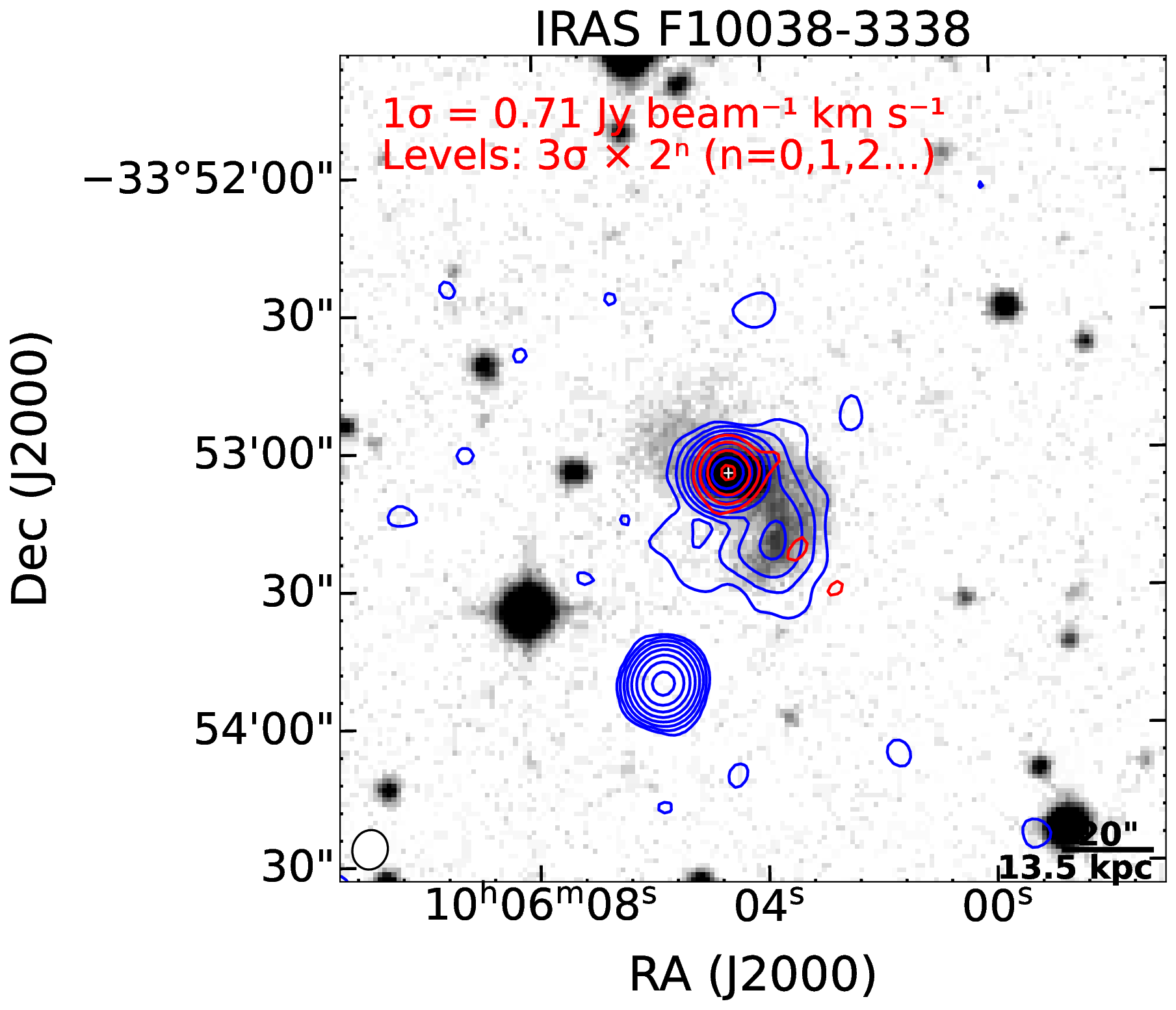}
\end{subfigure}
\hfill
\begin{subfigure}[t]{0.32\textwidth}
\centering
\includegraphics[width=\textwidth]{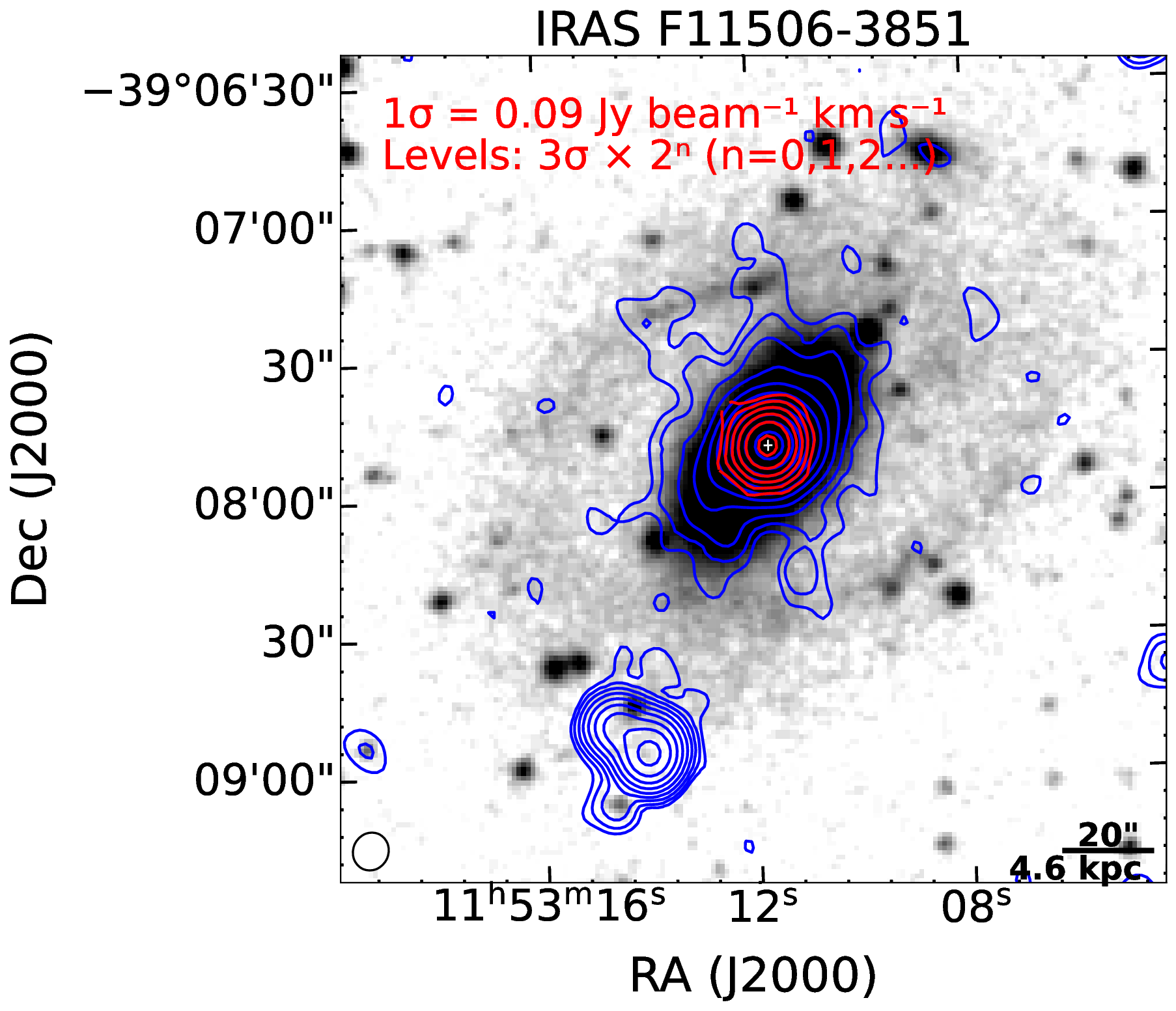}
\end{subfigure}
\hfill
\begin{subfigure}[t]{0.32\textwidth}
\centering
\includegraphics[width=\textwidth]{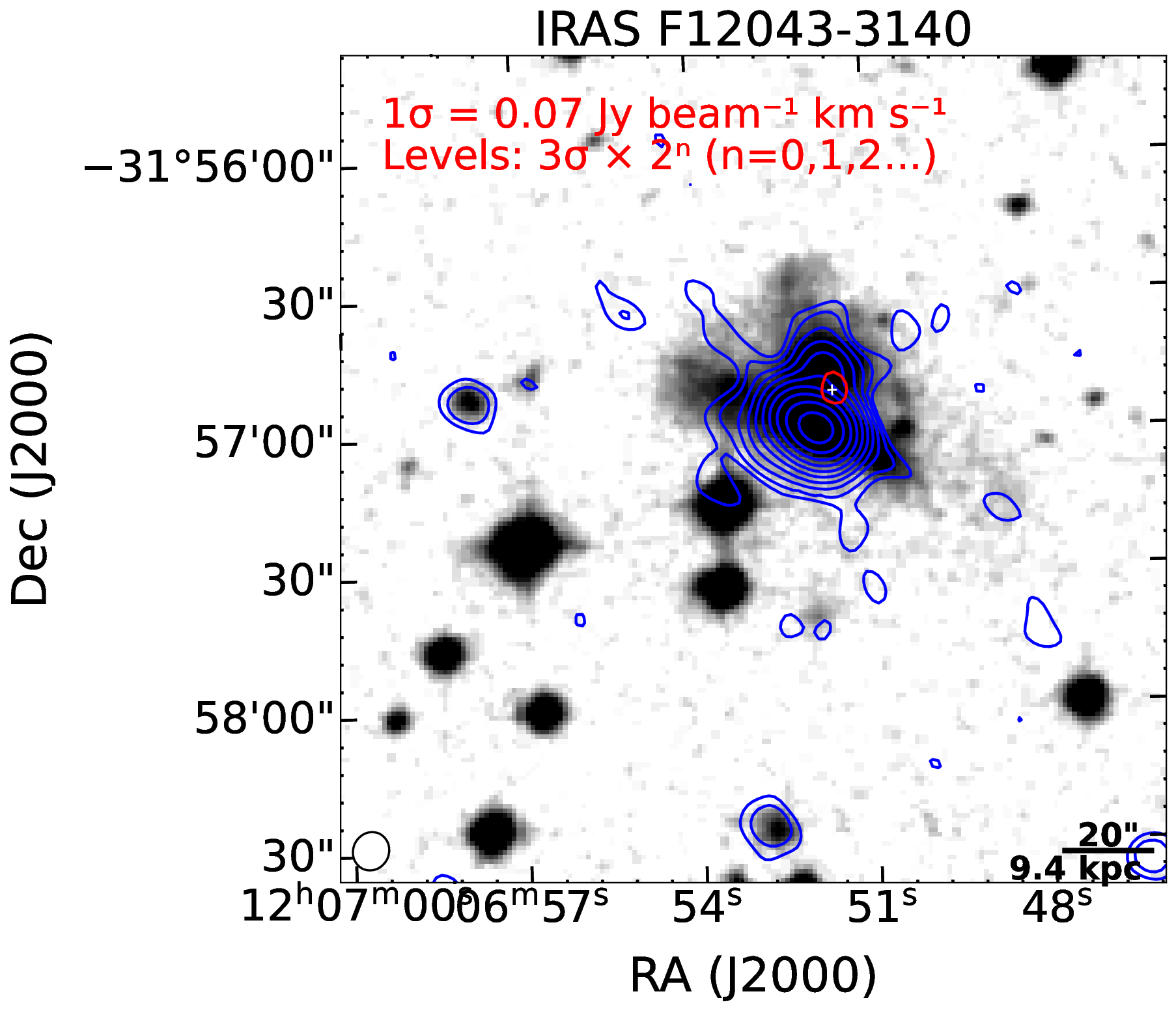}
\end{subfigure}

\vspace{0.5em}

\begin{subfigure}[t]{0.32\textwidth}
\centering
\includegraphics[width=\textwidth]{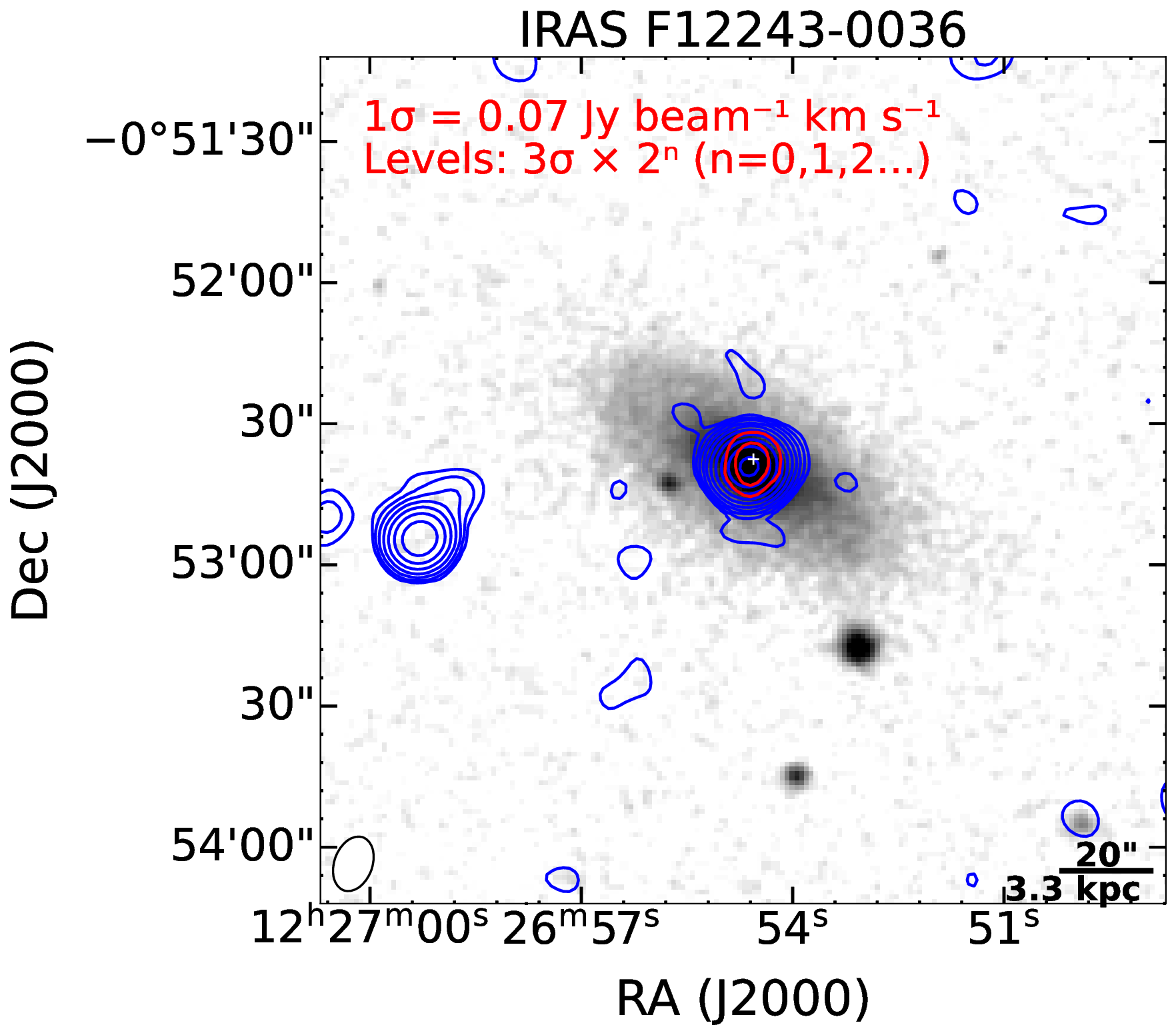}
\end{subfigure}
\hfill
\begin{subfigure}[t]{0.32\textwidth}
\centering
\includegraphics[width=\textwidth]{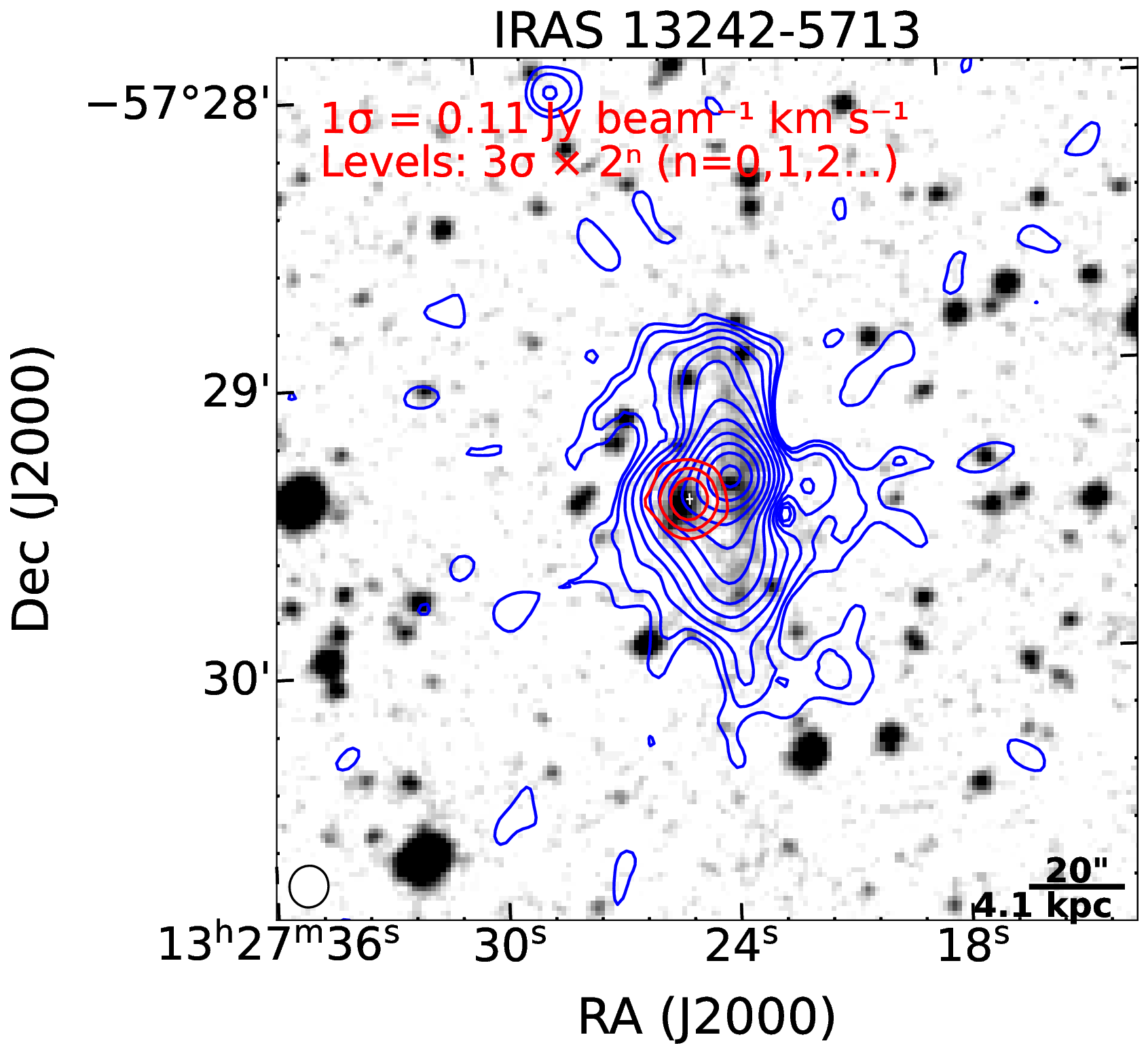}
\end{subfigure}
\hfill
\begin{subfigure}[t]{0.32\textwidth}
\centering
\includegraphics[width=\textwidth]{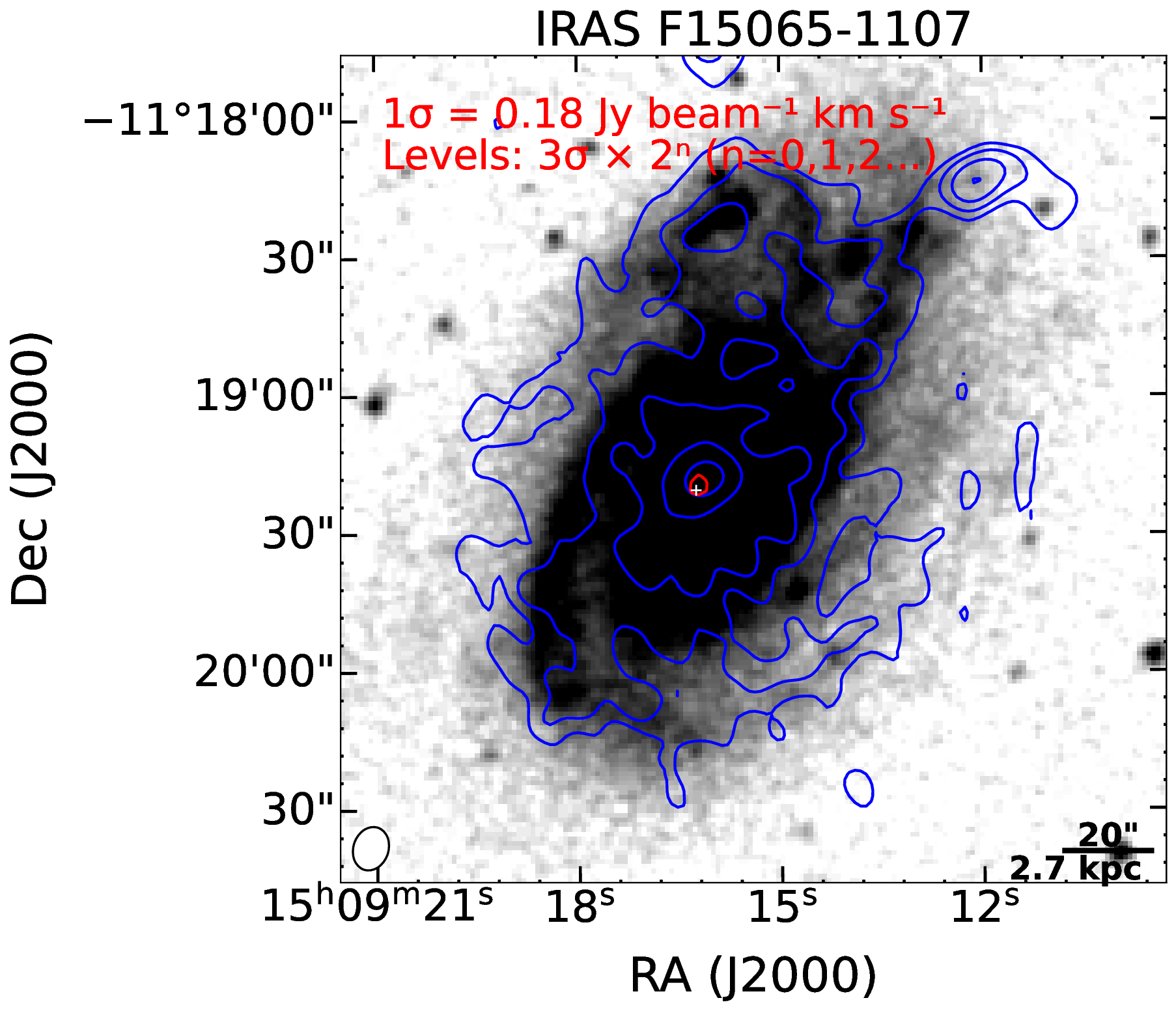}
\end{subfigure}

\vspace{0.5em}

\begin{center}
\begin{subfigure}[t]{0.32\textwidth}
\centering
\includegraphics[width=\textwidth]{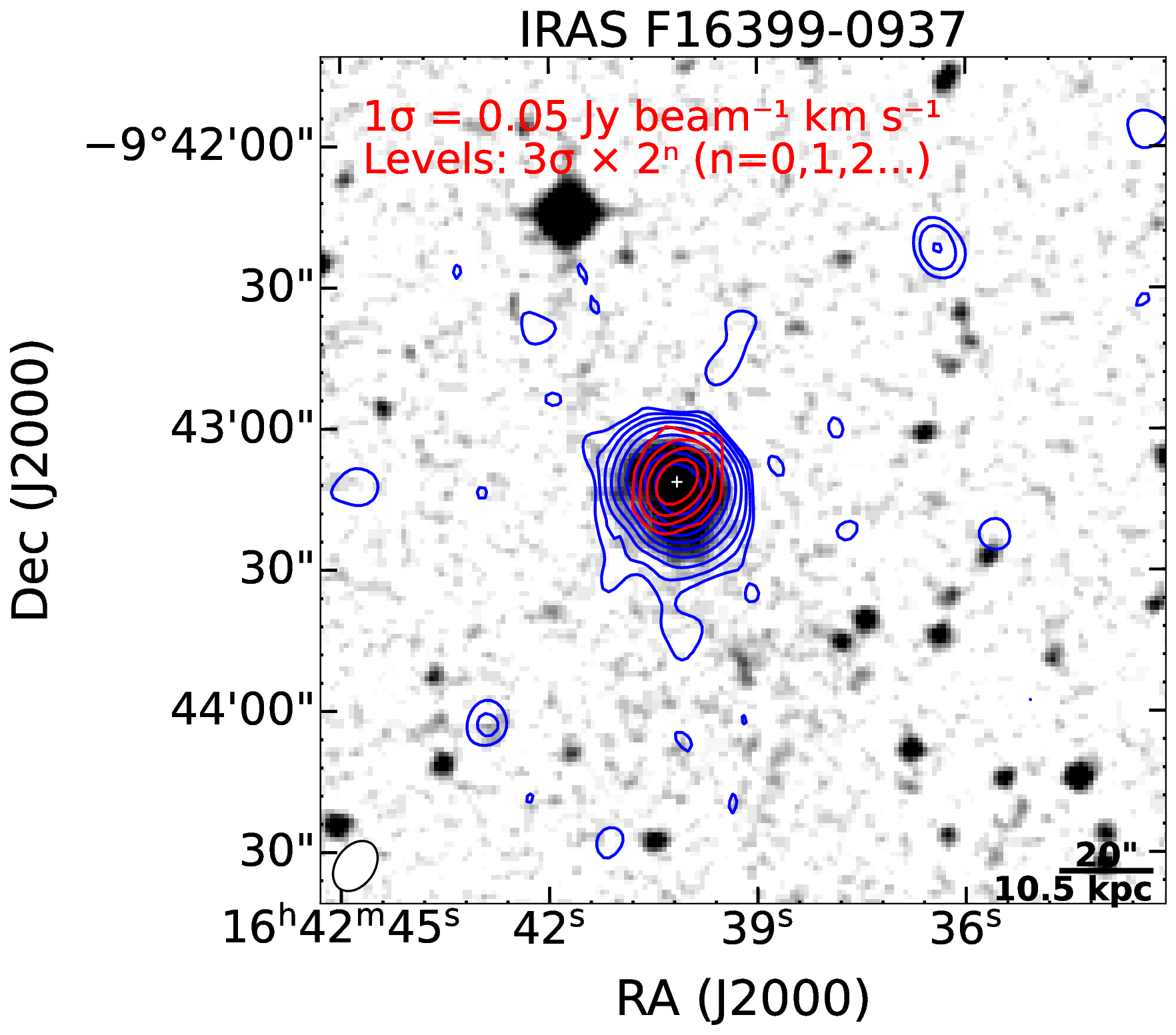}
\end{subfigure}
\hspace{0.05\textwidth}
\begin{subfigure}[t]{0.32\textwidth}
\centering
\includegraphics[width=\textwidth]{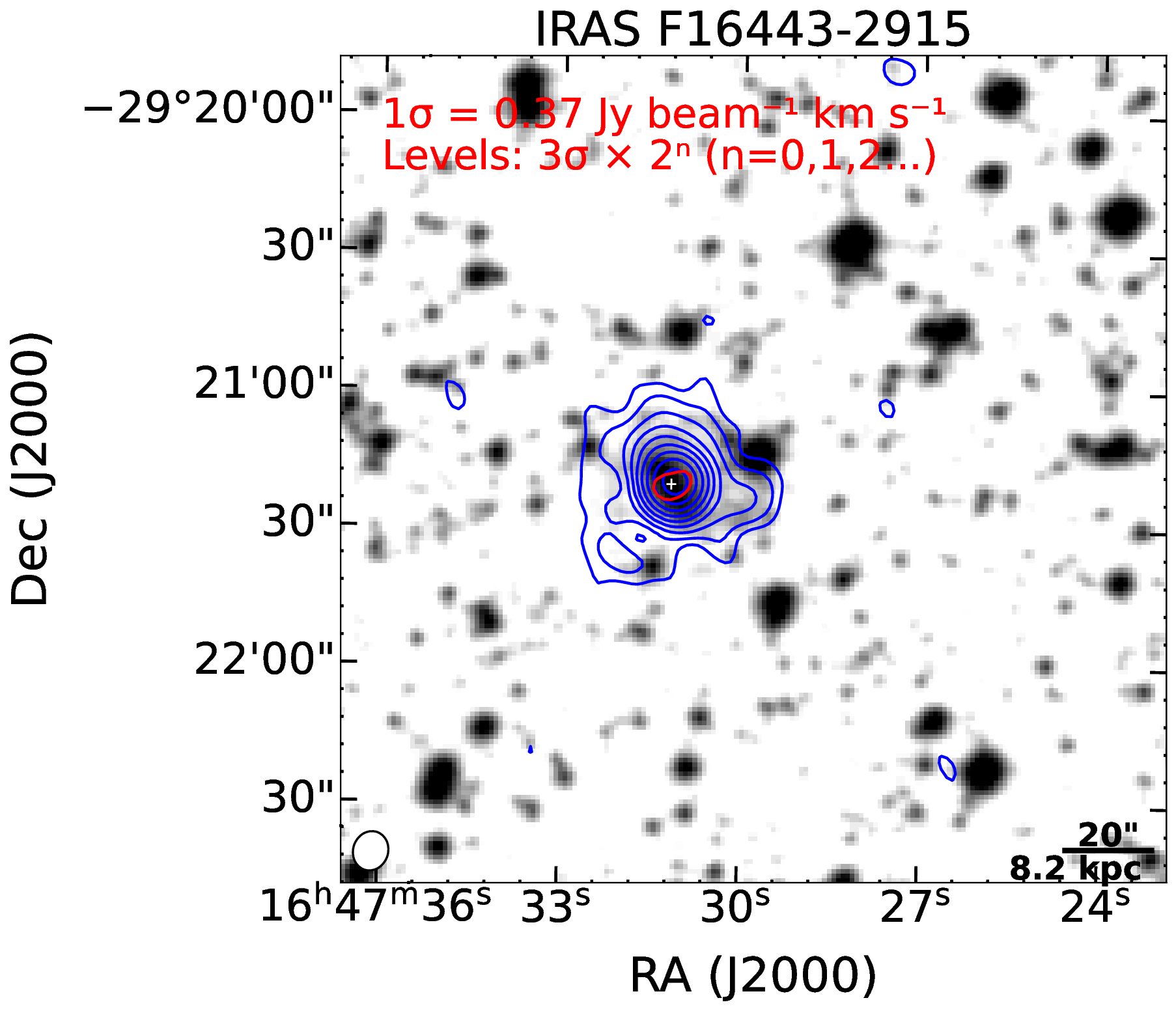}
\end{subfigure}
\end{center}

\caption{
Combined OH emission (moment~0) contour maps of eight OH maser galaxies,
overlaid on DSS $R$-band grayscale images.
Red contours show the OH emission detected in this work.
The $1\sigma$ rms noise levels and and contour levels of the
OH maps are indicated in the upper left corner of each panel.
Blue contours indicate the radio continuum emission from the full-band MeerKAT
data \citep{2021ApJS..257...35C}, with a typical $1\sigma$ rms noise level of
$\sim$0.02~mJy~beam$^{-1}$.
For all sources, contours start at $3\sigma$ and increase by factors of two.
White crosses mark the OH emission peaks.
Synthesized beam sizes and fitted OH source sizes are listed in
Table~\ref{tab:ohemission}.
}
\label{OH8image}

\end{figure*}

\section{Discussions}

\subsection{OH detection statistics of our IR-bright galaxy sample}
\label{sect:statis}
OH maser emission has been studied for more than four decades; however,
approximately $\sim$80\% of (U)LIRGs do not exhibit masing activity, resulting in
overall low detection rates of OH megamasers
\citep[see, e.g.,][]{2025ApJ...986...70R}. Nevertheless, the OH megamaser detection
rate is known to increase with infrared luminosity
\citep{2005ARA&A..43..625L}. The infrared luminosities of our IR-bright galaxy
sample span a range of $\log (L_{\rm IR}/L_\odot)$ from 9.1 to 11.8. Following common practice,
IR-bright galaxies are classified into LIRGs with
$\log (L_{\rm IR}/L_\odot) > 11$ and non-LIRG IR-bright galaxies with
$\log (L_{\rm IR}/L_\odot) < 11$ \citep{1996ARA&A..34..749S}. We therefore present
our statistical analysis based on this classification.

\subsubsection{LIRGs ($\log L_{\rm IR}/L_\odot \geq 11$)}

Based on the infrared luminosities, 62 galaxies in our sample are classified as LIRGs. Among these, we detect OH emission in 8 sources, including 7 galaxies with clear OH emission and 1 systems showing weak OH emission accompanied by strong absorption (Table~\ref{tab:ohemission}). The OH emission in two complex systems (IRAS F03135-0236 and IRAS F10257-4339, see Fig. \ref{oheands}) are classified as candidates. In addition, three LIRGs with $\log (L_{\rm IR}/L_\odot) > 11$ (IRAS~00085$-$1223, 02071$-$1023, 17222$-$5953, see parameters of OH non-detection availabe on Zenodo: (\href{https://doi.org/10.5281/zenodo.19904867}{doi:10.5281/zenodo.19904867}) ) exhibit tentative OH emission features ($\sim$$2\sigma$). Higher-sensitivity observations are required to confirm these candidates. Accounting for the candidates and these tentative detections, the OH emission detection rate among LIRGs is approximately $\sim$13--19\%. Because only about five of the OH maser galaxies have Log $L_{OH}/L_\odot >1$, thus, the detection rate for OH megamaser galaxies in our sample is only about 8 \%.

Among the 62 LIRGs in our sample, OH absorption is detected in 19 galaxies (Table~\ref{tab:ohabsorption}), including 17 secure detections and 2 candidates. However, only 47 of these LIRGs have peak radio continuum flux densities at 1.28~GHz (from \citealt{2021ApJS..257...35C}) above 20~mJy. This value can be regarded as an approximate lower limit for achieving a $\sim$3$\sigma$ OH absorption detection under typical conditions. The estimate is based on adopting a representative peak optical depth of $\tau_{\rm peak} \sim 0.05$ (Fig.~\ref{fig:stacked-tau}) and a characteristic rms noise level of $\sim0.4$~mJy~beam$^{-1}$ inferred from the non-detections (availabe on Zenodo: (\href{https://doi.org/10.5281/zenodo.19904867}{doi:10.5281/zenodo.19904867})), although both quantities may vary from source to source.
Restricting the analysis to these continuum-bright systems, the implied OH absorption detection rate among LIRGs is therefore $\sim$36--40\%, with the exact value depending on how candidate detections are treated.

\subsubsection{Non-LIRG IR-bright galaxies ($\log L_{\rm IR}/L_\odot < 11$)}

The remaining 124 galaxies in our sample are IR-bright but do not satisfy the
LIRG criterion. Among these systems, we detect OH emission in 2 sources,
including 1 galaxy with weak OH emission and 1 source dominated by OH
emission (Table~\ref{tab:ohemission}).  And also a OH emission/absorption candidates. This corresponds to an OH emission detection rate of only
$\sim$1.6--2.4\%.

In contrast, OH absorption is detected in 17 non-LIRG galaxies, including 13 secure detections and 4 candidates (Table~\ref{tab:ohabsorption}). Among the 124 non-LIRGs in the sample, 48 sources have peak radio continuum flux densities at 1.28~GHz (from \citealt{2021ApJS..257...35C}) exceeding 20~mJy. As discussed in the above subsection, this threshold can be regarded as an approximate lower limit for enabling a $\sim$3$\sigma$ detection of OH absorption under typical conditions. Restricting the analysis to this continuum-bright subsample, the implied OH absorption detection rate among non-LIRG IR-bright galaxies is therefore $\sim$27--35\%, with the range reflecting the inclusion or exclusion of candidate detections.

\subsubsection{Comparing with the known OH emission and absorption detections in RBGS sample}
We cross-matched the RBGS sample compiled by \cite{2025MNRAS.543.2463M} with the catalog of known OH megamaser (OHM) galaxies from \cite{2002AJ....124..100D}. Based on this comparison, we find that $\sim$16 OH emission sources have been reported in the northern RBGS sample. In the southern RBGS sample, we detect 8 OH emission sources, including five previously known OHM galaxies, and identify an additional five OH emission systems reported in the literature. Including two high-S/N OH emission+absorption systems newly identified in this work, the total number of OH emission sources in the southern RBGS sample is $\sim$15, broadly comparable to that in the northern sample.

For OH absorption, we identify $\sim$15 confirmed detections in the northern RBGS sample from the literature, excluding sources without available OH absorption spectra. In comparison, we find $\sim$30 OH absorption systems in the southern RBGS sample, including $\sim$6 previously reported absorbers. This difference likely reflects, at least in part, the higher sensitivity of the MeerKAT observations, although differences in sample selection, data quality, analysis methods, and RFI conditions may also contribute. Overall, this comparison suggests that the current census of OH absorption in the RBGS—particularly in the northern sample—may be incomplete, and that additional sensitive spectral-line observations could reveal further systems.

However, radio frequency interference (RFI) in the frequency range 1554–1626 MHz reduces sensitivity to OH lines over the redshift range $z \sim 0.025$–0.073. This effect can impact the detectability of both OH emission and absorption features. As a result, the current census of OH detections in the RBGS may be incomplete, and the true number of OH-emitting and absorbing galaxies could be somewhat higher.

\subsection{Physical Parameters Affecting OH Line Emission Detections}

In this work, we report only three newly identified OH maser emission sources (see Table \ref{tab:ohemission}); however, our investigations provide a large and uniformly selected sample of OH non-detections obtained with high sensitivity, allowing us to place robust upper limits on the OH luminosity ($L_{\rm OH}$) of these galaxies. Despite more than four decades of observational effort, the physical reasons behind the relatively low detection rate of OH line emission remain unclear \citep[see, e.g.,][]{2024AAS...24340607R}. Previous studies have suggested that several physical parameters—including far-infrared (FIR) luminosity, FIR color, and the amount of dense molecular gas—may play key roles in determining whether OH maser emission is present. In the following subsections, we examine the influence of these parameters using our sample.

\subsubsection{Correlation between $L_{\rm OH}$ and $L_{\rm FIR}$}

\begin{figure}[htbp]
\centering

\includegraphics[width=0.48\textwidth]{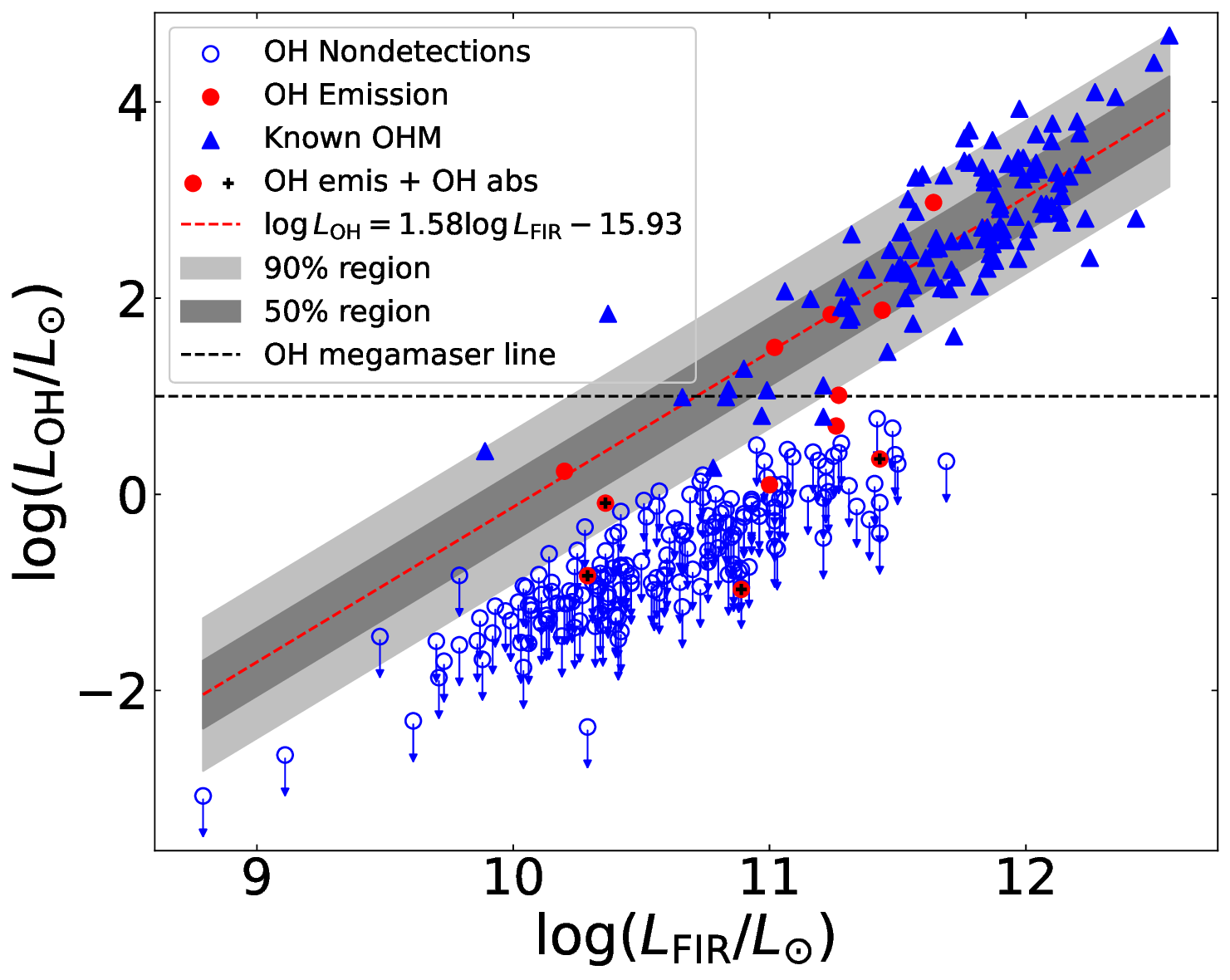}

\caption{Relation between OH and FIR luminosities for OH maser galaxies. The blue triangles denote 114 known OH megamasers compiled from the literature \citep{2024ApJ...971..131Z,2002AJ....124..100D,2006AJ....132.2596D,2012IAUS..287..345W,2016MNRAS.459..220S,2021A&A...647A.193H,2022ApJ...931L...7G,2024MNRAS.529.3484J}. The red filled circles, blue open circles, and plus symbols represent OH emission, non-detection, and absorption galaxies in this work, respectively. The five red filled circles overlaid with black plus symbols indicate the five galaxies showing both OH emission and absorption. The gray and dark-gray shaded regions enclose 50\% and 90\% of the known OH megamasers in our selected sample, respectively. Arrows mark upper limits on the OH luminosity. The horizontal line at $\log (L_{\rm OH}/L_\odot)=1$ indicates the typical threshold for OH megamaser emission.}

  \label{fig:firandoh}
  
\end{figure}

The correlation between OH and far-infrared (FIR) luminosities has been
extensively investigated in the literature
\citep[e.g.,][]{1996Ap.....39..237K,2002AJ....124..100D,2024ApJ...971..131Z},
where a tight empirical relation has been established primarily for samples of
known OH megamaser (OHM) galaxies.
Fig.~\ref{fig:firandoh} shows the distribution of $L_{\rm OH}$ versus
$L_{\rm FIR}$ for OHMs compiled from the literature, together with the OH
detections and non-detections from this work.
The eight OH emission sources identified in our sample broadly follow the
established $L_{\rm OH}$--$L_{\rm FIR}$ relation. Three of these sources lie
slightly below the 90\% confidence region defined by previously known OHMs
(Fig.~\ref{fig:firandoh}), but they appear to extend the distribution toward
lower FIR luminosities. This continuity may suggest that OH kilomasers in
lower-$L_{\rm FIR}$ galaxies are related to the same general class of
phenomena as classical OH megamasers, with their lower OH luminosities
potentially linked to weaker FIR radiation fields.

In contrast, four galaxies exhibiting weak OH emission (two detections and two
candidates; see Fig.~\ref{oheands}) superposed on strong absorption, as well as
systems showing only OH absorption or no OH detection, lie well below the
best-fit $L_{\rm OH}$--$L_{\rm FIR}$ relation. For absorption-dominated
systems, any intrinsic maser emission may be partially or significantly
suppressed by foreground absorbing gas, which could lead to an
underestimation of the observed OH luminosity. For OH non-detections, the absence of measurable maser
emission may instead reflect intrinsically weak or absent masing activity,
possibly indicating that the physical conditions required for efficient OH
amplification are not satisfied.

Overall, our results suggest that galaxies with relatively low FIR
luminosities tend to fall below the canonical $L_{\rm OH}$--$L_{\rm FIR}$
relation and may have a lower likelihood of hosting detectable OH maser
emission. This trend is broadly consistent with theoretical models in which
strong FIR radiation fields facilitate, but may not uniquely determine,
the radiative pumping of OH masers \citep{2008ApJ...677..985L}.

\subsubsection{The FIR color index, q-ratio, and dense gas properties of OH non-detections}

Using the FIR-to-radio flux density ratio ($q$) and the FIR spectral index ($\alpha_{\rm FIR}$; 25–60~$\mu$m), \citet{2022MNRAS.510.2495S} showed that OH megamasers (OHMs) tend to have higher $q$ values and steeper FIR spectra. In this work, we derive $\alpha_{\rm FIR}$ and adopt $q_{\rm TIR}$ from \citet{2025MNRAS.543.2463M}, based on total infrared luminosities. As shown in Fig.~\ref{fig:ohnon-prop}, OH non-detections with $\log L_{\rm IR} > 11$ broadly follow the parameter space reported in previous studies.
However, systematic differences are evident. OH-emitting galaxies span a wider range of infrared luminosities, with most at $\log L_{\rm IR} \gtrsim 11.5$, and cover a broad range in both $q_{\rm TIR}$ and $\alpha_{\rm FIR}$. In contrast, OH non-detections are more concentrated at $\log L_{\rm IR} \lesssim 11.5$ and tend to show lower $q_{\rm TIR}$ and flatter FIR spectra, although significant overlap remains.
At $\log L_{\rm IR} \lesssim 11.5$, the separation becomes more apparent: OH-emitting galaxies preferentially occupy higher $q_{\rm TIR}$ and steeper $\alpha_{\rm FIR}$, while non-detections cluster at lower $q_{\rm TIR}$ and flatter $\alpha_{\rm FIR}$. Although not a strict division, these parameters together provide a useful empirical discriminator in this regime. This trend may reflect differences in ISM conditions.

The role of dense molecular gas is further explored using the relation between
normalized OH maser luminosity and HCN luminosity presented by
\citet{2018JApA...39...34H}, with our sources highlighted in
Fig.~\ref{fig:ohnon-dense}. Both OH non-detections and OH absorption galaxies
occupy a region characterized by lower OH luminosities and reduced normalized
HCN luminosities compared to typical OH megamasers. However, the number of
galaxies with both OH and HCN measurements remains limited, both in our sample
and in the literature compilation. A larger and more complete dense-gas sample
is therefore required to robustly assess the role of dense molecular gas in
regulating OH maser activity.

\subsection{OH Absorption as a Probe of Cold Molecular Gas}

OH absorption traces foreground cold molecular gas against a bright radio
continuum background, and thus provides a sensitive probe of diffuse and
cold molecular material that may be difficult to detect in emission alone.
The high detection rate of OH absorption in our IR-bright galaxy sample
suggests that a substantial fraction of these systems host significant
reservoirs of cold molecular gas along the line of sight to the radio
continuum source.  As disscussed Section~\ref{sect:statis},  we find that galaxies with peak
continuum flux densities below $\sim$20~mJy~beam$^{-1}$ are generally not
well suited for reliable OH absorption detection at the sensitivity achieved
in this work. We adopt 20~mJy~beam$^{-1}$ as a conservative threshold for inclusion in the OH
absorption analysis. In addition, we exclude the eight OH maser emission galaxies and two OH emission candidates and three tenatative detections (see Section~\ref{sect:statis}). Applying these criteria yields a
final sample of 30 galaxies with detected OH absorption and 51 non-detections. 
We also compile $\sim$15 additional Northern RBGS sources from the literature with
reported OH absorption and published line profiles, resulting in a total
sample of 45 OH absorbers. Using these two subsamples (45 absorbers and 51
non-detections; available on Zenodo \zenodosite), we investigate which physical and
observational parameters most strongly affect the detectability of OH
absorption, as discussed in the following sections.

\subsubsection{Parameters Affecting OH Absorption}
Since both the 45 OH absorbers and 51 non-detections analyzed here are drawn from the RBGS, we adopt galaxy parameters from \citet{2025MNRAS.543.2463M}, which provides a homogeneous compilation for the full sample. We further quantify the compactness of the radio continuum emission using the compactness factor $C$, defined as
\begin{equation}
C = \frac{S_{\rm peak}}{S_{\rm int}},
\end{equation}
where $S_{\rm peak}$ and $S_{\rm int}$ are the peak and integrated flux densities measured from the MeerKAT data \citep{2021ApJS..257...35C}.
For the additional 15 OH absorbers from the northern RBGS, $S_{\rm peak}$ and $S_{\rm int}$ are primarily taken from the VLA FIRST survey, with a subset measured from archival VLA images available through the NRAO VLA Archive Survey (NVAS)\footnote{\url{http://www.vla.nrao.edu/astro/nvas/}}
. The corresponding angular resolutions ($\sim$4\arcsec–10\arcsec) are comparable to those of the MeerKAT and FIRST data, minimizing systematic uncertainties due to differing beam sizes.
Values of $C \sim 1$ indicate compact sources, while lower values correspond to more extended emission.

We further consider parameters related to neutral atomic gas and galaxy
structure. Following \citet{2020ApJ...898..102Y}, we adopt the \HI
concentration parameter $C_V = V_{85}/V_{25}$, where $V_{85}$ and $V_{25}$ are
the velocity widths enclosing 85\% and 25\% of the integrated \HI\ flux,
respectively. This parameter has been shown to be useful for distinguishing
merging from non-merging systems \citep{2022ApJ...929...15Z}, although some
overlap between populations is expected. \HI\ line profiles for 25 OH absorption galaxies and 27 non-detections were retrieved from the NASA/IPAC Extragalactic Database (NED), from which \HI
masses and $C_V$ values were derived. Galaxy inclination angles and axial
ratios ($b/a$) were taken from the HyperLeda database.

A two-sample Kolmogorov--Smirnov test was performed to compare the distributions of these parameters between OH absorbers and non-detections (see Table~\ref{tab:kstest}). We find that radio continuum flux densities and WISE fluxes (W1-4) are, on average, higher in OH absorption galaxies (see Table~\ref{tab:kstest} and data used to generate this table available on Zenodo \zenodosite). This likely reflects a selection effect, as stronger background radio continuum enhances the detectability of absorption with low optical depth.
In contrast, mid-infrared colors show more distinctive differences. In particular, the W2--W3 color is significantly lower in OH absorption galaxies than in non-detections, while no statistically significant differences are found for W1--W2 or W3--W4. The W2--W3 color is commonly used as a tracer of star formation activity, with larger values indicating stronger star formation \citep{2010AJ....140.1868W}. Since the W3 (12\,$\mu$m) emission is dominated by young stellar populations ($<0.6$ Gyr; \citep{2012ApJ...748...80D}), the elevated W2--W3 values in non-detections suggest that more intense star formation may suppress the detectability of OH absorption. Consistently, OH emission galaxies in the RBGS also tend to exhibit higher W2--W3 colors (see Fig \ref{fig:ohabs-prop}), indicating that the physical environments of OH emitters and absorbers are likely different. We also find that non-detections show slightly higher $q$-parameters (derived from both IRAS and WISE; see Table~\ref{tab:kstest}), which are commonly interpreted as indicators of star formation dominance \citep[e.g.,][]{2006A&A...449..559B}. Lower $q$ values correspond to increasing AGN contributions, while higher values indicate stronger star formation. Taken together, these results suggest that extreme star formation may reduce the likelihood of detecting OH absorption.

No statistically significant differences are found for infrared luminosity, \HI\ mass, \HI\ concentration parameter, or inclination-related quantities. Although previous \HI\ absorption studies have reported a dependence on radio source compactness, with higher detection rates toward compact sources \citep{2017A&A...604A..43M}, we find no significant difference in the compactness factor $C$ between OH absorbers and non-detections. This may indicate that the OH absorbing gas has a relatively high covering factor against the background continuum, such that orientation effects are not the dominant factor governing OH absorption detectability in our sample.

\subsubsection{Stacking of the OH Absorption Line Spectra}

Beyond orientation effects, \citet{2017A&A...604A..43M} showed through spectral stacking that \HI\ absorption remains undetected in samples of individual non-detections, implying that intrinsic differences are required to explain its presence or absence. Motivated by this result, we perform a similar stacking analysis of our OH spectra to test whether galaxies classified as non-detections may host absorption below the sensitivity limits of individual observations. 

Given that OH emission is much less frequently detected than OH absorption in our sample, absorption is expected to dominate any stacked signal. We therefore construct stacked spectra for two subsamples: 45 galaxies with detected OH absorption and 51 OH non-detections (available on Zenodo \zenodosite). For the 15 northern RBGS OH absorbers, we digitized the published OH line profiles from the following references: \cite{1985ApJ...293..394B,1982ApJ...252..147R,1986AJ.....92.1291S,1976A&A....52..467N,1985Natur.314..144H,1992AJ....103..728B,2010AJ....139.2066F}. The stacked mean optical-depth spectrum is computed following \citet{2011MNRAS.411..993F}:
\begin{align}
\tau_{\mathrm{stack}} &= 
\frac{\sum_{i=1}^{N} \tau_i \, w_i}{\sum_{i=1}^{N} w_i}, \\
w_i &= \frac{1}{\mathrm{rms}_i^{2}},
\end{align}
where $\tau_i$ is the optical-depth spectrum of the $i$-th source, defined as
\begin{equation}
\tau(\nu) = -\ln\!\left(1 - \frac{S(\nu)}{S_{\rm cont}} \right),
\end{equation}
and $w_i$ is the inverse-variance weight based on the rms noise of each spectrum. The continuum flux density $S_{\rm cont}$ is taken as the peak flux density measured from the MeerKAT continuum images \citep{2021ApJS..257...35C}. Using the same weight of each spectrum, the stacke median optical-depth spectrum was generated by calculating the weighted median at each velocity channel across all individual spectra.

\begin{figure}
\centering

\begin{subfigure}[t]{0.48\textwidth}
\centering
\includegraphics[width=\textwidth]{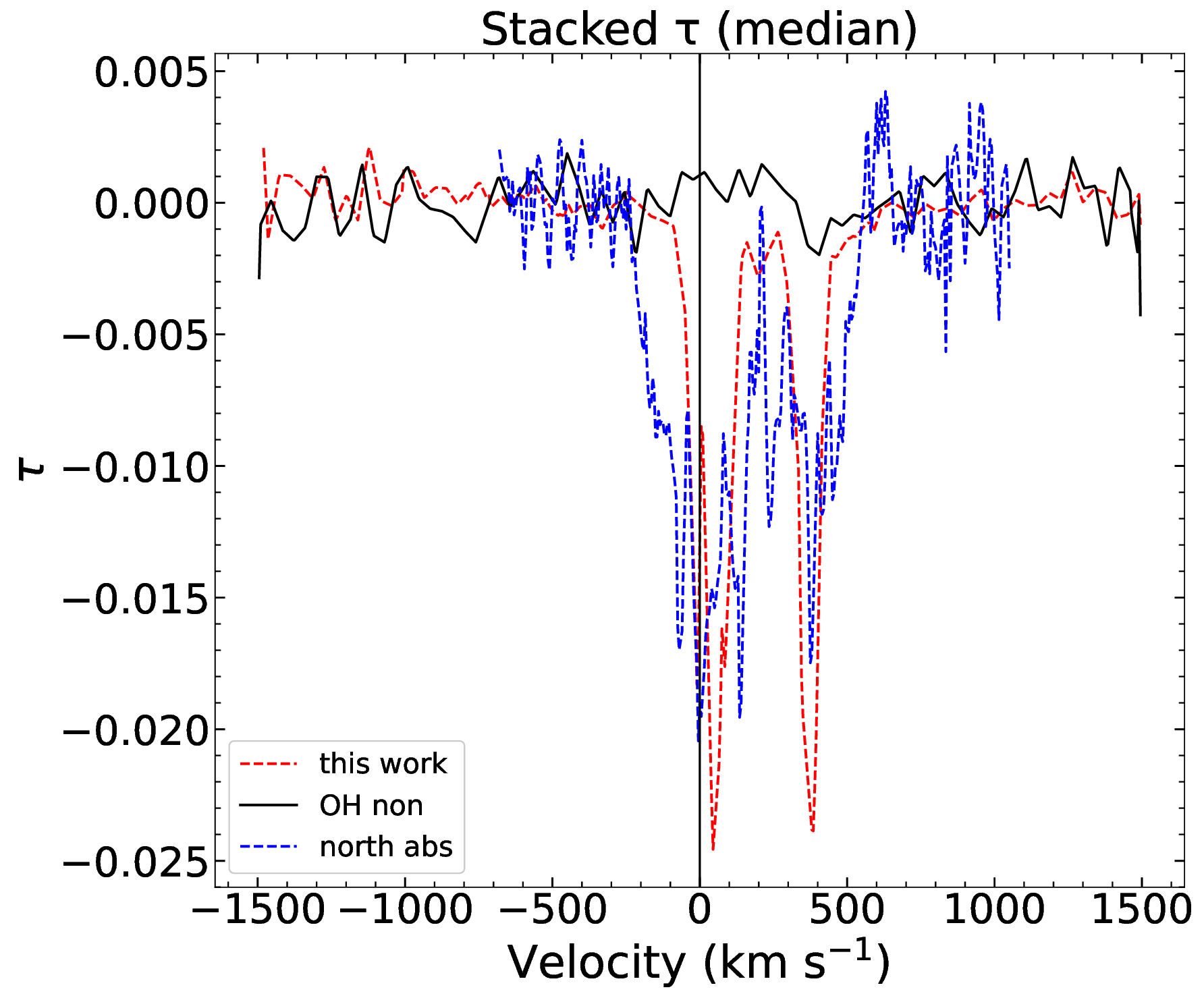}
\end{subfigure}
\hfill
\begin{subfigure}[t]{0.48\textwidth}
\centering
\includegraphics[width=\textwidth]{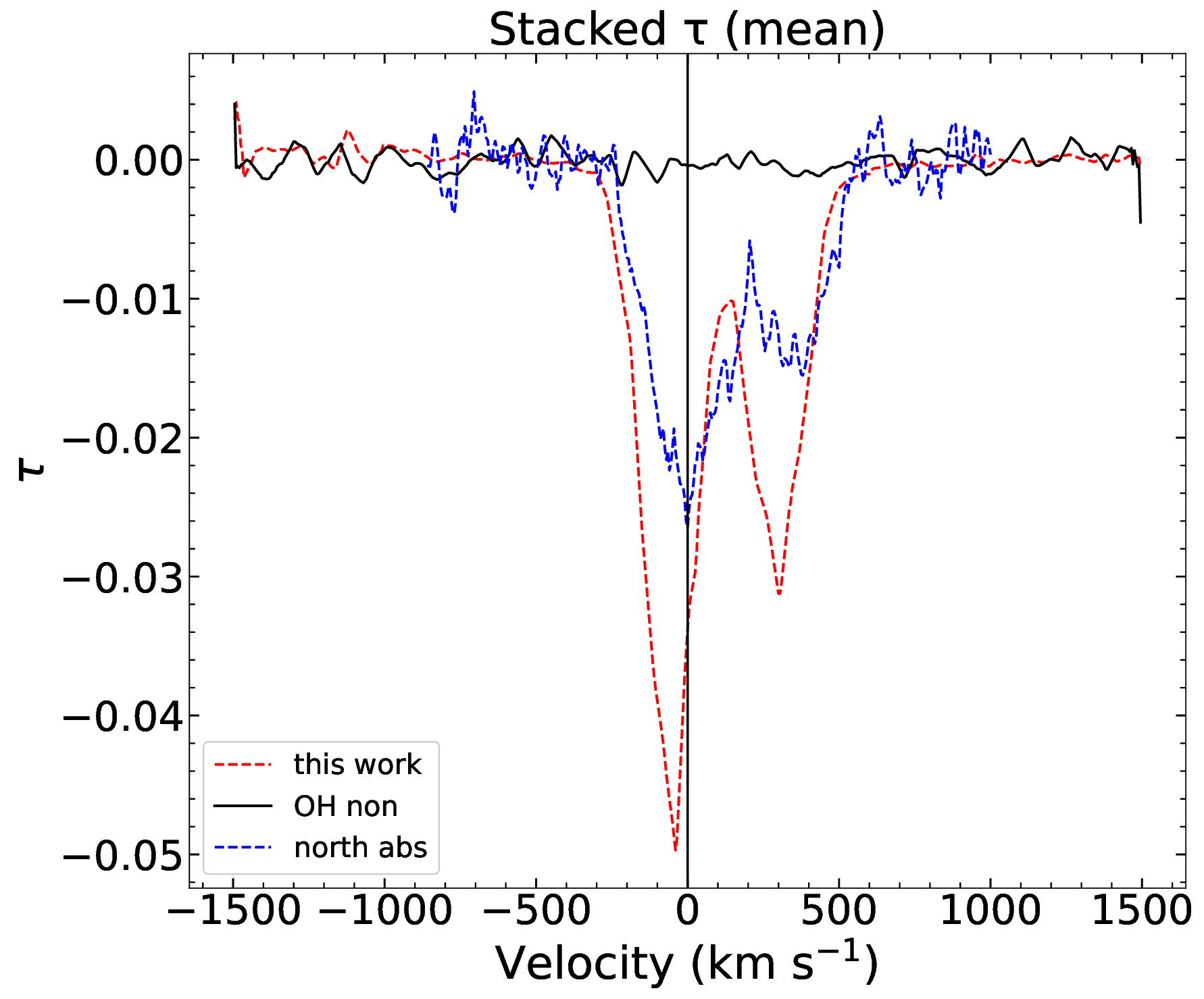}
\end{subfigure}

\begin{subfigure}[t]{0.48\textwidth}
\centering
\includegraphics[width=\textwidth]{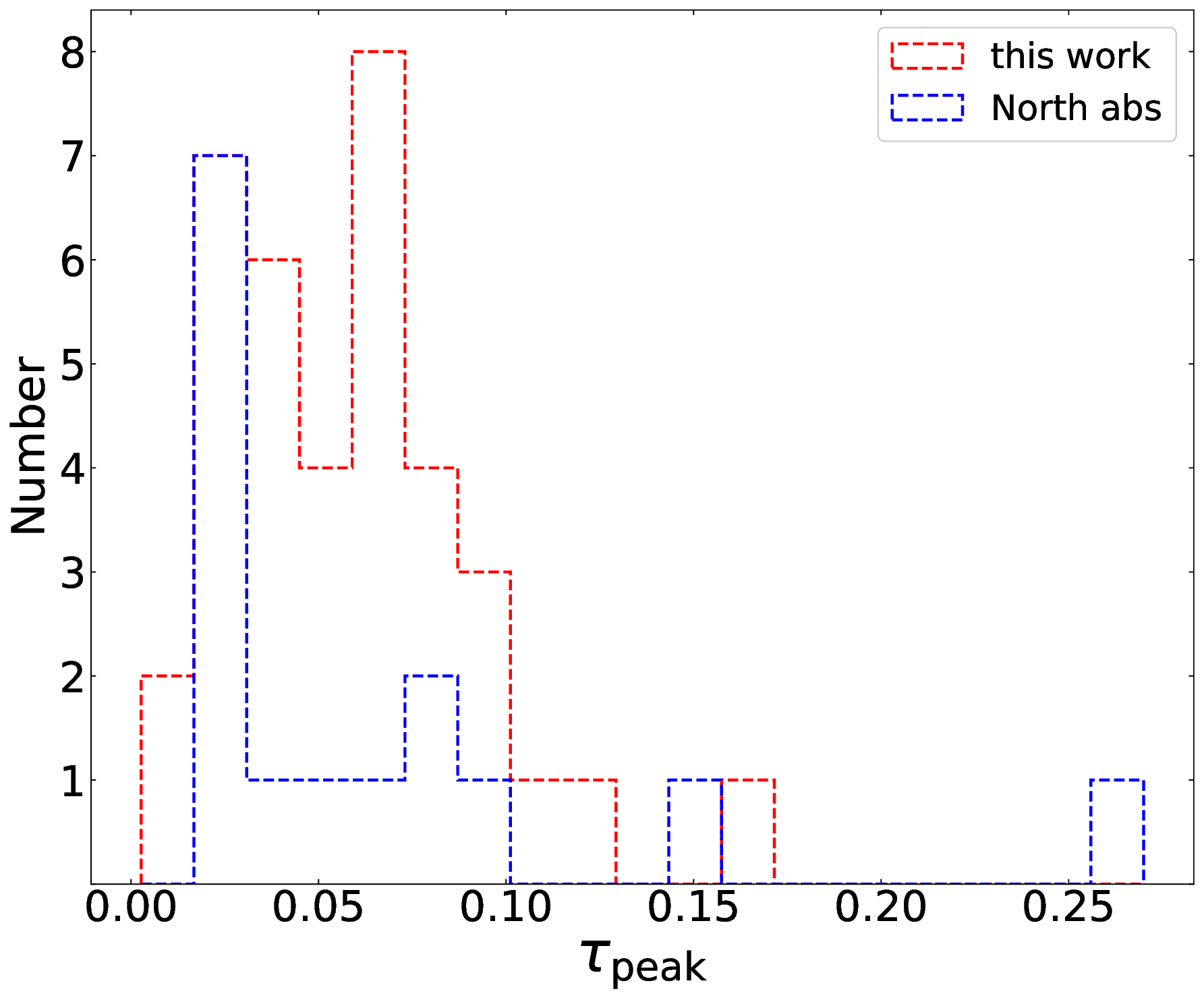}
\end{subfigure}

\caption{Distributions of the peak optical depth $\tau$ and the stacked $\tau$ spectra for the OH absorption and OH non-detection galaxies. The OH absorbers including 30 absorbers in this work and 15 northern OH absorbers (North abs) from RBGS sample described in the text. The member sources of both samples are available in the Zenodo data repository \zenodosite. The non-detection sample includes only sources with continuum flux densities $S_{\rm cont} > 20$~mJy.}
\label{fig:stacked-tau}
\end{figure}

For OH absorption detections, spectra are extracted over regions defined by the Moment-0 maps. For non-detections, we adopt circular apertures of $\sim$ 10\arcsec\ diameter centered on the peak radio continuum position, consistent with the typical extent of the OH absorption regions (see Table~\ref{tab:ohabsorption}). 
The resulting stacked spectra and the distributions of $\tau_{\rm peak}$ are shown in Fig.~\ref{fig:stacked-tau}. The stacked spectrum of the OH absorption sample shows a clear signal, with an average optical depth of $\sim$0.05 for the southern subsample and $\sim$0.025 for the 15 northern absorbers. In contrast, the stacked spectrum of the non-detections does not reveal any statistically significant absorption or emission features. The corresponding $3\sigma$ upper limit is $\tau \sim 0.002$ for both the stacked mean and median spectra (see Fig. \ref{fig:stacked-tau}), more than an order of magnitude below the mean optical depth of the absorption sample. These results suggest that the absence of OH absorption in the non-detection sample is unlikely to be solely due to sensitivity limitations. Instead, it points to intrinsic differences in the physical conditions of the molecular gas as an important factor governing the detectability of OH absorption.

\subsubsection{Dense Gas Content and IR Colors of OH-absorbing Galaxies}
\label{sect:denseandabs}

It has long been proposed that the dense gas fraction, commonly parameterized by the luminosity ratio $L_{\rm HCN}/L_{\rm CO}$, provides a key discriminator among different classes of OH line detections. \citet{2007ApJ...669L...9D} first suggested that $L_{\rm HCN}/L_{\rm CO}$ can effectively distinguish OH megamaser galaxies from non-detections and absorbers. Subsequently, \citet{2015MNRAS.447..392M} studied OH lines in early-type galaxies and found that OH absorption preferentially occurs in systems with the highest $L_{\rm HCN}/L_{\rm CO}$ ratios. Similarly, \citet{2018JApA...39...34H} compiled a large HCN database for local galaxies and argued that OH megamaser formation is closely linked to the presence of dense molecular gas. Despite these advances, the number of confirmed OH absorption systems in previous samples remains limited.

\begin{figure}[htbp]
\centering

\includegraphics[width=0.48\textwidth]{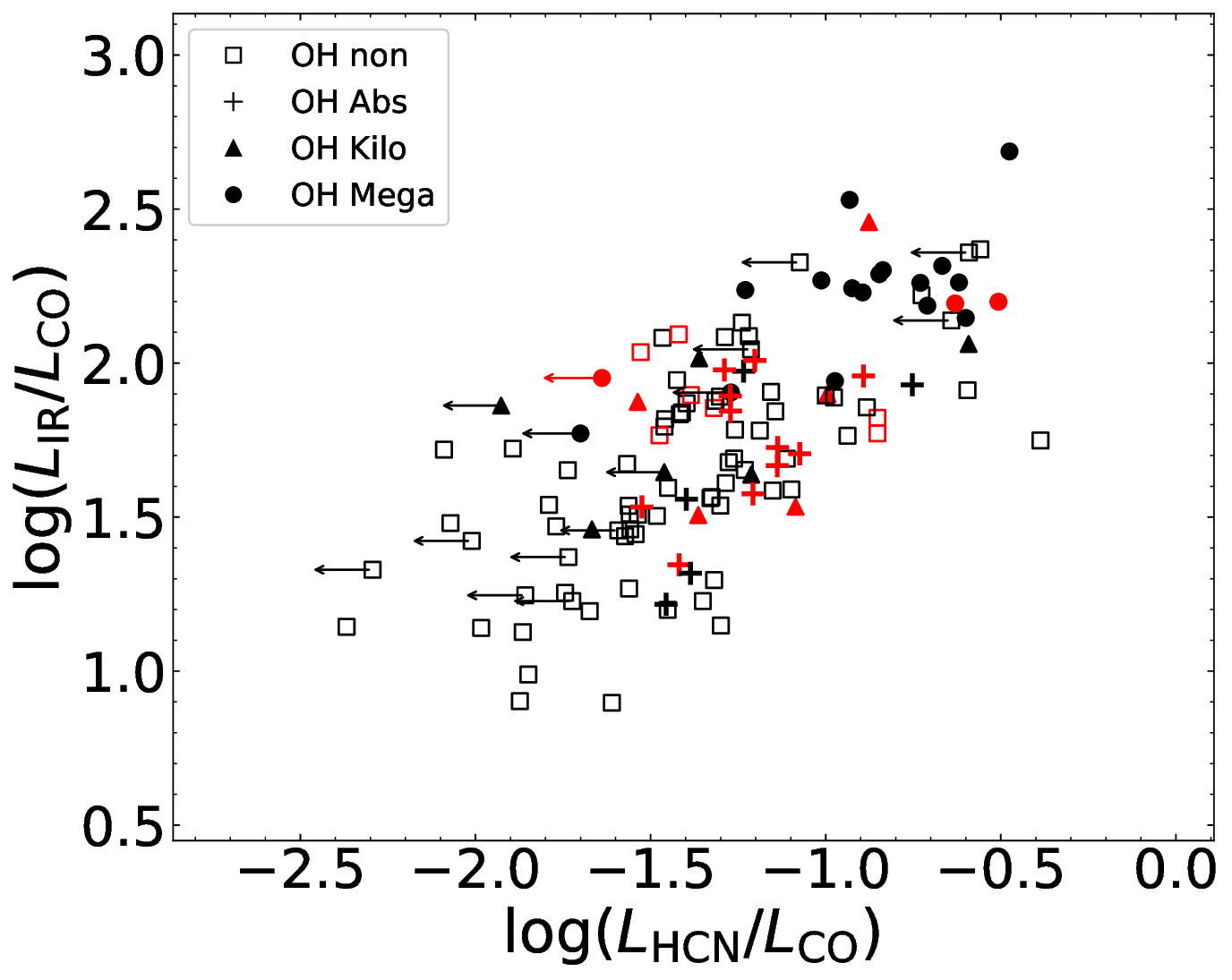}

\caption{Normalized infrared luminosity versus HCN luminosity for OH maser galaxies. This figure is a replot of a subpanel from Figure~1 of \citet{2018JApA...39...34H}, updated to include our new results for a subsample of these galaxies. The symbols are adopted from the original figure (as shown in the upper-left corner), with red colors used to indicate the revised classifications of these galaxies as OH megamasers, kilomasers, absorbers, and non-detections. Arrows mark sources with upper limits on the HCN luminosity.}
\label{fig:ohabs-dens}
\end{figure}

To improve the statistics, we cross-matched our sample with the compilation of \citet{2018JApA...39...34H} and find that $\sim$11 galaxies classified as OH non-detections in their work are in fact OH absorbers in our sample (see Fig.~\ref{fig:ohabs-dens}). In contrast, only three of our OH non-detections have available measurements of dense gas tracers. Compared with the OH non-detections, OH absorbers, and OH megamasers compiled by \citet{2018JApA...39...34H}, the OH absorption galaxies identified here tend to occupy an intermediate regime in $L_{\rm HCN}/L_{\rm CO}$, lying between OH megamasers and non-detections (see Fig.~\ref{fig:ohabs-dens}). This result supports the conclusion of \citet{2015MNRAS.447..392M} that elevated dense gas fractions are important not only for OH megamaser activity, but also for the detectability of OH absorption.

In addition to dense gas content, far-infrared (FIR) properties have been used to characterize OH line populations. \citet{1989ApJ...338..804B} examined FIR luminosities and IRAS 25/100~$\mu$m colors of extragalactic OH megamasers and absorbers, concluding that both populations are associated with edge-on molecular disks but differ in infrared luminosity. Fig.~\ref{fig:LFIR-color} shows the distribution of infrared luminosity and 25/100~$\mu$m color for OH megamasers in the RBGS reported in the literature, together with the OH absorption galaxies identified in this work and 15 additional OH absorbers in the northern RBGS from the literature. Consistent with previous studies, OH megamasers are preferentially found in galaxies with higher infrared luminosities, whereas OH absorption is more common in systems with lower $L_{\rm IR}$. However, neither infrared luminosity nor the 25/100~$\mu$m color alone provides a clear separation between OH absorption galaxies and OH non-detections, consistent with the KS-test results (Table~\ref{tab:kstest}). This suggests that while FIR properties contribute to the observed OH phenomenology, additional factors—such as the distribution, excitation, and geometry of dense molecular gas—are required to fully explain the occurrence of OH absorption.



\subsection{Comparison with Previous Studies and Physical Interpretations of OH Line Detections in IR-bright Galaxies}

Our detections of the OH main lines are based on a sample of nearby IRAS galaxies at low redshift ($z < 0.035$) and relatively modest infrared luminosities ($\log L_{\rm IR}/L_\odot < 11.8$). We find that the OH maser detection rate increases to $\sim 13$--$19\%$ once the infrared luminosity exceeds the LIRG threshold, in good agreement with previous surveys \citep[e.g.,][]{2005ARA&A..43..625L}. In contrast, the OH absorption detection rate in our sample is significantly higher than those reported in earlier single-dish studies, such as the NRAO 300-ft telescope survey \citep{1992AJ....103..728B} and Arecibo observations \citep{1986AJ.....92.1291S,2010AJ....139.2066F}. The most directly comparable work is the Arecibo survey of \citet{2010AJ....139.2066F}, which targeted 85 galaxies from the 2~Jy IRAS--NVSS sample and detected OH lines in only seven sources. A key difference lies in the infrared selection: their sample adopts $F_{60\,\mu{\rm m}} > 2$~Jy, whereas the RBGS sample used here typically satisfies $F_{60\,\mu{\rm m}} \gtrsim 5.24$~Jy. As a result, the overall OH detection fraction in our sample (38/186 $\sim 20\%$) is higher by a factor of $\sim 2-3$ compared to the Arecibo survey (7/85). This higher detection rate can be attributed to several factors. First, the higher far-infrared flux densities of our sample likely play a key role. Since the OH main lines are pumped by far-infrared radiation, stronger radiation fields may enhance not only OH maser emission but also the detectability of OH absorption. Second, the radio continuum flux densities in our sample are systematically higher. The OH absorption detections have mean and median continuum flux densities of $\sim 760$ mJy and $\sim 174$ mJy, respectively (Table~\ref{tab:kstest}), compared to $\sim 95$ mJy and $\sim 61$ mJy for non-detections, and only $\sim 40$ mJy (mean) and $\sim 23$ mJy (median) for the sample of \citet{2010AJ....139.2066F}. Given that OH absorption features are typically shallow, a stronger background continuum directly improves their detectability. Third, differences in observing technique may also play an important role. Interferometric observations are generally more sensitive to absorption against compact background continuum sources than single-dish measurements, owing to their higher angular resolution and improved bandpass stability \citep{2026A&A...707A.350M}. In contrast, the large beams of single-dish telescopes (typically arcminute scale) include substantial diffuse continuum emission that may not be fully covered by the absorbing gas, thereby diluting absorption signatures, particularly in systems with low covering factors.

Early theoretical and observational studies suggest that OH absorption predominantly arises in molecular disks within a few hundred parsecs of galactic nuclei \citep{1982ApJ...252..147R}. In our data, the spatial extent of the OH absorption, defined by the $3\sigma$ detection regions, spans $\sim8$--$40^{\prime\prime}$, corresponding to physical scales of $\sim0.5$--10~kpc over the distance range of our sample. These values should be regarded as upper limits set by the angular resolution of the MeerKAT observations. They are nevertheless broadly consistent with higher-resolution studies that locate OH absorption within the central kiloparsec regions of IR-bright galaxies \citep[e.g.,][]{2007ApJ...661..173B}. Observations of dense molecular gas show that such material is preferentially concentrated in the inner (few-kpc) regions of galaxies \citep{2019ApJS..241...19I}, and a strong association between dense gas content and OH absorption has been reported both in previous work \citep{2015MNRAS.447..392M} and in Section~\ref{sect:denseandabs}. Together, these results support a scenario in which OH absorption primarily traces molecular gas in the central regions of galaxies, although not necessarily confined to sub-kiloparsec scales. At the same time, alternative scenarios propose that OH absorption may arise from dense molecular gas in compact ($\sim10$~pc) circumnuclear disks or tori, which would require VLBI-scale observations to resolve \citep{2000A&A...360...49H}. To further examine the origin of the absorbing gas, we compare the OH absorption profiles with HCN line profiles from the literature (available on Zenodo \zenodosite). For most sources, the velocity ranges of the OH 1667~MHz absorption are consistent with those of the HCN lines, suggesting that both tracers probe similar kinematic components of the molecular gas.
One notable exception is IRAS~F13025$-$4912, where the HCN line exhibits a significantly broader profile. This discrepancy is likely due to the much larger beam size of the HCN(1--0) observations ($\sim53$--$63^{\prime\prime}$; \citealt{2004A&A...422..883W}) compared to the $\sim8^{\prime\prime}$ beam of the MeerKAT data, as well as the contribution of more extended HCN emission. Overall, this comparison indicates that OH absorption and HCN emission generally trace similar nuclear regions, although differences in spatial resolution can lead to variations in the observed line widths.
These results suggest that OH absorption traces the bulk kinematics of nuclear molecular gas, while its exact spatial origin remains unresolved at the current angular resolution.

In contrast to OH absorption, OH maser emission is generally concentrated in compact nuclear structures on parsec to sub-kiloparsec scales, with VLBI observations of several well-studied OH megamaser galaxies revealing characteristic sizes of $\lesssim100$~pc \citep[see][and references therein]{1994Natur.370..117L,1999ApJ...511..178D,2005ARA&A..43..625L,2005ApJ...618..705P}. On these scales, up to $\sim$30\% of the emission may be diffuse, while the majority is believed to arise from compact regions of order $\sim$10~pc \citep[e.g.,][]{2021A&A...647A.193H,2005ARA&A..43..625L}. Despite extensive observational efforts, it remains unclear why only a subset of major galaxy mergers host OH megamasers. Recent studies suggest that the dense gas tracer ratio $L_{\rm HCN(1-0)}/L_{\rm CO(1-0)}$ alone is insufficient to account for the presence or absence of OH megamaser activity \citep{2024AAS...24340607R}. In our sample, OH maser emission is detected primarily in LIRGs, with an overall detection rate of $\sim$13\%. At the same time, a substantial fraction of LIRGs with relatively strong radio continuum emission ($\sim$36\%; see Section~\ref{sect:statis}) exhibit OH absorption instead, which may contribute to the non-detection of maser emission in these systems. The Arecibo OH megamaser surveys \citep{2000AJ....119.3003D,2001AJ....121.1278D,2002AJ....124..100D}, which mainly targeted more distant $(U)$LIRGs ($z > 0.1$), also include a small number of OH absorbers. Among sources with radio continuum flux densities exceeding 20~mJy, approximately $13\%$ (2/15) are reported as OH absorbers. This suggests that a subset of luminous infrared galaxies may host physical conditions that are more conducive to absorption than to masing. Consistent with this picture, the dense gas ratios ($L_{\rm HCN}/L_{\rm CO}$) in OH absorbers tend to fall between those of OH megamaser galaxies and OH non-detections (Section~\ref{sect:denseandabs}). Taken together, these results indicate that the absence of OH maser emission in a significant fraction of (U)LIRGs may reflect a diversity of interstellar medium conditions, with some systems favoring absorption rather than maser amplification.
Future high-resolution observations of both dense molecular gas and OH line emission or absorption will be essential for clarifying the physical differences between masing, absorbing, and non-masing systems.
 
\begin{deluxetable*}{l c c c c c c c c c c c c c c c c c c}
\tabletypesize{\scriptsize}
\tablewidth{0pt} 
\tablecaption{Parameters of the OH maser emission \label{tab:ohemission}}
\tablehead{
\colhead{IRAS Name} & \colhead{$z$} & \colhead{Position} & \colhead{beam} & \colhead{PA} & \colhead{Major} & \colhead{Minor} &\colhead{PA} & \colhead{rms$_{\rm img}$} & \colhead{rms$_{\rm line}$} & \colhead{$V_{\rm c}$} & \colhead{Amp} & \colhead{FWHM} & \colhead{$\int S\,dv$} & \colhead{$\log L_{\rm OH}$} & \colhead{$\log L^{\rm pred}_{\rm OH}$} &\colhead{$\log L_{\rm FIR}$} & \colhead{$\log L_{\rm IR}$} & \colhead{GOALS} \\
& &\colhead{(J2000)} & \colhead{($''\!\times\!''$)} & \colhead{(deg)} & \colhead{($''$)} & \colhead{($''$)} & \colhead{(deg)} &\colhead{(mJy/b)} & \colhead{(mJy)} & \colhead{(km/s)} & \colhead{(mJy)} & \colhead{(km/s)} & \colhead{(Jy.km/s)} & \colhead{($L_\odot)$}&\colhead{($L_\odot$)} & \colhead{$(L_\odot)$}&\colhead{($L_\odot)$} & \\
}
\startdata
\textcolor{green}{F10038-3338} & 0.034 & 100604.64 & $8.7 \times 7.7$ & -15 & $<$7.8 & $<$6.9 & &   1.11 & 0.99 & $9694 \pm 5$ & $39 \pm 5$ & $72 \pm 11$ & $29 \pm 2$ & 2.98 & 2.04& 11.64 &11.73& * \\
 &  & -335304.91& & & &  &  &  &  & $10091 \pm 1$ & $179 \pm 7$ & $83.5 \pm 3.4$ &  &  &  &  &  \\
 \addlinespace
 &  & & & & &  &  &  &  & $10114 \pm 41$ & $15 \pm 5$ & $527 \pm 134$ &  & & &  &  &  \\
 \addlinespace
 &  & & & & &  &  &  &  & $10203 \pm 8$ & $28 \pm 7$ & $57 \pm 21$ &  &  & & &  &  \\
\addlinespace
\textcolor{green}{F11506-3851} & 0.011 & 115311.73& $8.4 \times 7.7$ & -25 & $2.0\pm0.4$ & $0.2\pm0.9$ & 171 & 0.46 & 0.62 & $2970 \pm 2$ & $16 \pm 4$ & $50 \pm 16$ & $13.0 \pm 0.2$ & 1.50  & 1.19 & 11.02 &11.1& * \\
 &  &  -390748.92& & &  &  &  &  &  & $3093 \pm 1 $ & $86 \pm 1$ & $105 \pm 2 $ &  &  &  &  & & \\
 &  & & & &  &  &  &  &  & $3461\pm 4$ & $14 \pm 1$ & $166 \pm 10$ &  &  &  &  & & \\
\addlinespace
F12043-3140 & 0.023 & 120651.62& $8.5\times7.8$ & -25 & $<$2.4 & $<$2.2 &  & 0.43 & 0.86 & $6263 \pm 21$ & $2.0 \pm 0.5$ & $182 \pm 51$ & $0.6 \pm 0.1$ & 1.01 & 1.53 & 11.27 &11.36& * \\
 &  &-315651.29 & & & &  &  &  &  & $6639 \pm 44$ & $1 \pm 0.5$ & $200 \pm 105$ &  &  &  &  & & \\
\addlinespace
\textcolor{green}{F12243-0036}  & 0.007 & 122654.57 & $11\times8$& 179 &$<$8.1 & $<$5.8 &  & 0.53 & 0.56 & $2035$ & $1.1$ & $10.8$ & $0.72$ & $0.097$ & 1.16 & 11 &11.08& * \\
& &-005237.47& & & & & & & & $2223 \pm 5$ & $6 \pm 1$ & $85 \pm 12 $ & & &  &  & &  \\
&  & & & &  &  &  &  &  & $2597\pm 7$ & $4 \pm 1$ & $52 \pm 20$ &  &  &  &  & & \\
\addlinespace
13242-5713  & 0.009 &132724.87& $8.7\times8.1$ & -11 & $<$8.5 & $<$8 &  & 0.73 & 0.71 & $2941 \pm 7$ & $-6 \pm 1$ & $78 \pm 22$ & $2.8 \pm 0.1$ & 0.70  & 1.52 & 11.26 &11.34& * \\
 &  &-572925.94& & &  &  &  &  &  & $3027 \pm 1$ & $56 \pm 3$ & $47 \pm 3$ &  &  &  &  & & \\
 &  & & & &  &  &  &  &  & $3311\pm 10$ & $-3 \pm 1$ & $64 \pm 18$ &  &  &  &  & & \\
\addlinespace
\textcolor{green}{F15065-1107} & 0.006 & 150916.20& $9.6\times7.7$ & 98 &$<$6.2 & $<$5.0 &  & 0.80 & 0.91 & $1829\pm 5$ & $12 \pm 1$ & $142 \pm 12$ & $1.8 \pm 0.1$ & 0.23  & 0.06 & 10.2 &10.34&  \\
& &-111921.11 & & & & & & & & & & & & & & & & \\
\addlinespace
\textcolor{green}{F16399-0937} & 0.027 & 164240.16& $11.7\times8.2$ & -36 &$<$9.4 & $<$6.6 &  & 0.51 & 0.59 & $8007 \pm 5 $ & $26 \pm 2$ & $103 \pm 11$ & $3.5 \pm 0.4$ & 1.88  & 1.77 & 11.44 &11.56& * \\
 &  &-094311.29  & & &  &  &  &  &  & $8405 \pm 24$ & $5 \pm 2$ & $114 \pm 56$ &  & & &  &  &  \\
\addlinespace
F16443-2915  & 0.021 & 164731.12& $8.7\times7.6$ & -18 & $<$7.8 & $<$6.8 &  & 1.07 & 0.97 & $6270\pm 9$ & $13 \pm 1$ & $232 \pm 22$ & $4.9 \pm 0.4$ & 1.83 & 1.49 & 11.24 &11.29& * \\
 &  & -292120.345 & & & &  &  &  &  & $6664 \pm 16$ & $7 \pm 1$ & $239 \pm 41$ &  &  &  &  & & \\
\addlinespace
\textcolor{green}{F00450-2533} & 0.0008 & 004733.20& $10.6\times 8.5$ & -26 & - & - & - & 0.81 & 1.34 &  $258\pm 3$ & $125\pm 19$ & $66\pm 9$ & $9\pm 2$ & -0.83 & 0.18 & 10.29 &10.44&  \\
& &-251715.14  & & & & & & & & & & & & & & & & \\
\addlinespace
\textcolor{green}{F02401-0013} & 0.0038 & 024241.05& $9.9\times 8.8$ & -7 & - & - & - & 0.55 & 0.97 & $833\pm 19$ & $5\pm 3$ & $68\pm 42$ & $0.3\pm 0.2$ & -0.97 & 1.01 & 10.89 &11.27& * \\
& & -000048.97 & & & & & & & & & & & & & & & & \\
\addlinespace
\textcolor{green}{F03135-0236} & 0.007 & 031600.25& $10\times 8$ & -18 & - & - & - & 0.64 & 1.06 &  $1985\pm 4$ & $9\pm 1$ & $58\pm 13$ & $0.6\pm 0.2$ & -0.09 & 0.28 & 10.36 &10.46&  \\
& &-022535.39 & & & & & & & & & & & & & & & & \\
\addlinespace
F10257-4339 & 0.009 & 102751.31& $8.3\times 7.8$ & -16 & - & - & - & 0.57 & 0.91 &  $2724\pm 3$ & $14\pm 2$ & $71\pm 9$ & $1.1\pm 0.3$ & 0.36 & 1.75 & 11.43 &11.56& * \\
& &-435418.58 & & & & & & & & & & & & & & & & \\
\enddata
\tablecomments{ Columns:
(1)IRAS source name; source names shown in green indicate OH maser galaxies
previously reported in the literature (see the notes for individual
sources in Section~\ref{notesonind}).
(2) redshift, $z$;
(3) OH emission position (J2000) derived from Gaussian fitting;
(4)--(5) synthesized beam FWHM and position angle;
(6)--(8) deconvolved major axis, minor axis, and position angle of the Gaussian-fitted OH emission component;
(9) $1\,\sigma$ rms noise level of the single-channel image with a bandwidth of $\sim$200~kHz;
(10) $1\,\sigma$ rms noise level of the extracted OH line profile;
(11)--(14) parameters of the Gaussian-fitted OH line profile, including central velocity, peak amplitude, FWHM, and integrated flux density;
(15) logarithmic OH luminosity;
(16) predicted OH luminosity from the FIR--OH correlation, $\log L_{\mathrm{OH}}^{\mathrm{pred}} = (1.38 \pm 0.14)\,\log L_{\mathrm{FIR}} - (14.02 \pm 1.66)$ \citep{2002AJ....124..100D};
(17) far-infrared luminosity from \citet{2021ApJS..257...35C};
(18) total infrared luminosity from \citet{2021ApJS..257...35C};
(19) an asterisk indicates sources included in the GOALS sample \citep{2021ApJS..257...35C}.}
\end{deluxetable*}

\begin{longrotatetable}
\setlength{\tabcolsep}{1pt}
\begin{deluxetable}{lcccccccccccccccccccc}
\tabletypesize{\scriptsize}
\tablecaption{Parameters of the OH absorption \label{tab:ohabsorption}}
\tablehead{
\colhead{IRAS Name} & \colhead{$z$} & \colhead{OH Position} & \colhead{beam} &\colhead{PA} & \colhead{Major} & \colhead{Minor} & \colhead{PA} &\colhead{rms$_{\rm image}$} & \colhead{std$_{\rm line}$} & \colhead{line} &\colhead{$V_{\rm center}$} & \colhead{Amp} & \colhead{FWHM} & \colhead{$S_{\rm cont}$} &
\colhead{$\tau_{\rm peak}$} & \colhead{$\int \tau\,dv$} &\colhead{$N_{\rm OH}/T_{\rm ex}\times10^{14}$} &
\colhead{$\log L_{\rm FIR}$} & \colhead{$\log L_{\rm IR}$} & \colhead{GOALS}\\
& &\colhead{(J2000)} & \colhead{($''\!\times\!''$)} & \colhead{(deg)} & \colhead{($''$)} & \colhead{($''$)} & \colhead{(deg)} &\colhead{(mJy/b)} & \colhead{(mJy)} & \colhead{} &\colhead{(km/s)} & \colhead{(mJy)} & \colhead{(km/s)} &\colhead{(mJy)} &\colhead{($\times10^{2}$)} & \colhead{(km/s)} & \colhead{(${\rm cm}^{-2}\,{\rm K}^{-1}$)}&\colhead{($L_\odot)$}&\colhead{($L_\odot$)} &\colhead{} \\
}
\startdata
\textcolor{green}{$F00450-2533$} & 0.0008 & 004733.01  & 10.6$\times$8.5 & -27 & 24.4$\pm$0.7 & 6.8$\pm$0.5 & 60 & 0.81 & 1.13 & 1667 & 184 $\pm$ 6 & -145 $\pm$ 13 & 133 $\pm$ 14 & 2736.8 & 5.5 $\pm$ 0.5 & 8 $\pm$ 1 & 18$\pm$ 2 & 10.29 & 10.44 &  \\
 & &-251717.95 & & & & & & & & 1665 & 551 $\pm$ 10 & -102 $\pm$ 11 & 198 $\pm$ 23 & & 3.8 $\pm$ 0.4 & 8 $\pm$ 1 & 19 $\pm$ 3 & & &  \\
\addlinespace
F01053-1746 & 0.02 & 010747.27& 10.4$\times$8.1 & 0 & $<$8.4 & $<$6.6 &  & 0.45 & 0.86 & 1667 & 5955$\pm$ 16& -6 $\pm$ 1 & 157$\pm$ 41 & 201.9& 2.8 $\pm$ 0.9 & 5 $\pm$ 2 & 11 $\pm$ 4 & 11.5 & 11.65 & * \\
 & &-173025.72 & & & & & & & &  & 6138 $\pm$ 25 & -3 $\pm$ 1 & 101 $\pm$ 64 & & 1.6 $\pm$ 0.7 & 2 $\pm$ 1 & 4 $\pm$ 3 & & &  \\
 \addlinespace
 & & & & & & & & & & 1665 & 6336 $\pm$ 26 & -4.5 $\pm$ 0.9 & 209 $\pm$ 67 & & 2.3 $\pm$ 0.4 & 5 $\pm$ 2 & 12  $\pm$ 4 & & &  \\
\addlinespace
$F01325-3623^{c}$ & 0.016 & 013451.73  & 9.3$\times$7.8 & -56 & $<$4.9 & $<$4.1 &  & 0.60 & 1.60 & 1667 & 4758 $\pm$ 33 & -4.7 $\pm$ 0.9 & 267 $\pm$ 84 & 115.9 & 4.2 $\pm$ 0.8 & 12 $\pm$ 4 & 28 $\pm$ 10 & 10.9 & 11 & * \\
 & &-360823.47& & & & & & & &  & 5005 $\pm$ 13 & -8 $\pm$ 1 & 134 $\pm$ 32 & & 7 $\pm$ 1 & 19 $\pm$ 3 & 23 $\pm$ 7 & & &  \\
 \addlinespace
 & & & & & & & & & & 1665 & 5322 $\pm$ 24 & -4.5 $\pm$ 0.9 & 240 $\pm$ 59 & & 4.0 $\pm$ 0.8 & 10 $\pm$ 3 & 24 $\pm$ 8 & & &  \\
\addlinespace
F02315-3915 & 0.007 & 023334.44  & 8.8$\times$7.9 & -45 & $<$6.3 & $<$5.7 &  & 0.51 & 1.40 & 1667 & 1957 $\pm$ 8 & -10 $\pm$ 2 & 90 $\pm$ 19 & 91.2 & 12 $\pm$ 2 & 12 $\pm$ 4 & 27 $\pm$ 8 & 10.66 & 10.78 &  \\
 & &-390239.24& & & & & & & & 1665 & 2316 $\pm$ 10 & -9 $\pm$ 2 & 90 $\pm$ 23 & & 10 $\pm$ 2 & 7 $\pm$ 3 & 23 $\pm$ 8 & & &  \\
\addlinespace
\textcolor{green}{$F02401-0013$} & 0.0038 & 024240.63  & 10$\times$8.8 & -7 & 5.2$\pm$2.4 & 3.2$\pm$2.0 & 65 & 0.55 & 1.53 & 1667 & 1159 $\pm$ 20 & -13 $\pm$ 2 & 245 $\pm$ 50 & 4791 & 0.27 $\pm$ 0.04 & 0.7 $\pm$ 0.2 & 1.7 $\pm$ 0.4 & 10.89 & 11.27 &  \\
 & &-000050.21& & & & & & & & 1665 & 1501 $\pm$ 31 & -9 $\pm$ 2 & 250 $\pm$ 75 & & 0.18 $\pm$ 0.04 & 0.5 $\pm$ 0.2 & 1.2 $\pm$ 0.4 & & &  \\
\addlinespace
\textcolor{green}{F03135-0236} & 0.007 & 031600.79 & 10$\times$8 & -18 & $<$5.3 & $<$4.1 &  & 0.64 & 0.82 & 1667 & 2155 $\pm$ 10 & -6.2 $\pm$ 0.7 & 184 $\pm$ 25 & 93.7 & 6.9 $\pm$ 0.8 & 13 $\pm$ 2 & 32 $\pm$ 6 & 10.36 & 10.46 &  \\
 & & -022539.27 & & & & & & & & 1665 & 2488 $\pm$ 10 & -6.3 $\pm$ 0.7 & 184 $\pm$ 24 & & 6.9 $\pm$ 0.8 & 14 $\pm$ 2 & 32 $\pm$ 6 & & &  \\
\addlinespace
F03316-3618 & 0.005 & 033336.48  & 9.7$\times$7.7 & -68 & $<$6.3 & $<$5.0 &  & 0.85 & 1.34 & 1667N & 1547 $\pm$ 7 & -10 $\pm$ 1 & 132 $\pm$ 22 & 327.4 & 3.2 $\pm$ 0.4 & 5 $\pm$ 1 & 11 $\pm$ 2 & 10.86 & 11 & * \\
 & &-360823.02& & & & & & & & 1665N & 1887 $\pm$ 33 & -7.4 $\pm$ 0.5 & 616 $\pm$ 70 & & 2.3 $\pm$ 0.2 & 15 $\pm$ 2 & 35 $\pm$ 5 & & &  \\
 \addlinespace
 & & & 9.7$\times$7.7 & -68 & $<$3.8 & $<$3.0 &  &  & 1.55 & 1667S & 1731 $\pm$ 15 & -8.4 $\pm$ 0.7 & 370 $\pm$ 39 & 300.3 & 2.8 $\pm$ 0.2 & 11 $\pm$ 2 & 26 $\pm$ 4 & & &  \\
 \addlinespace
 & & & & & & & & & & 1665S & 2109 $\pm$ 8 & -9 $\pm$ 1 & 110 $\pm$ 19 & & 3.0 $\pm$ 0.4 & 3.5 $\pm$ 0.8 & 8 $\pm$ 2 & & &  \\
\addlinespace
F04210-4042 & 0.020 & 042242.93 & 10$\times$7.8 & -74 & $<$3.8 & $<$3.0 &  & 0.62 & 1.06 & 1667 & 6073 $\pm$ 20 & -3.8 $\pm$ 0.9 & 179 $\pm$ 49 & 50.4 & 8 $\pm$ 2 & 15 $\pm$ 5 & 35 $\pm$ 13 & 11.17 & 11.24 & * \\
 & &-403600.43& & & & & & & & 1665 & 6419 $\pm$ 57 & -2.5 $\pm$ 0.5 & 414 $\pm$ 142 & & 5 $\pm$ 1 & 22 $\pm$ 9 & 52 $\pm$ 21 & & &  \\
\addlinespace
 & & & & & & & & & & 1665 & 5219 $\pm$ 96 & -0.7 $\pm$ 0.2 & 598 $\pm$ 249 & & 0.7 $\pm$ 0.2 & 4 $\pm$ 2 & 10 $\pm$ 5 & & &  \\
\addlinespace
$F06070-6147^{c}$ & 0.004 & 060729.64  & 11.3$\times$8.0 & 45 & $<$5.1 & $<$3.6 &  & 0.36 & 0.58 & 1667 & 1221 $\pm$ 6 & -4.7 $\pm$ 0.6 & 49 $\pm$ 7 & 69.9 & 7 $\pm$ 1 & 3.6 $\pm$ 0.7 & 8 $\pm$ 2 & 9.59 & 9.7 &  \\
 & &-614827.31& & & & & & & & 1665 & 1576 $\pm$ 8 & -2.6 $\pm$ 0.7 & 66 $\pm$ 22 & & 4 $\pm$ 1 & 3 $\pm$ 1 & 6 $\pm$ 3 & & &  \\
\addlinespace
F07160-6215 & 0.011 & 071637.18  & 91$\times$7.4 & -37 & $<$2.4 & $<$1.9 &  & 0.94 & 1.23 & 1667 & 3335 $\pm$ 27 & -6.5 $\pm$ 0.9 & 364 $\pm$ 69 & 151.8 & 4.4 $\pm$ 0.6 & 17 $\pm$ 4 & 40 $\pm$ 10 & 11 & 11.1 & * \\
 & &-622036.57& & & & & & & & 1665 & 3748 $\pm$ 36 & -4 $\pm$ 1 & 194 $\pm$ 86 & & 2.4 $\pm$ 0.8 & 5 $\pm$ 3 & 11 $\pm$ 6 & & &  \\
\addlinespace
F10225-3903 & 0.009 & 102442.42  & 9.3$\times$8.6 & -70 & $<$5.6 & $<$5.1 &  & 0.45 & 0.76 & 1667 & 2745 $\pm$ 7 & -6.1 $\pm$ 0.5 & 174 $\pm$ 17 & 69 & 9.3 $\pm$ 0.8 & 17 $\pm$ 2 & 40 $\pm$ 5 & 10.67 & 10.74 &  \\
 & &-391819.82 & & & & & & & & 1665 & 3133 $\pm$ 12 & -3.6 $\pm$ 0.5 & 159 $\pm$ 27 & & 5.3 $\pm$ 0.8 & 9 $\pm$ 2 & 21 $\pm$ 5 & & &  \\
\addlinespace
F10257-4339 & 0.009 & 102751.13  & 8.3$\times$7.8 & -17 & 5.4$\pm$0.6 & 3.1$\pm$0.8 & 128 & 0.57 & 0.83 & 1667 & 2839 $\pm$ 3 & -20 $\pm$ 1 & 139 $\pm$ 8 & 458.5 & 4.4 $\pm$ 0.2 & 6.5 $\pm$ 0.5 & 15 $\pm$ 1 & 11.43 & 11.56 & * \\
 & &-435417.77 & & & & & & & & 1665 & 3190 $\pm$ 7 & -11.6 $\pm$ 0.8 & 197 $\pm$ 17 & & 2.6 $\pm$ 0.2 & 5.3 $\pm$ 0.6 & 13 $\pm$ 1 & & &  \\
\addlinespace
F11143-7556 & 0.0056 & 111603.84  & 9.7$\times$7.8 & 1.6 & $<$6.9 & $<$5.6 &  & 0.45 & 1.00 & 1667 & 1687 $\pm$ 16 & -3.4 $\pm$ 0.7 & 150 $\pm$ 37 & 160.9 & 2.1 $\pm$ 0.5 & 3 $\pm$ 1 & 8 $\pm$ 3 & 10.59 & 10.7 &  \\
 & &-761257.66& & & & & & & & 1665 & 2009 $\pm$ 6 & -7 $\pm$ 1 & 87 $\pm$ 17 & & 4.3 $\pm$ 0.7 & 4 $\pm$ 1 & 9 $\pm$ 2 & & &  \\
 \addlinespace
 & & & & & & & & & &  & 22512 $\pm$ 28 & -3.8 $\pm$ 0.5 & 337 $\pm$ 72 & & 2.4 $\pm$ 0.3 & 9 $\pm$ 2 & 20 $\pm$ 5 & & &  \\
\addlinespace
$F11290-3001^{c}$ & 0.0059 & 113131.62  & 8.9$\times$7.9 & -27 & $<$8.4 & $<$7.4 &  & 0.40 & 0.51 & 1667 & 1762 $\pm$ 67 & -1.7 $\pm$ 0.4 & 343 $\pm$ 149 & 49.1 & 3.6 $\pm$ 0.9 & 13 $\pm$ 7 & 31 $\pm$ 15 & 10.18 & 10.29 &  \\
 & &-301825.02& & & & & & & & 1665 & 1971 $\pm$ 14 & -3 $\pm$ 1 & 92 $\pm$ 46 & & 5 $\pm$ 2 & 5 $\pm$ 3 & 12 $\pm$ 8 & & &  \\
\addlinespace
F13001-2339 & 0.022 & 130252.05  & 9.8$\times$8.7 & -77 & $<$2.3 & $<$2.1 &  & 0.79 & 1.67 & 1667 & 6440 $\pm$ 13 & -7 $\pm$ 1 & 161 $\pm$ 31 & 85.9 & 8 $\pm$ 2 & 14 $\pm$ 4 & 34 $\pm$ 9 & 11.42 & 11.49 & * \\
 & &-235518.40 & & & & & & & & 1665 & 6810 $\pm$ 11 & -6 $\pm$ 2 & 97 $\pm$ 27 & & 8 $\pm$ 2 & 8 $\pm$ 3 & 18 $\pm$ 7 & & &  \\
\addlinespace
\textcolor{green}{$F13025-4912$} & 0.002 &130527.66 & 8.4$\times$7.8 & -31 & 7.5 & 4.0$\pm$0.1 & 45 & 0.50 & 5.56 & 1667 & 602 $\pm$ 3 & -690 $\pm$ 23 & 182 $\pm$ 7 & 4411.4 & 17.0 $\pm$ 0.7 & 33 $\pm$ 2 & 78 $\pm$ 4 & 10.41 & 10.48 &  \\
 & &-492803.68& & & & & & & & 1665 & 950 $\pm$ 5 & -430 $\pm$ 24 & 174 $\pm$ 11 & & 10.3 $\pm$ 0.6 & 19 $\pm$ 2 & 45 $\pm$ 4 & & &  \\
\addlinespace
13052-5711 & 0.021 & 130818.70 & 8.7$\times$8.1 & -15 & $<$2.1 & $<$2.1 &  & 0.55 & 0.93 & 1667 & 6600 $\pm$ 18 & -4.3 $\pm$ 0.6 & 222 $\pm$ 45 & 58.8 & 8 $\pm$ 1 & 18 $\pm$ 5 & 43 $\pm$ 11 & 11.28 & 11.34 & * \\
 & & -572727.87& & & & & & & & 1665 & 6947 $\pm$ 32 & -2.4 $\pm$ 0.6 & 226 $\pm$ 81 & & 4 $\pm$ 1 & 10 $\pm$ 5 & 24 $\pm$ 11 & & &  \\
\addlinespace
F13097-1531 & 0.01 & 131226.63& 9.2$\times$8.2 & -17 & $<$2.3 & $<$2.1 &  & 0.52 & 0.37 & 1667 & 2355 $\pm$ 24 & -1.0 $\pm$ 0.4 & 135 $\pm$ 57 & 50.9 & 2.1 $\pm$ 0.8 & 3 $\pm$ 2 & 7 $\pm$ 4 & 11.41 & 11.5 &  \\
 & &-154751.36& & & & & & & &  & 2753 $\pm$ 20 & -1.2 $\pm$ 0.4 & 113 $\pm$ 48 & & 2.5 $\pm$ 0.9 & 3 $\pm$ 2 & 7 $\pm$ 4 & & &  \\
 \addlinespace
 & & & & & & & & & &  & 3091 $\pm$ 40 & -1.3 $\pm$ 0.3 & 324 $\pm$ 125 & & 2.6 $\pm$ 0.5 & 9 $\pm$ 4 & 21 $\pm$ 9 & & &  \\
 \addlinespace
 & & & & & & & & & & 1665 & 3438 $\pm$ 50 & -1.0 $\pm$ 0.3 & 187 $\pm$ 142 & & 1.9 $\pm$ 0.7 & 4 $\pm$ 3 & 9 $\pm$ 7 & & &  \\
 \addlinespace
 & & & & & & & & & &  & 3691 $\pm$ 76 & -0.7 $\pm$ 0.3 & 204 $\pm$ 175 & & 1.3$\pm$ 0.6 & 3 $\pm$ 2 & 7 $\pm$ 6 & & &  \\
\addlinespace
F13123-1541 & 0.007 &131502.11  & 10.9$\times$8.4 & -28 & $<$4.0 & $<$3.0 &  & 0.47 & 0.78 & 1667 & 2276 $\pm$ 11 & -3.4 $\pm$ 0.6 & 133 $\pm$ 26 & 50.5 & 7 $\pm$ 1 & 10 $\pm$ 3 & 23 $\pm$ 6 & 10.25 & 10.39 &  \\
 & &-155707.46 & & & & & & & & 1665 & 2613 $\pm$ 15 & -2.4 $\pm$ 0.6 & 123 $\pm$ 36 & & 5 $\pm$ 1 & 6 $\pm$ 4 & 15 $\pm$ 6 & & &  \\
\addlinespace
$F13166-1434^{c}$ & 0.009 & 131920.60  & 11.1$\times$8.4 & -29 & $<$1.7 & $<$1.2 &  & 0.59 & 0.63 & 1667 & 2841 $\pm$ 27 & -3 $\pm$ 1 & 182 $\pm$ 99 & 31.4 & 9 $\pm$ 5 & 18 $\pm$ 13 & 42 $\pm$ 32 & 10.47 & 10.6 &  \\
 & &-145032.44 & & & & & & & & 1665 & 3143 $\pm$ 164 & -1.7 $\pm$ 0.5 & 542 $\pm$ 308 & & 6 $\pm$ 2 & 33 $\pm$ 21 & 77 $\pm$ 49 & & &  \\
\addlinespace
F13170-2708 & 0.007 & 131950.17  & 10.2$\times$8.5 & -28 & $<$3.7 & $<$3.0 &  & 0.53 & 0.61 & 1667 & 2215 $\pm$ 10 & -4.0 $\pm$ 0.6 & 142 $\pm$ 24 & 140.1 & 2.9 $\pm$ 0.4 & 4 $\pm$ 1 & 10 $\pm$ 2 & 10.44 & 10.5 &  \\
 & &-272435.93 & & & & & & & & 1665 & 2565 $\pm$ 12 & -2.3 $\pm$ 0.9 & 66 $\pm$ 31 & & 1.6 $\pm$ 0.7 & 1.2 $\pm$0.7 & 3 $\pm$ 2 & & &  \\
\addlinespace
F13229-2934 & 0.014 & 132544.25  & 8.7$\times$8.1 & -24 & $<$5.7 & $<$5.3 &  & 0.53 & 0.83 & 1667 & 4133 $\pm$ 7 & -6.5 $\pm$ 0.8 & 130 $\pm$ 18 & 159.2 & 4.2 $\pm$ 0.5 & 6 $\pm$ 1 & 14 $\pm$ 2 & 11.06 & 11.17 & * \\
 & & -295002.07& & & & & & & & 1665 & 4471 $\pm$ 9 & -5.4 $\pm$ 0.7 & 139 $\pm$ 22 & & 3.5 $\pm$ 0.5 & 5 $\pm$ 1 & 12 $\pm$ 3 & & &  \\
\addlinespace
F13286-3432 & 0.008 &133128.24  & 8.7$\times$7.9 & -14 & $<$3.7 & $<$3.3 &  & 0.60 & 0.83 & 1667 & 2478 $\pm$ 10 & -5.2 $\pm$ 0.6 & 173 $\pm$ 25 & 81.8 & 6.6 $\pm$ 0.8 & 12 $\pm$ 2 & 29 $\pm$ 5 & 10.6 & 10.72 &  \\
 & & -344741.90& & & & & & & & 1665 & 2793 $\pm$ 16 & -3.2 $\pm$ 0.6 & 163 $\pm$ 39 & & 4.0$\pm$ 0.8 & 7 $\pm$ 2 & 16 $\pm$ 5 & & &  \\
\addlinespace
\textcolor{green}{F14106-0258} & 0.006 & 141314.95  & 11.7$\times$8.3 & -25 & $<$8.2 & $<$5.8 &  & 0.47 & 0.81 & 1667 & 1811 $\pm$ 6 & -6.9 $\pm$ 0.8 & 106 $\pm$ 14 & 337.1 & 2.1 $\pm$ 0.3 & 2.4 $\pm$ 0.4 & 6 $\pm$ 1 & 10.14 & 10.49 &  \\
 & & -031226.98& & & & & & & & 1665 & 2164 $\pm$ 9 & -5.3 $\pm$ 0.7 & 128 $\pm$ 21 & & 1.6 $\pm$ 0.2 & 2.2 $\pm$ 0.5 & 5 $\pm$ 1 & & &  \\
\addlinespace
F14544-4255 & 0.016 & 145741.23  & 8.8$\times$7.7 & -45 & $<$1.3 & $<$1.1 &  & 1.10 & 1.16 & 1667 & 4846 $\pm$ 4 & -13 $\pm$ 1 & 87 $\pm$ 10 & 168.2 & 8.3 $\pm$ 0.9 & 8 $\pm$ 1 & 18 $\pm$ 3 & 10.98 & 11.13 & * \\
 & &-430748.25& & & & & & & & 1665 & 5194 $\pm$ 7 & -9 $\pm$ 1 & 85 $\pm$ 15 & & 6 $\pm$ 1 & 5 $\pm$ 1 & 12 $\pm$ 3 & & &  \\
\addlinespace
\textcolor{green}{F14565-1629} & 0.012 & 145924.75  & 9.9$\times$7.7 & -19 & $<$9.5 & $<$7.4 &  & 0.52 & 0.78 & 1667 & 3512 $\pm$ 1 & -58$\pm$ 2 & 75 $\pm$ 2& 983.3 & 6.1 $\pm$ 0.2 & 4.8 $\pm$ 0.2 & 11.4 $\pm$ 0.4 & 10.56 & 10.65 &  \\
 & & -164135.90& & & & & & & & 1665 & 3870 $\pm$ 1 & -46 $\pm$ 1 & 81 $\pm$ 3 & & 4.8 $\pm$ 0.2 & 4.1 $\pm$ 0.2 & 10 $\pm$ 1 & & &  \\
\addlinespace
F16164-0746 & 0.023 & 161911.92  & 9.5$\times$7.9 & -9 & $<$3.1 & $<$2.5 &  & 0.43 & 0.93 & 1667 & 7062 $\pm$ 19 & -3.7 $\pm$ 0.9 & 163 $\pm$ 44 & 63.7 & 6 $\pm$ 2 & 11 $\pm$ 4 & 25 $\pm$ 9 & 11.48 & 11.55 & * \\
& &-075407.54& & & & & & & & & & & & & & & & & & \\
\addlinespace
17578-0400 & 0.014 & 180031.69  & 9.7$\times$7.7 & -17 & $<$4.6 & $<$3.7 &  & 0.49 & 0.92 & 1667 & 4038 $\pm$ 19 & -4 $\pm$ 1 & 166 $\pm$ 44 & 69.9 & 6 $\pm$ 2 & 11 $\pm$ 4 & 26 $\pm$ 9 & 11.31 & 11.35 & * \\
 & &-040053.12& & & & & & & & 1665 & 4382 $\pm$ 16 & -4 $\pm$ 1 & 137 $\pm$ 38 & & 7 $\pm$ 2 & 10 $\pm$ 4 & 22 $\pm$ 8 & & &  \\
\addlinespace
F18293-3413 & 0.018 & 183241.22  & 8.5$\times$7.5 & -29 & $<$5.5 & $<$4.7 &  & 0.67 & 0.94 & 1667 & 5399 $\pm$ 29 & -4 $\pm$ 2 & 179 $\pm$ 66 & 206 & 2 $\pm$ 1 & 4 $\pm$ 3 & 9 $\pm$ 6 & 11.69 & 11.81 & * \\
 & &-341128.57 & & & & & & & & 1665 & 5671 $\pm$ 60 & -3.7 $\pm$ 0.6 & 349 $\pm$ 274 & & 1.8 $\pm$ 0.3 & 7 $\pm$ 5 & 16 $\pm$ 13 & & &  \\
 \addlinespace
 & & & & & & & & & &  & 5960 $\pm$ 30 & -3 $\pm$ 2 & 140 $\pm$ 83 & & 1.2 $\pm$ 0.7 & 2 $\pm$ 1 & 4 $\pm$ 3 & & &  \\
\addlinespace
F18341-5732 & 0.016 & 183825.13  & 8.9$\times$7.2 & -25 & $<$4.3 & $<$3.5 &  & 0.50 & 1.07 & 1667 & 4704 $\pm$ 17 & -5.4 $\pm$ 0.8 & 204 $\pm$ 40 & 64.6 & 9 $\pm$ 1 & 19 $\pm$ 5 & 44 $\pm$ 11 & 11.22 & 11.3 & * \\
 & & -572922.99& & & & & & & & 1665 & 5050 $\pm$ 25 & -3.5 $\pm$ 0.8 & 208 $\pm$ 62 & & 6 $\pm$ 1& 12 $\pm$ 5 & 29 $\pm$ 11 & & &  \\
\addlinespace
$F20482-5715^{c}$ & 0.011 & 205202.29  & 9.2$\times$7.9 & 37 & $<$1.8 & $<$1.6 &  & 0.43 & 0.99 & 1667 & 2781 $\pm$ 16 & -3.1 $\pm$ 0.9 & 114 $\pm$ 38& 1140.2 & 0.3 $\pm$ 0.1 & 0.3 $\pm$ 0.2 & 0.8 $\pm$ 0.3 & 10.47 & 10.85 &  \\
 & &-570408.97 & & & & & & & & 1665 & 3162 $\pm$ 22 & -2.6 $\pm$ 0.8 & 153 $\pm$ 52 & & 0.2 $\pm$ 0.1 & 0.4 $\pm$ 0.2 & 1.0 $\pm$ 0.4 & & &  \\
\addlinespace
$F21008-4347^{c}$ & 0.017 & 210411.44  & 8.8$\times$8.1 & 37 & $<$1.7 & $<$1.6 &  & 0.57 & 0.62 & 1667 & 5253 $\pm$ 27 & -1.9 $\pm$ 0.4 & 274 $\pm$ 63 & 42.2 & 5 $\pm$ 1 & 14 $\pm$ 4 & 32 $\pm$ 10 & 11 & 11.13 & *\\
& &-433537.72& & & & & & & & & & & & & & & & & & \\ 
\addlinespace
F21453-3511 & 0.016 & 214819.63  & 8.7$\times$8.0 & 20 & $<$7.0 & $<$6.5 &  & 0.63 & 0.63 & 1667 & 4896 $\pm$ 4 & -8.5 $\pm$ 0.9 & 87 $\pm$ 11 & 151.6 & 5.8 $\pm$ 0.6 & 5 $\pm$ 1 & 13 $\pm$ 2 & 11.23 & 11.35 & * \\
 & &-345706.74& & & & & & & & 1665 & 5256 $\pm$ 6 & -5.9 $\pm$ 0.9 & 89 $\pm$ 15 & & 4.0 $\pm$ 0.6 & 4 $\pm$ 1 & 9 $\pm$ 2 & & &  \\
\addlinespace
F22359-2606 & 0.011 & 223841.33  & 10.0$\times$8.6 & 2 & $<$5.9 & $<$5.1 &  & 0.63 & 0.72 & 1667 & 3368 $\pm$ 8 & -7.5 $\pm$ 0.8 & 157 $\pm$ 20 & 71.7 & 11 $\pm$ 1 & 19 $\pm$ 3 & 43 $\pm$ 8 & 10.52 & 10.65 &  \\
 & &-255102.11 & & & & & & & & 1665 & 3730 $\pm$ 12 & -5.2 $\pm$ 0.8 & 160 $\pm$ 29 & & 8 $\pm$ 1 & 13 $\pm$ 3 & 30 $\pm$ 7 & & &  \\
\addlinespace
F23133-4251 & 0.005 & 231610.78  & 9.2$\times$7.9 & 21 & $<$6.4 & $<$5.5 &  & 0.48 & 0.91 & 1667 & 1605 $\pm$ 7 & -10.4 $\pm$ 0.9 & 159 $\pm$ 16 & 207.9 & 5.2 $\pm$ 0.5 & 9 $\pm$ 1 & 20 $\pm$ 2 & 10.88 & 11.03 & * \\
 & &-423504.72 & & & & & & & & 1665 & 1650$\pm$8 & -8$\pm$1 & 135 $\pm$ 19 & & 4.1 $\pm$ 0.5 & 6$\pm$ 1 & 14 $\pm$ 3 & & &  \\
\addlinespace
F23156-4238 & 0.005 & 231823.75  & 10.1$\times$7.5 & -17 & $<$5.2 & $<$3.9 &  & 0.53 & 0.65 & 1667 & 1620 $\pm$ 10 & -5.5 $\pm$ 0.5 & 193 $\pm$ 25 & 172.1 & 3.3 $\pm$ 0.3 & 7 $\pm$ 1 & 16 $\pm$ 3 & 10.73 & 10.87 &  \\
 & &-422212.25 & & & & & & & & 1665 & 1946 $\pm$ 19 & -3 $\pm$ 1 & 213 $\pm$ 48 & & 1.8 $\pm$ 0.3 & 4 $\pm$ 1 & 9 $\pm$ 3 & & &  \\
\addlinespace
\enddata
\hsize=\linewidth
\tablecomments{Columns: 
(1) IRAS source name; superscript ``c'' marks low-S/N OH absorption. Green labels indicate known OH absorbers from the literature: IRAS~F00450$-$2533 and IRAS~F13025$-$4912 \citep{1974ApL....15..211W}; OH emission in IRAS~F02401$-$0013 \citep{1996ApJ...462..740G}; IRAS~F03135$-$0236 and IRAS~F14565$-$1629 \citep{1992AJ....103..728B}; and IRAS~F14106$-$0258 \citep{1985AA152L9K}. 
(2) Redshift $z$; 
(3) OH position (J2000); 
(4)--(5) synthesized beam (FWHM, PA); 
(6)--(8) deconvolved size (major, minor, PA); 
(9) rms of single-channel image ($\sim$200~kHz); 
(10) rms of extracted spectrum; 
(11) OH transition (1667/1665~MHz); 
(12)--(14) Gaussian-fit parameters (velocity, peak, FWHM); 
(15) continuum flux density; 
(16) peak optical depth ($\tau \times 10^{2}$); 
(17) integrated optical depth $\int \tau dv$ (km~s$^{-1}$); 
(18) $N_{\mathrm{OH}}/T_{\mathrm{ex}} = 2.35 \times 10^{14} \int \tau dv$ \citep{2010AJ....139.2066F}; 
(19)--(20) $L_{\mathrm{FIR}}$ and $L_{\mathrm{IR}}$ \citep{2021ApJS..257...35C}; 
(21) asterisk denotes GOALS sources \citep{2021ApJS..257...35C}.}
\end{deluxetable}
\end{longrotatetable}


\begin{deluxetable*}{lcccccccc}
\tabletypesize{\small}
\tablecaption{K-S Test Results.\label{tab:kstest}}
\tablehead{
\colhead{Parameter} & \multicolumn{2}{c}{Sample Size} & \multicolumn{2}{c}{Abs Sample} & \multicolumn{2}{c}{Non Sample} & \colhead{} & \colhead{} \\
\colhead{} & \colhead{OH\_abs} & \colhead{OH\_Non} & \colhead{Mean} & \colhead{Median} & \colhead{Mean} & \colhead{Median} & \colhead{D} & \colhead{$p$-value}
}
\startdata
$C$ & 45 & 48 & 0.7 & 0.8 & 0.7 & 0.7 & 0.15 & $6.1\times 10^{-1}$ \\
$z$&45 &51 &0.01 &0.008 &0.01 &0.01 &0.20 &$1.8\times 10^{-1}$\\
$S_{1.4}$(mJy)&45 &51 & 759.8&174.4 &95.1 &60.9 &0.50 &$5.4\times 10^{-6}$ \\
$L_{1.4}$(W/Hz)& 45& 51& $7\times 10^{22}$&$2.5\times 10^{22}$& $3.3\times 10^{22}$&$1.9\times 10^{22}$ &0.30 &$6.3\times 10^{-2}$ \\
$W1$(mJy)& 45 & 51 &940.9 &155.5 &143.1 &71.0 &0.30 &$2.3\times 10^{-2}$ \\
$W2$(mJy)& 45 & 51 & 870.1&107.9 &101.1 &52.6 &0.29 &$3.2\times 10^{-2}$ \\
$W3$(mJy)& 45 & 51 & 5162.2&613.6 &544.5 &362.3 &0.30 &$5.2\times 10^{-3}$ \\
$W4$(mJy)& 45 & 51 &15506.6 &2065.4 &1835.5 &1221.8 &0.30 &$2.1\times 10^{-2}$ \\
$W1\_W2$(mag) & 45 & 51 & 0.4 & 0.3 & 0.3 & 0.3 & 0.15 & $6.3\times 10^{-1}$ \\
$W2\_W3$(mag) & 45 & 51 & 3.7 & 3.7 & 3.9 & 4.1 & 0.44 & $1.3\times 10^{-4}$ \\
$W1\_W3$(mag) & 45 & 51 & 4.0 & 4.0 & 4.2 & 4.4 & 0.38 & $1.4\times 10^{-3}$ \\
$W3\_W4$(mag) & 45 & 51 & 2.7 & 2.8 & 2.8 & 2.8 & 0.10 & $9.6\times 10^{-1}$ \\
$q_{\mathrm{TIR(IRAS)}}$ & 45 & 51 & 2.5 & 2.6 & 2.6 & 2.6 & 0.28 & $3.8\times 10^{-2}$ \\
$q_{\mathrm{TIR(WISE)}}$ & 45 & 51 & 2.4 & 2.4 & 2.6 & 2.6 & 0.37 & $2.2\times 10^{-3}$ \\
$\log (L\SI{12}{\micro\meter})(L_\odot)$ & 45 & 51 & 10.0 & 10.0 & 9.9 & 9.9 & 0.09 & $9.7\times 10^{-1}$ \\
$\log (L\SI{25}{\micro\meter})(L_\odot)$ & 45 & 51 & 10.2 & 10.2 & 10.1 & 10.1 & 0.13 & $7.9\times 10^{-1}$ \\
$\log (L\SI{60}{\micro\meter})(L_\odot)$ & 45 & 51 & 10.7 & 10.7 & 10.5 & 10.5 & 0.20 & $2.4\times 10^{-1}$ \\
$\log (L\SI{100}{\micro\meter})(L_\odot)$ & 45 & 51 & 10.7 & 10.7 & 10.5 & 10.6 & 0.28 & $4.0\times 10^{-2}$ \\
$\log(L_{\text{IR}})(L_\odot)$ & 45 & 51 & 11.0 & 11.0 & 10.8 & 10.9 & 0.18 & $3.7\times 10^{-1}$ \\
$\log(L_{\text{FIR}})(L_\odot)$ & 45 & 51 & 10.9 & 10.9 & 10.7 & 10.7 & 0.22 & $1.8\times 10^{-1}$ \\
$\log\left(\frac{S_{25\mu m}}{S_{100\mu m}}\right)$ & 45 & 51 & 0.09 & 0.07 & 0.12 & 0.09 & 0.25 & $8.8\times10^{-2}$ \\
$C_v$ & 25 & 27 & 3.7 & 3.5 & 3.4 & 3.5 & 0.20 & $5.8\times 10^{-1}$ \\
$\log M_{\mathrm{HI}}$(M$_\odot$) & 25 & 27 & 9.5 & 9.5 & 9.4 & 9.4 & 0.18 & $7.2 \times 10^{-1}$ \\
incl($^\circ$) & 45 & 51 & 62.7 & 62.6 & 59.0 & 54.8 & 0.21 & $2.1\times 10^{-1}$ \\
$b/a$ & 44 & 49 & 0.6 & 0.7 & 0.7 & 0.8 & 0.23 & $1.5\times 10^{-1}$ \\
\enddata
\tablenotetext{}{Columns: (1) Physical parameters.
$C$ is the compactness factor of the radio continuum emission, defined as $C = F_{\rm peak}/F_{\rm int}$.
$z$ is the redshift.
$S_{1.4}$ and $L_{1.4}$ are the flux density and luminosity at 1.4 GHz, respectively.
$W1$–$W4$ are the \textit{WISE} band flux densities in mJy.
$W1$–$W2$, $W2$–$W3$, $W1$–$W3$, and $W3$–$W4$ are the corresponding mid-infrared color indices in magnitudes.
$q_{\rm TIR(IRAS)}$ and $q_{\rm TIR(WISE)}$ are the total infrared-to-radio flux density ratios derived from \textit{IRAS} and \textit{WISE}, respectively.
$\log L(12,\mu{\rm m})$, $\log L(25,\mu{\rm m})$, $\log L(60,\mu{\rm m})$, and $\log L(100,\mu{\rm m})$ are the monochromatic infrared luminosities.
$\log L_{\rm IR}$ and $\log L_{\rm FIR}$ are the total infrared and far-infrared luminosities.
$\log (S_{25,\mu{\rm m}}/S_{100,\mu{\rm m}})$ is the infrared color.
$C_v$ is the velocity concentration parameter.
$\log M_{\rm HI}$ is the neutral hydrogen mass.
$\mathrm{incl}$ is the inclination angle of the galaxy.
$b/a$ is the axial ratio.
The data used to generate this table, including the parameter definitions and their detailed values for both samples, are available in the Zenodo data repository \zenodosite.}
\end{deluxetable*}

\subsection{Notes on individual sources}
\label{notesonind}

\subsubsection{IRAS F10038$-$3338}

IRAS F10038$-$3338 is a known OH megamaser (OHM) galaxy at a redshift of $z \approx 0.034$. It was first detected with the Nançay telescope by \citet{1988IAUC.4629....2K} and subsequently analyzed in more detail by \citet{1990A&A...237L...1K}. A later observation with the Parkes telescope by \citet{1992MNRAS.258..725S} re-detected the OH emission, reporting a peak flux density nearly a factor of two higher than that measured with Nançay. The origin of this discrepancy was not discussed, but it may suggest possible variability in the OH megamaser emission.

Our MeerKAT spectrum is broadly consistent with the ATCA observations reported by \citet{1996MNRAS.280.1143K}. However, minor differences are present: in particular, the MeerKAT spectrum shows an OH line flux density that is lower by $\sim$43\,mJy at a velocity of $\sim$10091~\kms\ (see Fig.~\ref{fig:oh_spectra_centered}). This discrepancy could be partly attributed to differences in spectral resolution, as the MeerKAT data have a velocity resolution more than an order of magnitude coarser than the ATCA data, although the spatial resolutions are comparable. Alternatively, variability in narrow, bright OH features—commonly reported in other OH megamaser galaxies \citep[e.g.,][]{2002ApJ...569L..87D,2023A&A...669A.148W}—may also contribute.

The OH maser emission is spatially associated with the radio continuum and the central region of the galaxy, as seen in the optical image (Fig.~\ref{OH8image}). The OH spectrum is dominated by a main-line component peaking at 10091~\kms, with two weaker features at 9694~\kms\ and 10203~\kms. In addition, there is tentative evidence for a shallow, broad component extending from $\sim$9600~\kms\ to $\sim$10561~\kms. Most OH features appear blueshifted relative to the systemic velocity reported in the literature. This could indicate that part of the OH maser emission is associated with outflowing gas driven by either an active galactic nucleus or a central starburst, similar to scenarios proposed for other systems such as IRAS~01298$-$0744 \citep{2024A&A...687A.193W}.

\subsubsection{IRAS F11506$-$3851}

IRAS F11506$-$3851 is a known OH megamaser galaxy, with OH emission previously reported from Parkes observations by \citet{1992MNRAS.258..725S}. The OH line profile obtained from our MeerKAT data is broadly consistent with the Parkes spectrum (see Fig.~\ref{fig:oh_spectra_centered}), with no significant differences apparent within the sensitivity and resolution limits.
Fig.~\ref{OH8image} shows that the OH emission is concentrated toward the central region of the galaxy and is spatially coincident with the peak of the radio continuum emission.


\subsubsection{IRAS F12043$-$3140}

IRAS F12043$-$3140 is identified here as a newly discovered OH megamaser, with an OH luminosity exceeding $10~L_{\odot}$ (see Table~\ref{tab:ohemission}). An upper limit of $\sim$35~mJy on the OH line flux density was previously reported by \citet{1992MNRAS.258..725S}; the flux density measured in this work remains below this limit and is therefore consistent with the earlier non-detection.

The OH line profile can be reasonably described by two Gaussian components (see Table~\ref{tab:ohemission}), which may correspond to the 1667~MHz and 1665~MHz OH main lines. Adopting an optical redshift of $z = 0.0232$ \citep{2003MNRAS.339..652K}, both components appear blueshifted by $\sim$700~\kms\ relative to the reported systemic velocity (see Fig.~\ref{fig:oh_spectra_centered}). 

This system is a merging galaxy pair, with northern and southern components separated by $\sim$6~kpc \citep{2017MNRAS.471.1634H}. The OH emission appears to originate from a region located between the two nuclei (Fig.~\ref{OH8image}), although it is more closely aligned with the northern nucleus, which has been classified as a LINER \citep[see][and references therein]{2017MNRAS.471.1634H}. However, given the current spatial resolution, a contribution from the southern component cannot be ruled out.
The region enclosed by the $3\sigma$ contour is smaller than the synthesized beam, suggesting that the emission is unresolved. Higher-sensitivity and higher-resolution observations will be required to better constrain the spatial distribution and kinematics of the OH-emitting gas.

\subsubsection{IRAS F12243$-$0036}

IRAS F12243$-$0036 is a known OH maser source, first detected with the Nançay telescope by \citet{1988A&A...201L..13M}, with an OH luminosity of $L_{\mathrm{OH}} \sim 0.73~L_{\odot}$. The OH spectrum obtained from our MeerKAT data is broadly consistent with the earlier Nançay results (see Fig.~\ref{fig:oh_spectra_centered}). The MeerKAT spectrum shows three distinct spectral features, with peaks at 2035~\kms, 2223~\kms, and 2597~\kms. The two lower-velocity components are plausibly associated with the 1667~MHz OH main line, while the higher-velocity feature at 2597~\kms\ may correspond to the 1665~MHz transition, although this identification remains somewhat uncertain.
The OH emission is spatially coincident with the radio continuum peak and appears to be concentrated toward the central region of the galaxy (Fig.~\ref{OH8image}), although it is not spatially resolved at the current resolution.

\subsubsection{IRAS F13242$-$5713}

IRAS F13242$-$5713 is identified here as a newly discovered OH megamaser galaxy. The OH spectrum shows a narrow emission component with a FWHM of $\sim$47~\kms, along with two additional narrow absorption features (see Table~\ref{tab:ohemission}). Optical and radio continuum images indicate that the system consists of a galaxy pair. The OH emission appears to be spatially associated with the western nucleus, which is separated from the eastern nucleus by $\sim$$20\arcsec$ (see Fig.~\ref{fig:oh_spectra_centered}), although the emission is not resolved in detail at the current resolution.

\subsubsection{IRAS F15065-1107}
This is a known OH maser galaxy, first detected with the NRAO 300-ft
telescope by \citet{1992AJ....103..728B}. The OH emission line profile
obtained from the MeerKAT observations is broadly consistent with that
from the NRAO 300-ft data, but appears to be broader by
$\sim$100--150~\kms at the high-velocity end
(see Fig.~\ref{fig:oh_spectra_centered}). The differences between the two OH spectra may
be related to the larger uncertainties of the NRAO 300-ft observations
($\sim4$~mJy) and/or the lower velocity resolution of the MeerKAT data
($\sim37$~\kms). The OH line emission is concentrated in the central
region of the galaxy, whereas the radio continuum emission is more
extended and distributed across a much larger area of the system (see
Fig.~\ref{fig:oh_spectra_centered}).

\subsubsection{IRAS F16399-0937}
\citet{1986IAUC.4248....2S} reported a detection of the 1667~MHz OH line with a
recession velocity of $\sim$8010~\kms and a peak flux density of 25~mJy; however,
no OH line profile has been published in the literature. The MeerKAT observations
recover a consistent recession velocity and peak OH line flux density
(see Fig.~\ref{fig:oh_spectra_centered}). Optical imaging reveals a double-nucleus system,
consisting of northern and southern components separated by a projected distance
of $\sim6\arcsec$ \citep[see Fig.~\ref{OH8image}]{2015ApJ...799...25S}. The MeerKAT
synthesized beam ($11.71\arcsec \times 8.23\arcsec$) is insufficient to fully
resolve the two nuclei. Nevertheless, the peak position of the OH line emission
(RA = 16:42:40, Dec = $-09$:43:11) is spatially coincident with the northern
nucleus (see Fig.~\ref{fig:oh_spectra_centered}). The positional uncertainty is approximately
$0.071\arcsec \times 0.085\arcsec$, estimated from Gaussian fitting using CASA,
indicating that the OH emission most likely originates from the northern nucleus.
This is consistent with the conclusion of \citet{2015ApJ...799...25S}, who argued
that the northern nucleus possesses all the necessary conditions for strong,
radiatively pumped OH maser emission.

The OH line profile exhibits two emission peaks and two apparent absorption
features (see Fig.~\ref{fig:oh_spectra_centered}). Examination of the channel maps indicates that
the absorption feature at $V = 7581$~\kms coincides with an elevated noise level
of $\sim1.5$~mJy~beam$^{-1}$, compared to an rms noise level of
$\sim0.2$~mJy~beam$^{-1}$ in line-free channels. Similar spurious absorption
features are present at other locations in the same channel map, strongly
suggesting that this feature is caused by RFI.
In contrast, the absorption feature at $V = 9130$~\kms reaches a depth of
$-7.2$~mJy~beam$^{-1}$, with a moderately enhanced rms noise level of
$\sim0.7$~mJy~beam$^{-1}$ in the corresponding channel image. Although the
increased noise introduces additional uncertainty, this feature does not show
obvious signatures of RFI and may therefore be real. The channel maps associated
with the OH emission features show no comparable anomalies and are consistent
with the noise characteristics of line-free channels.
We therefore conclude that the apparent absorption feature at
$V = 7581$~\kms is spurious, the weak absorption feature at
$V = 9130$~\kms is tentatively real but requires confirmation with
higher-sensitivity observations, and the two OH emission features represent
robust detections.

\subsubsection{IRAS F16443$-$2915}

IRAS F16443$-$2915 is identified here as a newly detected OH megamaser galaxy. Previous observations by \citet{1992MNRAS.258..725S} did not detect OH emission and reported an upper limit of $\sim$20~mJy on the line flux density. Our MeerKAT data reveal a peak flux density of $\sim$12.5~mJy, consistent with the earlier non-detection. The OH line profile can be described by two components, which are plausibly associated with the 1667~MHz and 1665~MHz transitions. The OH emission appears to be concentrated toward the central region of the galaxy and is spatially coincident with the radio continuum emission.

\section{Summary}

Using archival \textit{MeerKAT} snapshot survey data, we carried out a systematic study of OH main-line emission and absorption in 186 southern galaxies from the IRAS Revised Bright Galaxy Sample. The main results are summarized as follows.

(1) OH main-line features are detected in 38 galaxies, including eight sources with OH maser emission and 30 galaxies showing OH absorption; two of the absorption systems also exhibit weak OH emission. The occurrence of OH maser emission strongly depends on infrared luminosity: LIRGs show an OH emission detection rate of $\sim$13\%, whereas non-LIRG IR-bright galaxies exhibit significantly lower rates. When restricting the analysis to galaxies with radio continuum flux densities $\gtrsim 20$~mJy, OH absorption is detected in $\sim$36\% of LIRGs and $\sim$27\% of non-LIRGs. The majority of the OH absorption systems are newly identified, and the detection rate indicates that OH absorption occurs in a substantial fraction of IRAS-selected galaxies with sufficient background radio continuum.

(2) The detected OH emitters broadly follow the established $L_{\rm OH}$--$L_{\rm FIR}$ relation, supporting the requirement of strong far-infrared radiation for efficient maser pumping. In contrast, OH absorption systems and non-detections lie well below this relation, indicating that they occupy a different physical regime from OH megamaser galaxies.

(3) A statistical comparison between OH absorbers and non-detections shows no significant differences in radio continuum compactness, infrared luminosity, or \HI-related properties. In addition, stacking of the OH spectra for non-detections reveals no statistically significant absorption or emission features, suggesting that sensitivity and orientation effects, even when considered together, are unlikely to fully account for the absence of OH absorption. In contrast, mid-infrared color (W2--W3) and the $q_{\rm TIR}$ parameter show significant differences between OH absorbers and non-detections, indicating a link between OH absorption and star formation activity. We also find that OH absorption galaxies occupy an intermediate regime in $L_{\rm HCN}/L_{\rm CO}$ between OH megamasers and non-detections reported in the literature, suggesting that elevated dense gas fractions are important not only for OH megamaser activity but also for the detectability of OH absorption. These results further imply that extreme star formation conditions may reduce the likelihood of detecting OH absorption.

\section*{Data availability}

Appendix A: Supplementary figures and tables are available on the Zenodo database at the address \href{https://zenodo.org/records/19904867?preview=1&token=eyJhbGciOiJIUzUxMiJ9.eyJpZCI6IjBhZTM5ZDUwLWNhN2YtNGViNi04MmI4LTk4OGNhMWFlMGRhNyIsImRhdGEiOnt9LCJyYW5kb20iOiI4ODUwNzJhYjFiMDM5NGQ2MjJjNTUwZWY1ZWUwZWMxYSJ9.FFsKntsIcKJIy_M0DZn9fK-TsYQu01t_1k0E2MCS0L2CipmkTn7J6uyPwyupNxlDFSZPYmG3QjWeVUAq5xNpDw}{https://doi.org/10.5281/zenodo.19904867} (DOI:10.5281/zenodo.19904867.)

\begin{acknowledgements}
This work is supported by the National Natural Science Foundation of China (NSFC; Grant No. 12363001) and partly by the Guizhou Provincial Major Scientific and Technological Program (XKBF; Grant Nos. 2025-010 and 2025-011). 
The MeerKAT telescope is operated by the South African Radio Astronomy Observatory, which is a facility of the National Research Foundation, an agency of the Department of Science and Innovation.

\end{acknowledgements}
\bibliographystyle{aasjournalv7}
\bibliography{zhaohan}{}
\clearpage
\appendix




\setcounter{figure}{0}
\renewcommand{\thefigure}{A.\arabic{figure}}


\begin{figure}[!htbp]
  \centering
  \includegraphics[width=0.6\textwidth,height=0.5\textheight]{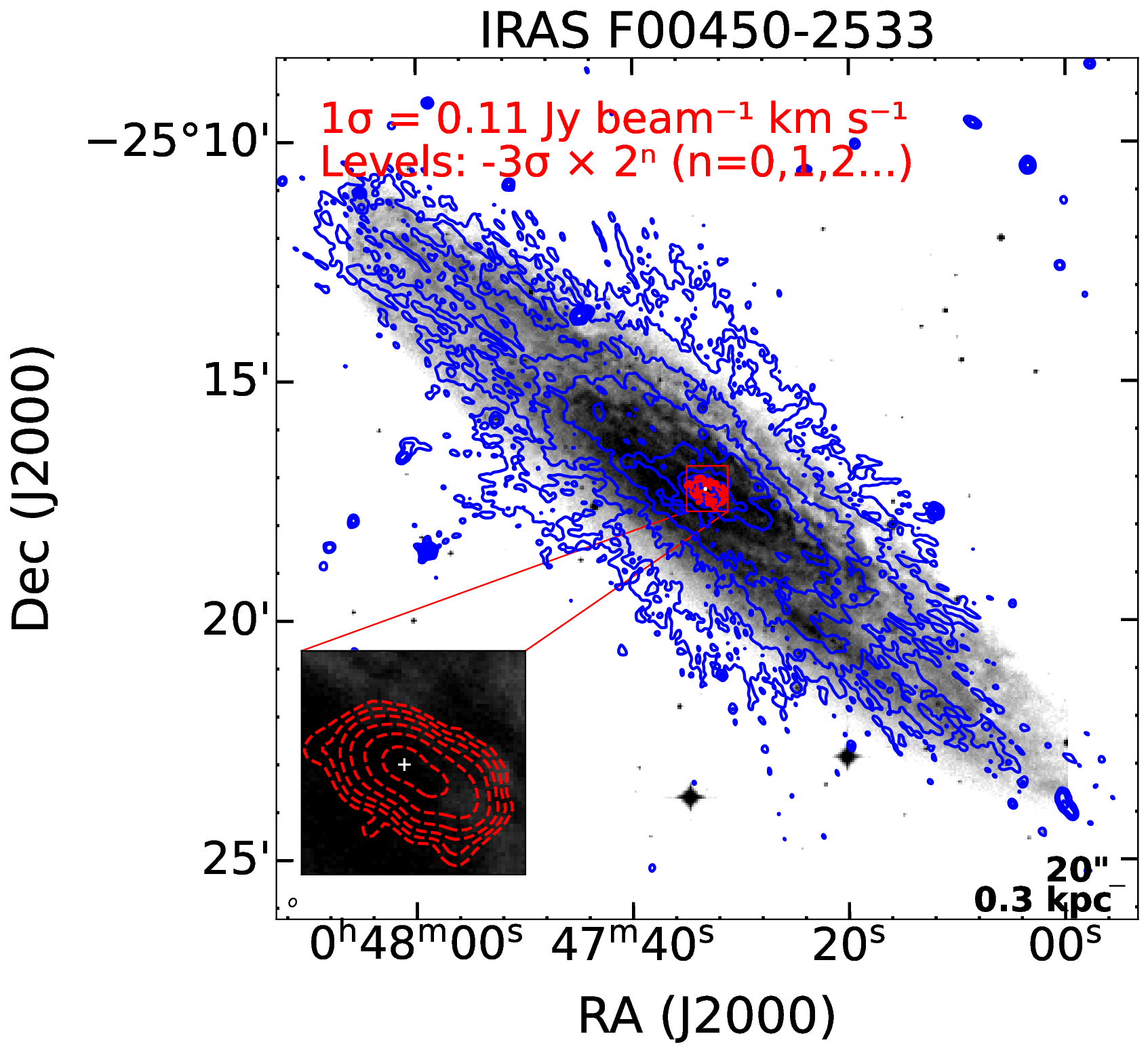}\hfill
  \caption{High S/N OH absorption-line (moment--0) maps overlaid on DSS \(R\)-band grayscale images.
Red contours show the integrated OH absorption derived from the MeerKAT data. The \(1\sigma\) rms noise levels of the OH absorption maps and contour levels of the
OH maps are indicated in the upper left corner of each panel.
Blue contours indicate the full-band radio continuum emission from \citet{2021ApJS..257...35C}. The contours start at \(3\sigma\) and increase by factors of two. White crosses mark the positions of peak OH absorption. The complete figure set (30 images) will be available in the online article upon publication of the paper, with an additional version hosted on Zenodo: \href{https://doi.org/10.5281/zenodo.19904867}{doi:10.5281/zenodo.19904867}. 
The six additional low-S/N OH absorption-line (moment--0) maps will only be available on Zenodo.
}
  \label{fig:radio-hst-1}
\end{figure}

\begin{figure*}[htbp]
  \centering
  \subcaptionbox*{}{%
    \begin{minipage}[t]{0.42\textwidth}
      \centering
      \includegraphics[width=\textwidth]{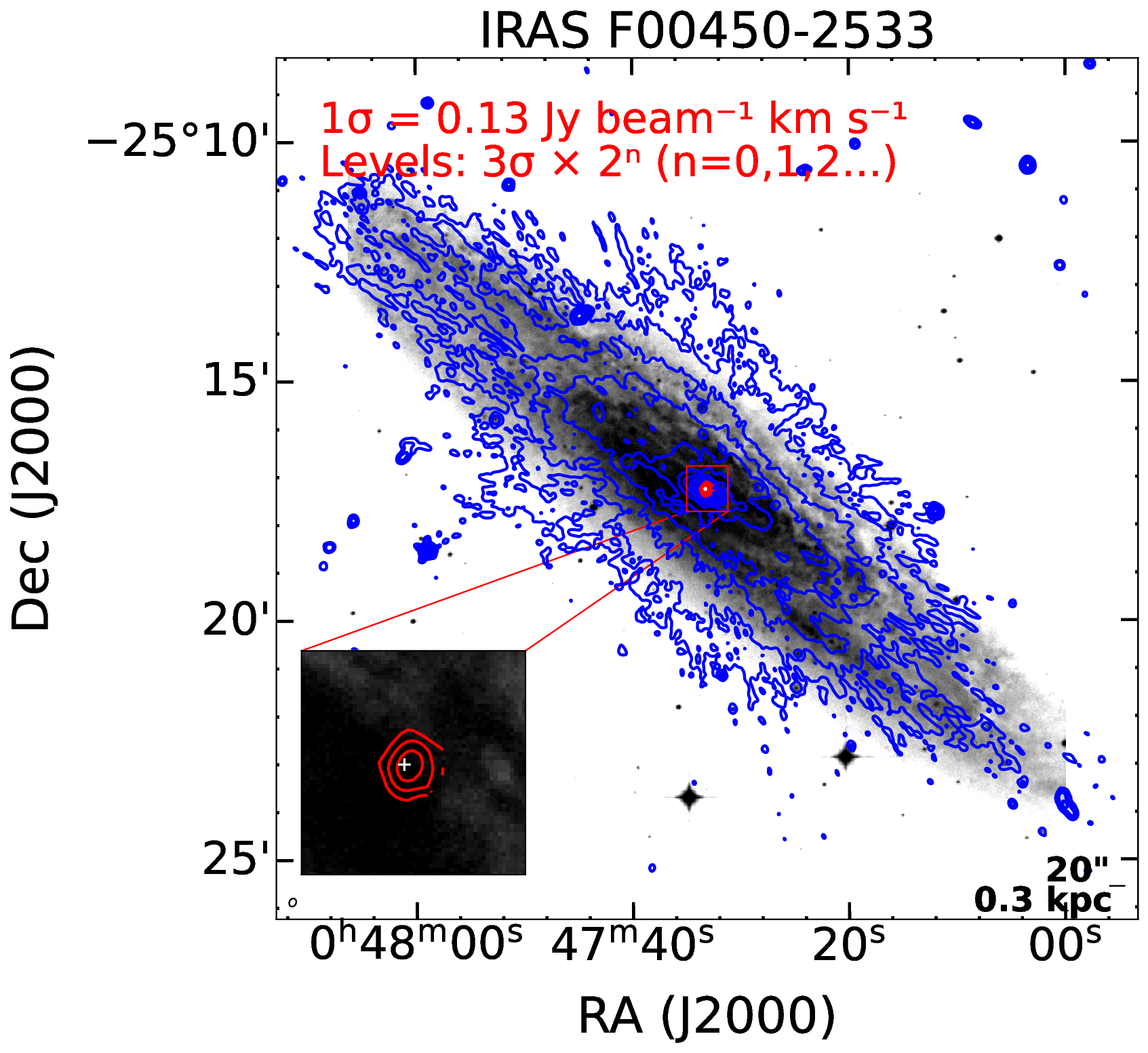}
    \end{minipage}%
    \hfill%
    \begin{minipage}[t]{0.42\textwidth}
      \centering
      \includegraphics[width=\textwidth]{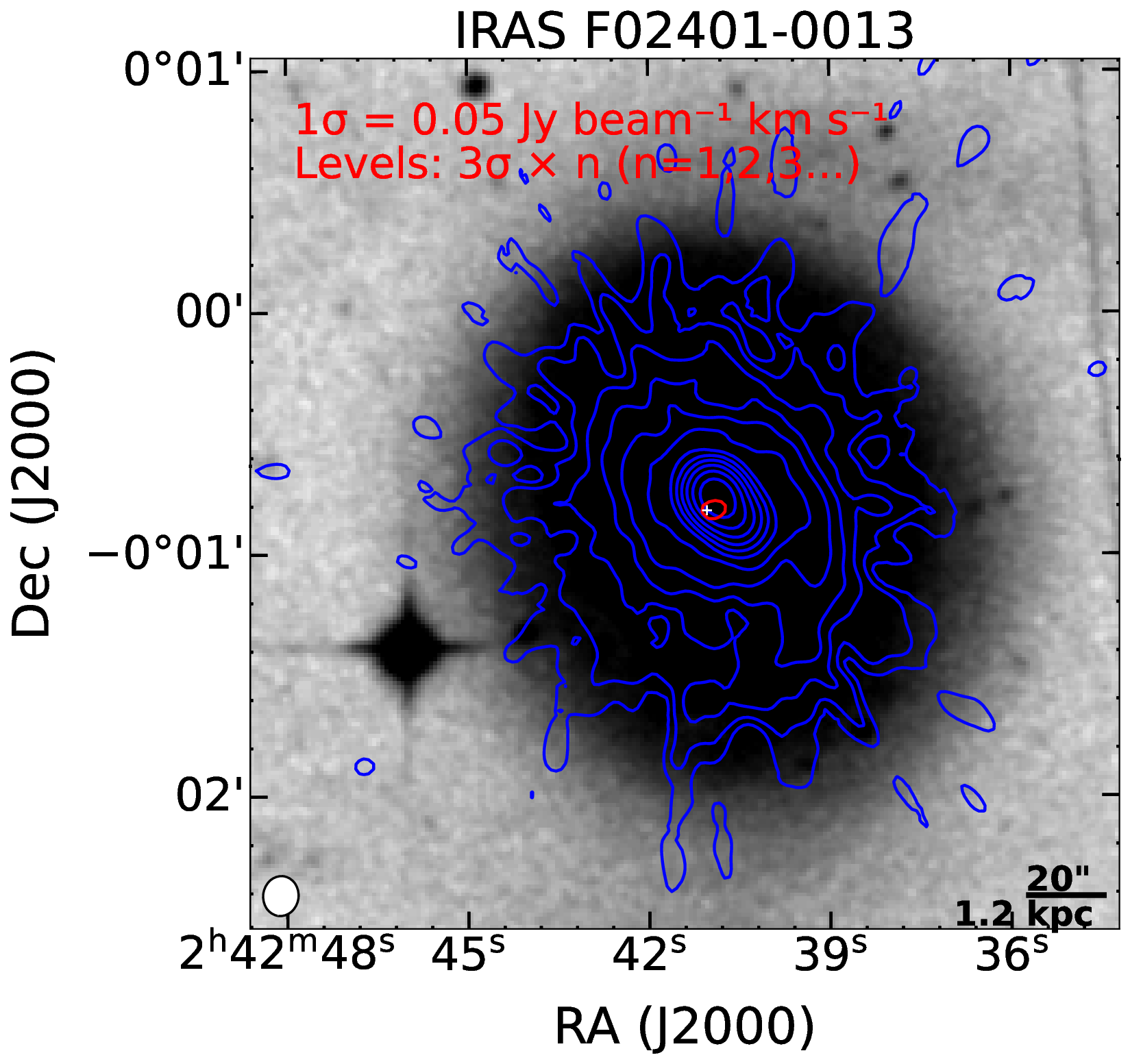}
    \end{minipage}%
  }
  \subcaptionbox*{}{%
    \begin{minipage}[t]{0.42\textwidth}
      \centering
      \includegraphics[width=\textwidth]{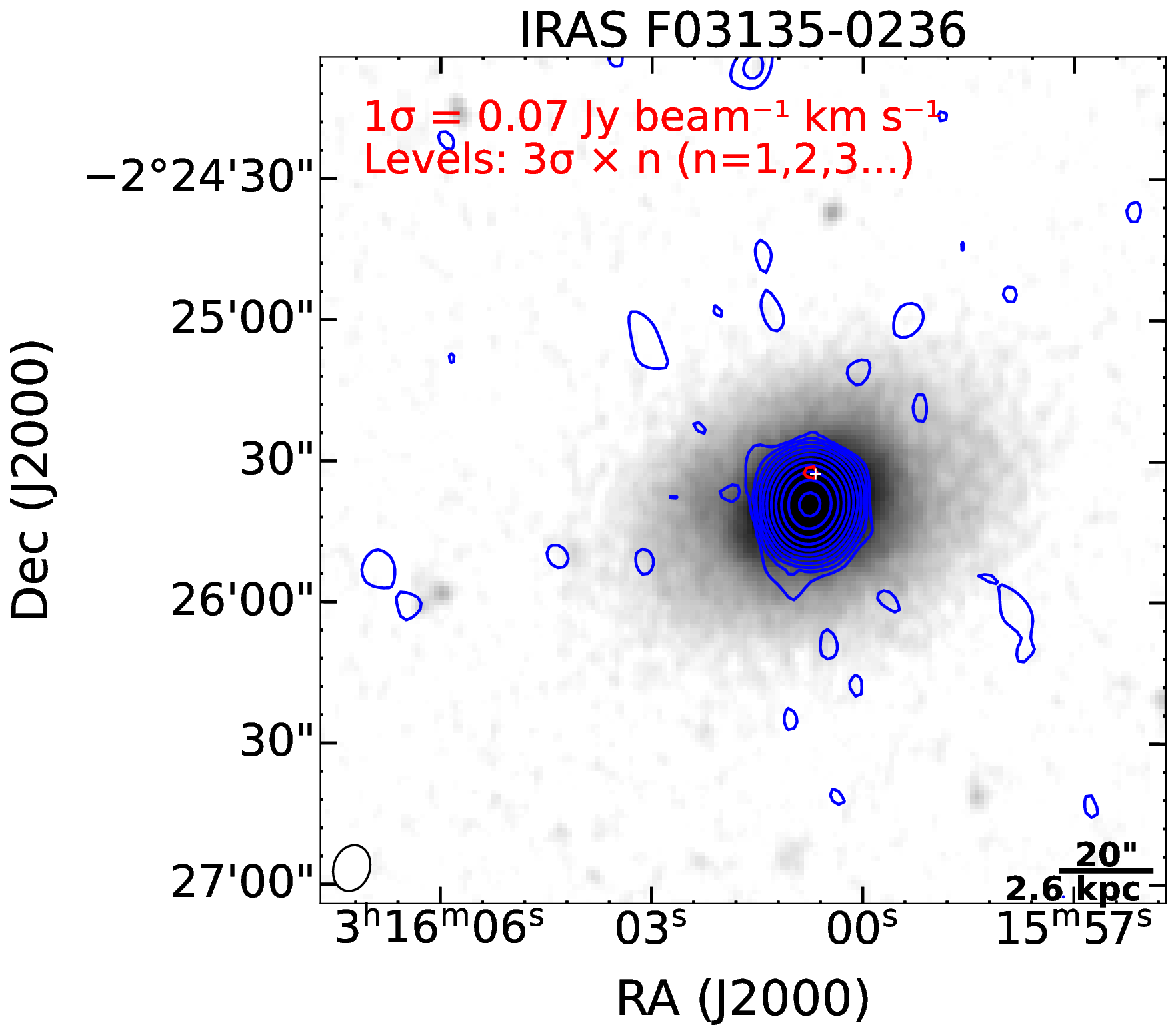}
    \end{minipage}%
    \hfill%
    \begin{minipage}[t]{0.42\textwidth}
      \centering
      \includegraphics[width=\textwidth]{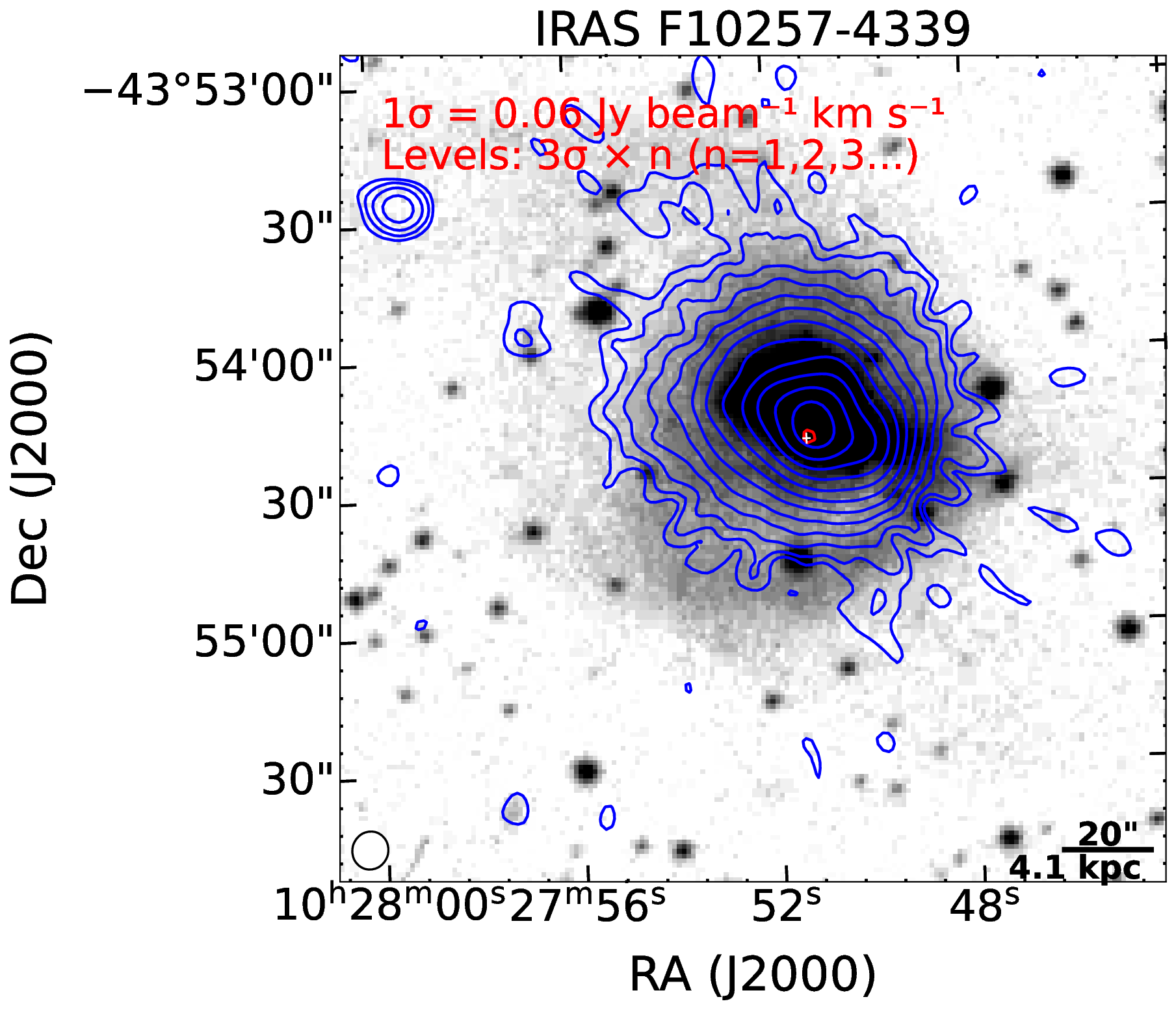}
    \end{minipage}%
  }
\caption{
Integrated OH emission (moment~0) contour maps of four OH maser galaxies, overlaid on DSS $R$-band grayscale images. 
Red contours show the OH emission detected in this work. The $1\sigma$ rms noise levels and contour levels of the OH maps are indicated in the upper left corner of each panel. 
Blue contours represent the radio continuum emission from the full-band MeerKAT data \citep{2021ApJS..257...35C}, with a typical $1\sigma$ rms noise level of $\sim$0.02~mJy~beam$^{-1}$. 
For IRAS 00450-2533 and IRAS 02401-0013, we initiate radio continuum contours at 4 $\sigma$ to avoid radial, spoke-like wide-band imaging artifacts, which are most pronounced in these two brightest continuum sources.
For the other two sources, the contours start at $3\sigma$ and increase by factors of two. 
White crosses mark the OH emission peaks. 
Synthesized beam sizes and fitted OH source sizes are listed in Table~\ref{tab:ohemission}.
}
\label{OH4image}
\end{figure*}

\begin{figure*}[htbp]
  \centering
  \begin{minipage}[t]{0.31\textwidth}
    \centering
    \includegraphics[width=\textwidth]{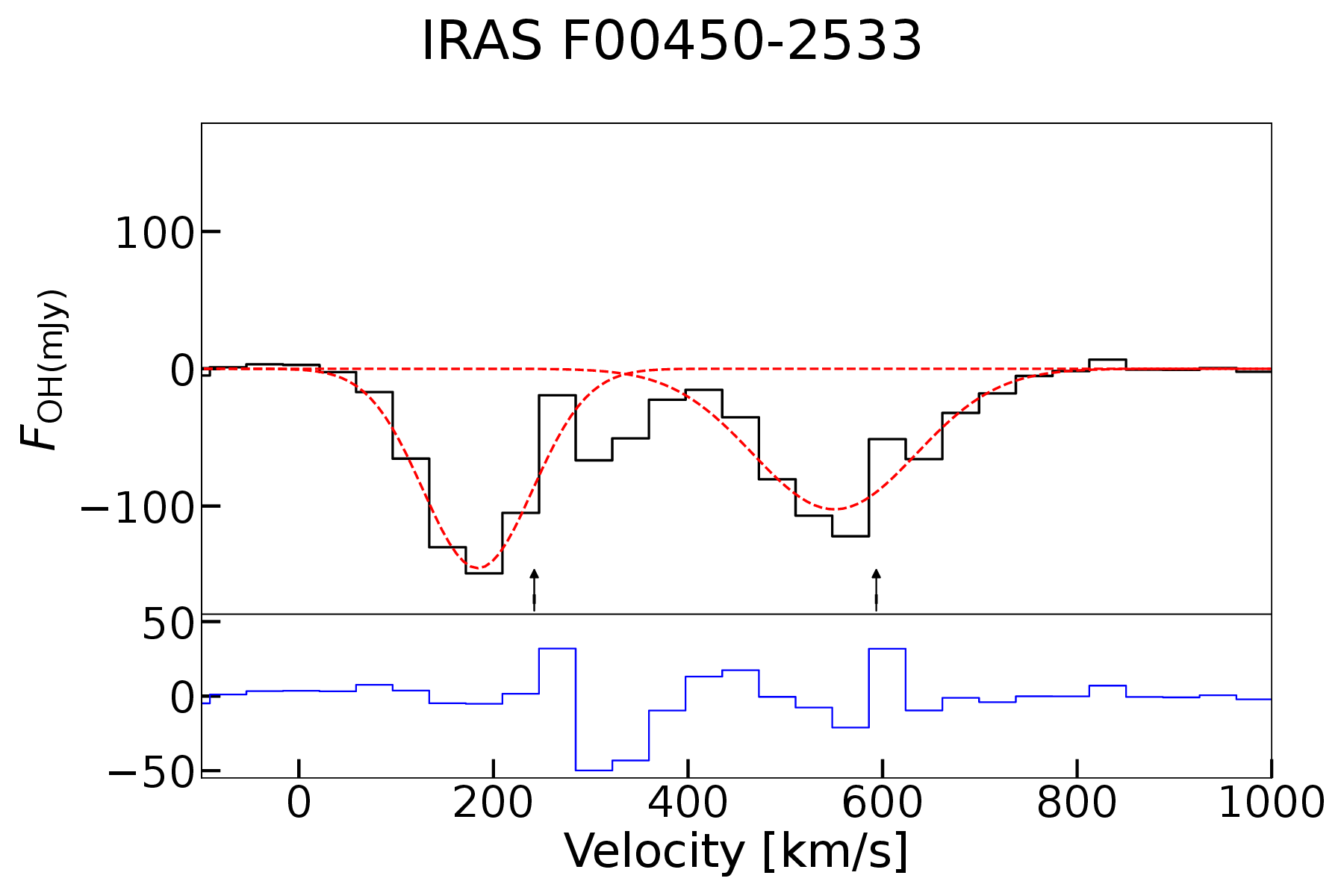}
  \end{minipage}\hfill
  \begin{minipage}[t]{0.31\textwidth}
    \centering
    \includegraphics[width=\textwidth]{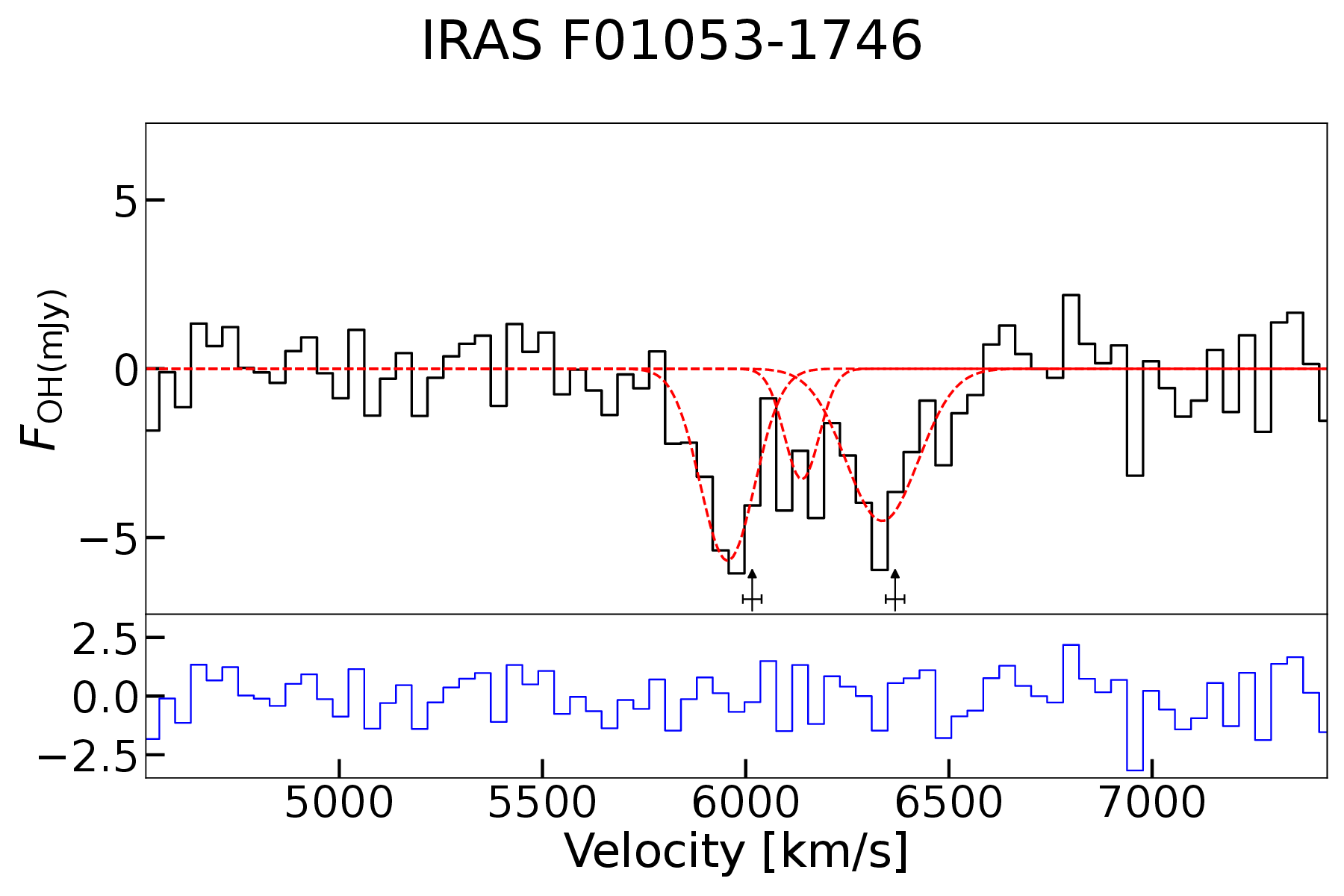}
  \end{minipage}\hfill
  \begin{minipage}[t]{0.31\textwidth}
    \centering
    \includegraphics[width=\textwidth]{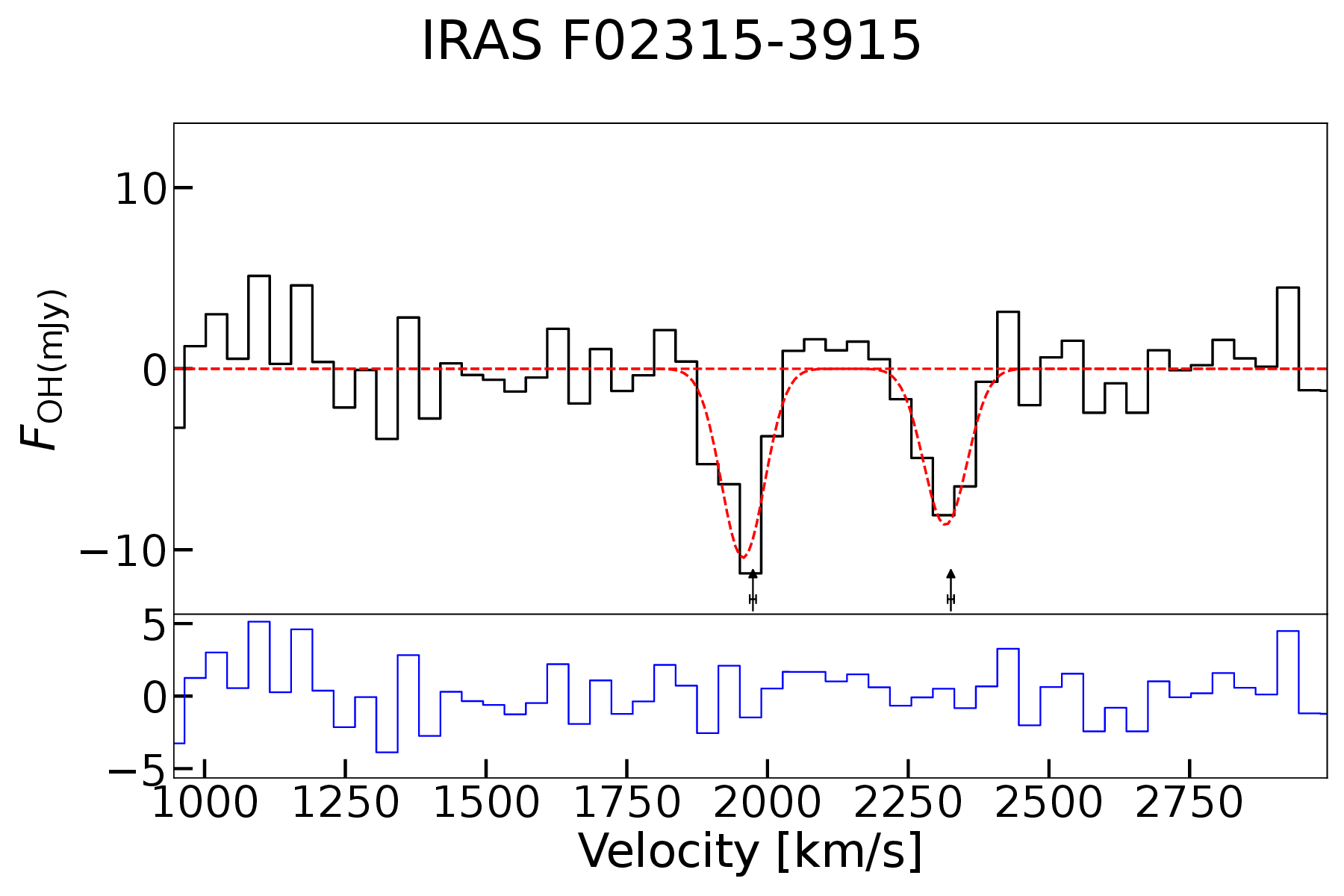}
  \end{minipage}\par
  \begin{minipage}[t]{0.31\textwidth}
    \centering
    \includegraphics[width=\textwidth]{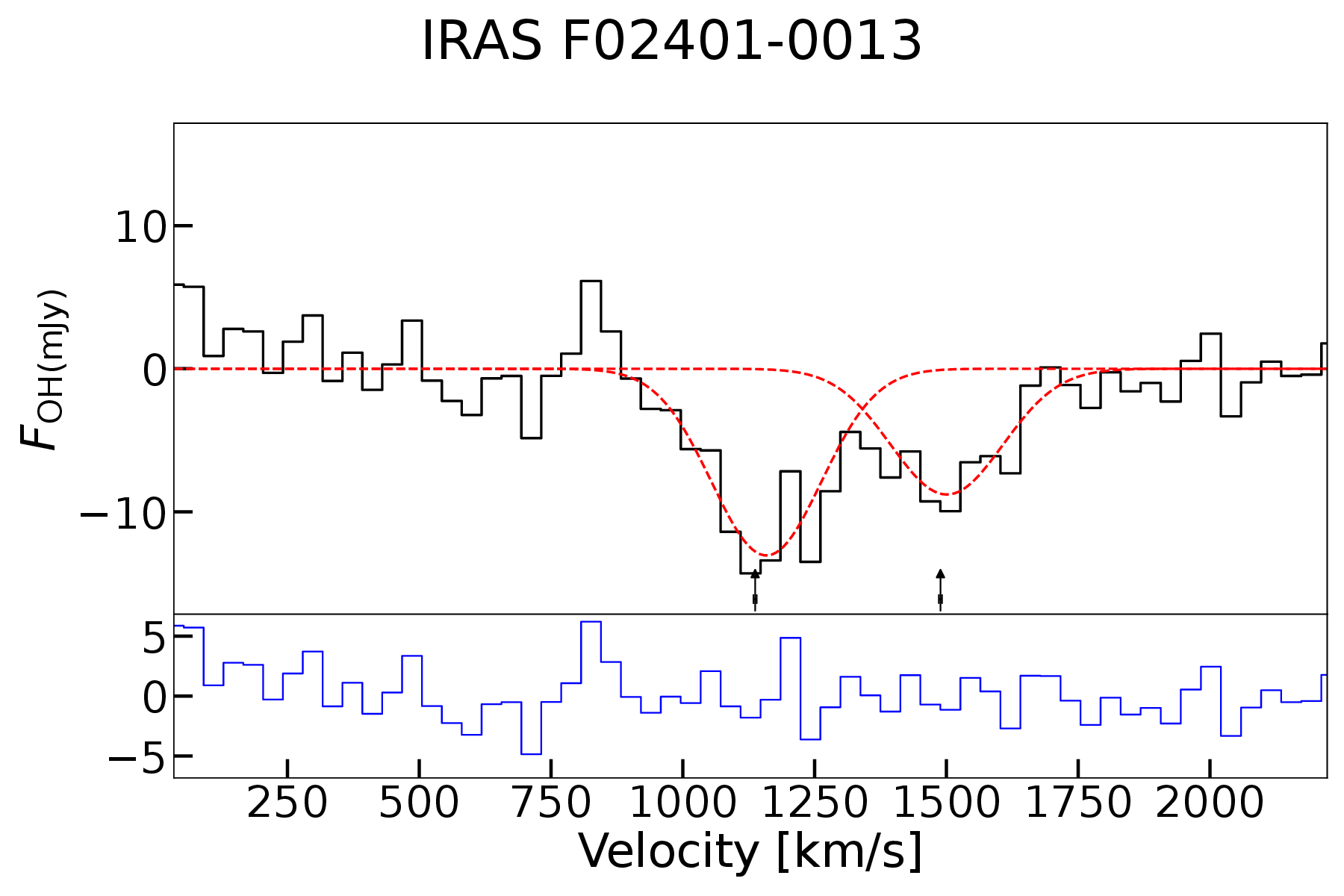}
  \end{minipage}\hfill
  \begin{minipage}[t]{0.31\textwidth}
    \centering
    \includegraphics[width=\textwidth]{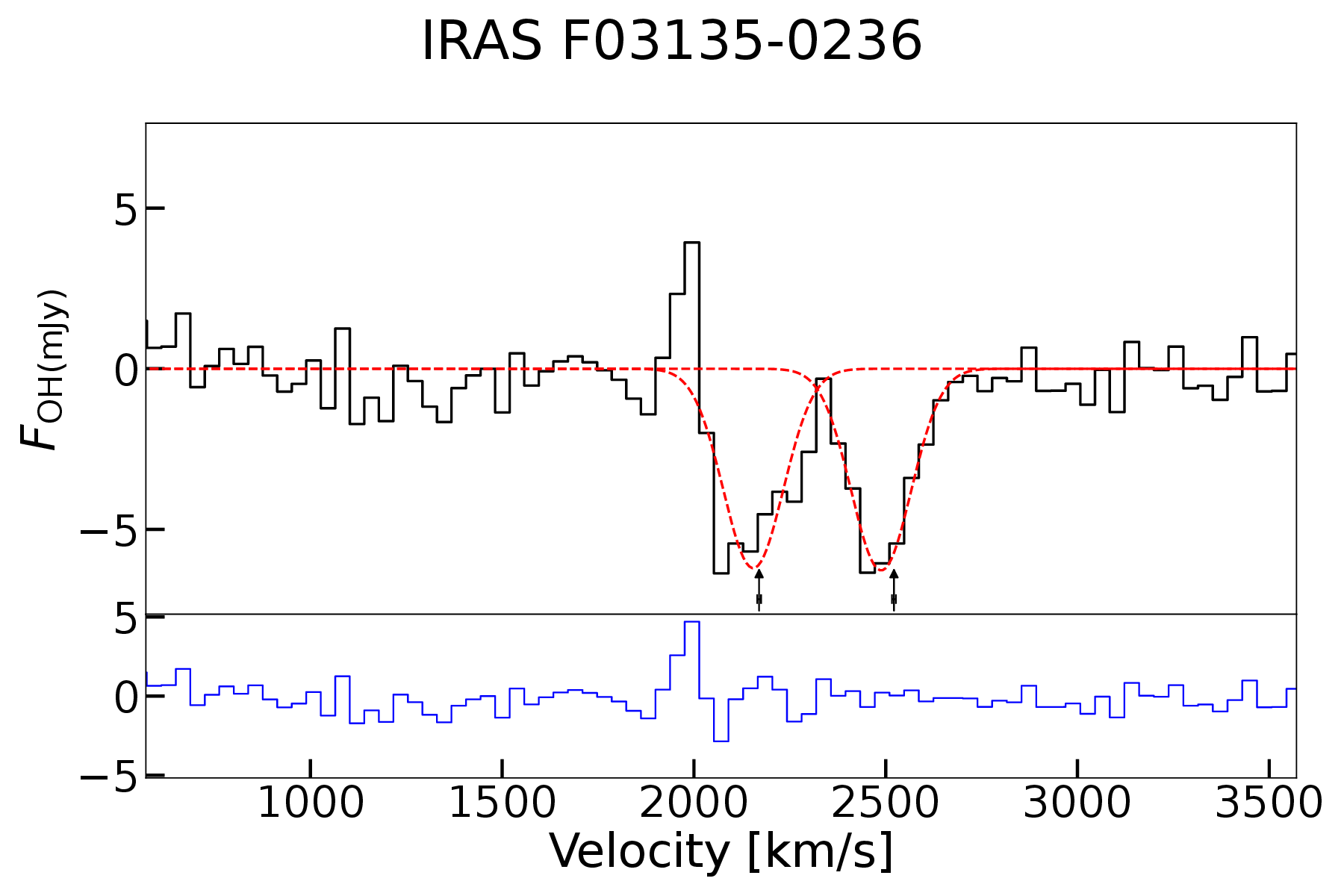}
  \end{minipage}\hfill
  \begin{minipage}[t]{0.31\textwidth}
    \centering
    \includegraphics[width=\textwidth]{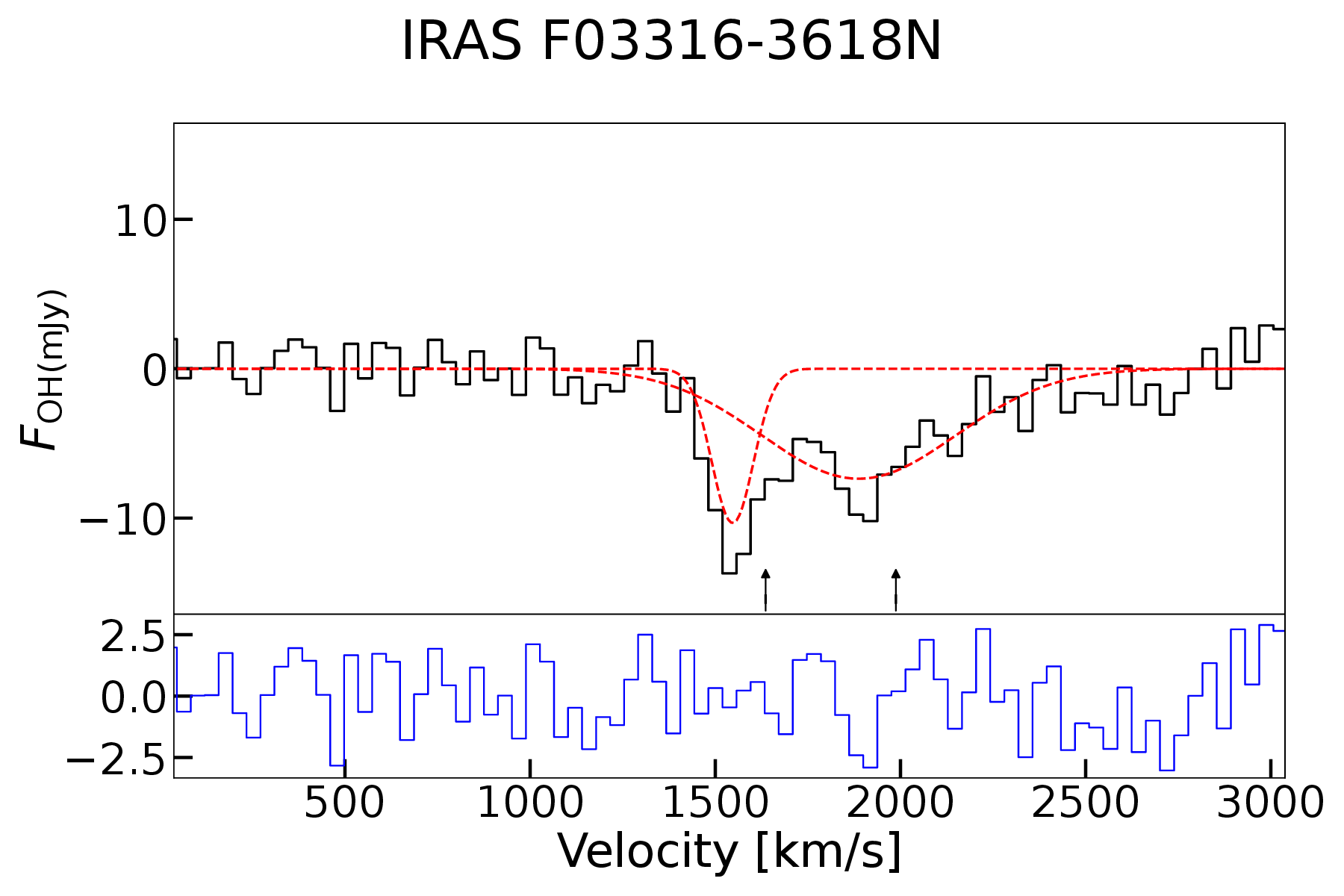}
  \end{minipage}\par
  \begin{minipage}[t]{0.31\textwidth}
    \centering
    \includegraphics[width=\textwidth]{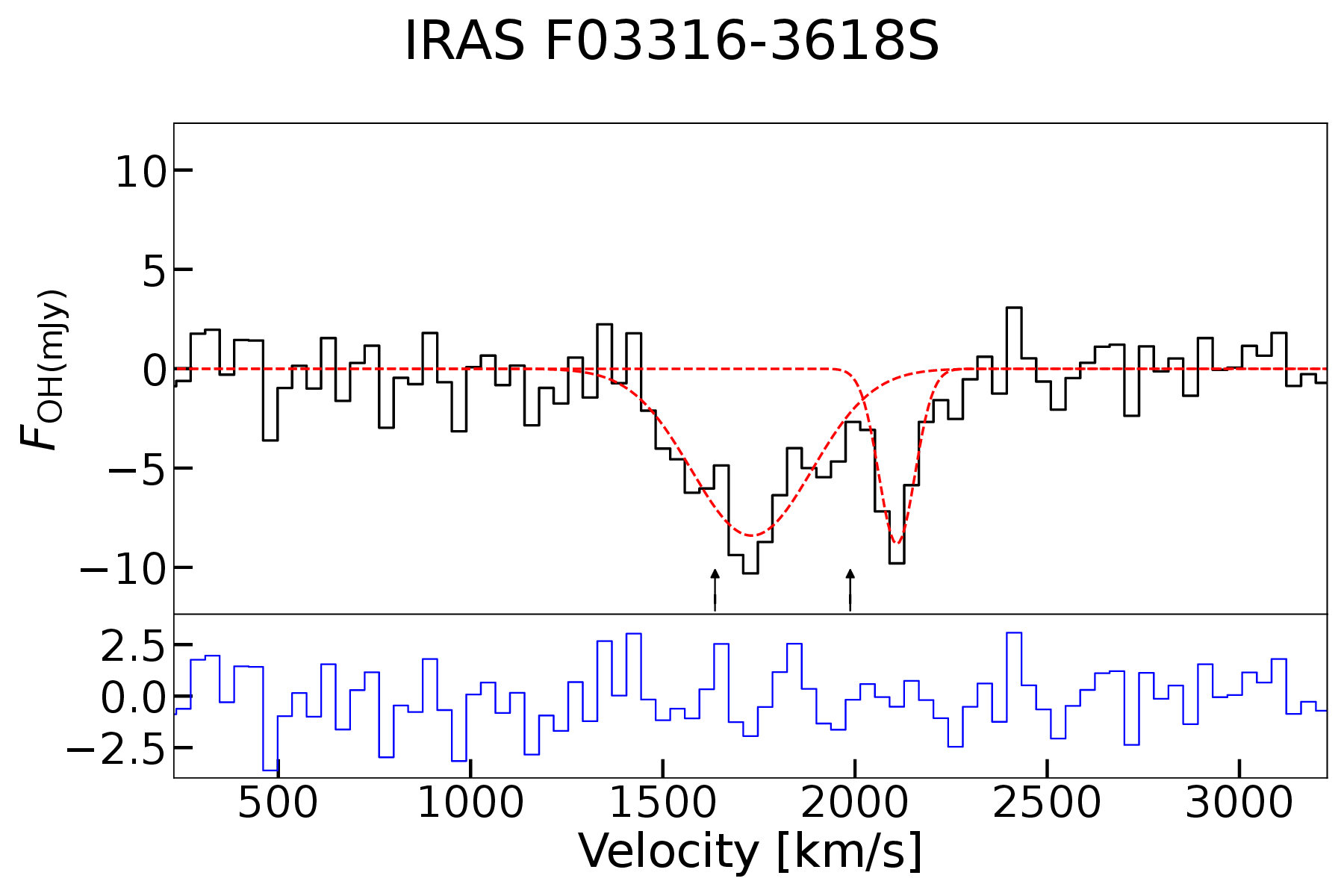}
  \end{minipage}\hfill
  \begin{minipage}[t]{0.31\textwidth}
    \centering
    \includegraphics[width=\textwidth]{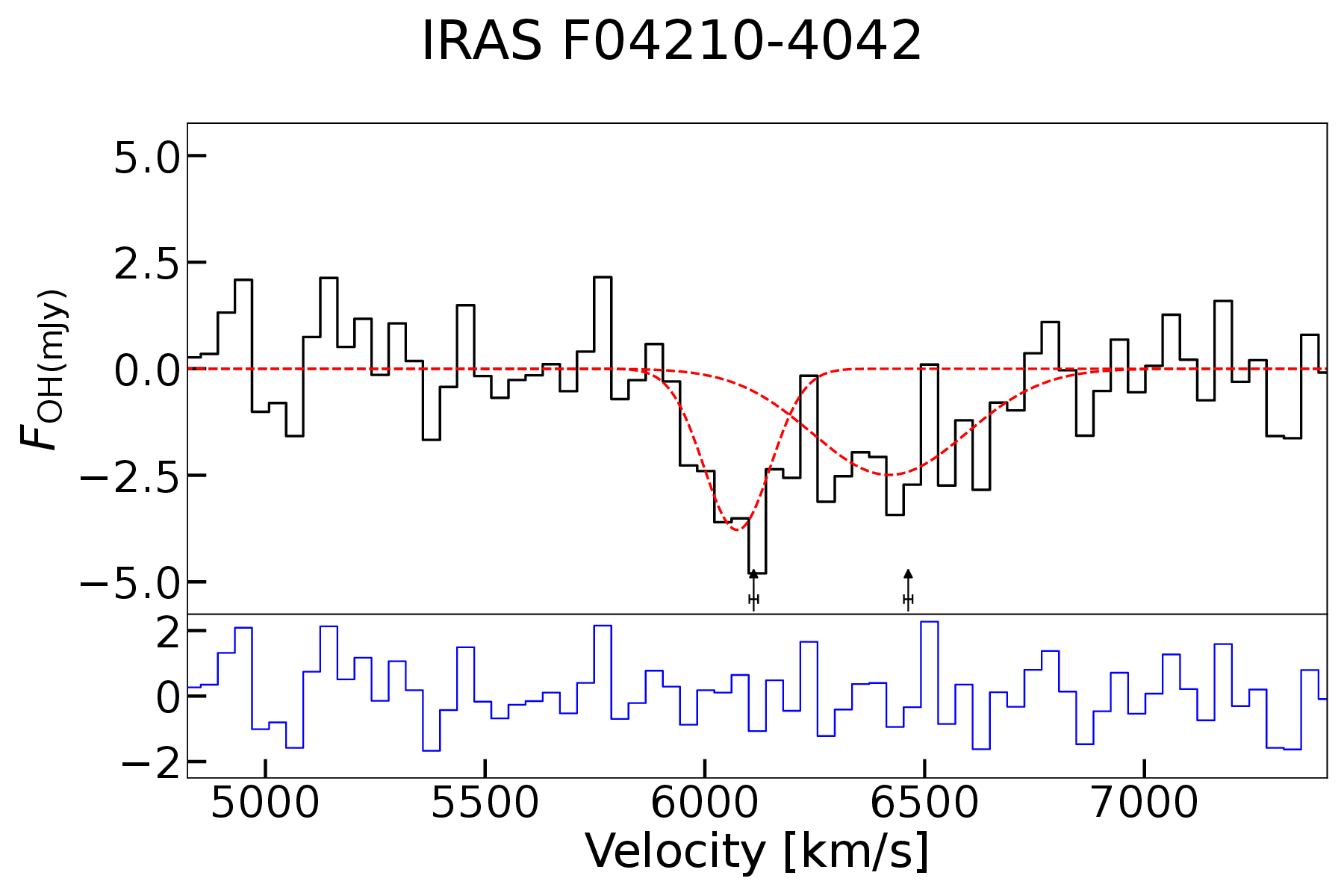}
  \end{minipage}\hfill
  \begin{minipage}[t]{0.31\textwidth}
    \centering
    \includegraphics[width=\textwidth]{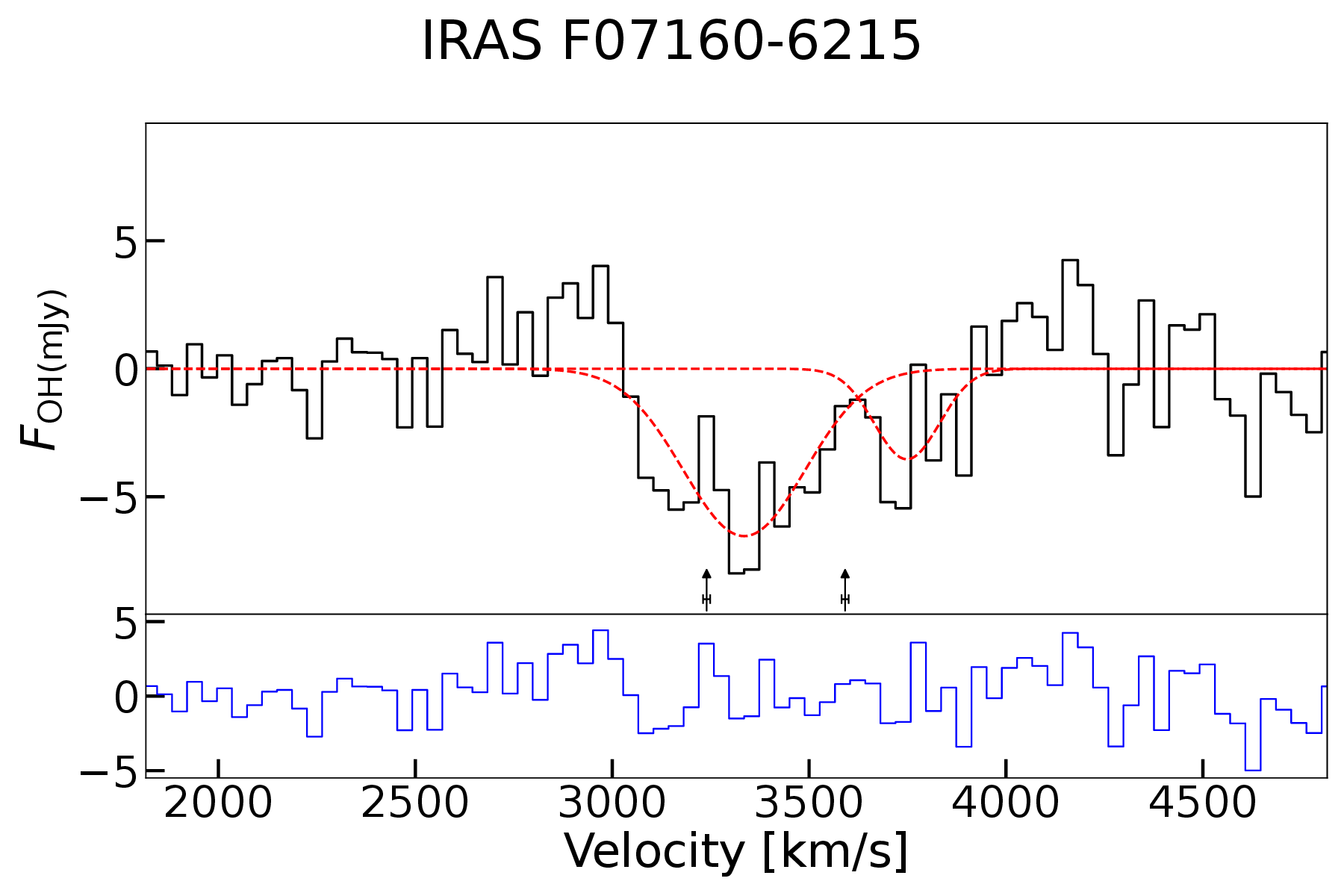}
  \end{minipage}\par
  \begin{minipage}[t]{0.31\textwidth}
    \centering
    \includegraphics[width=\textwidth]{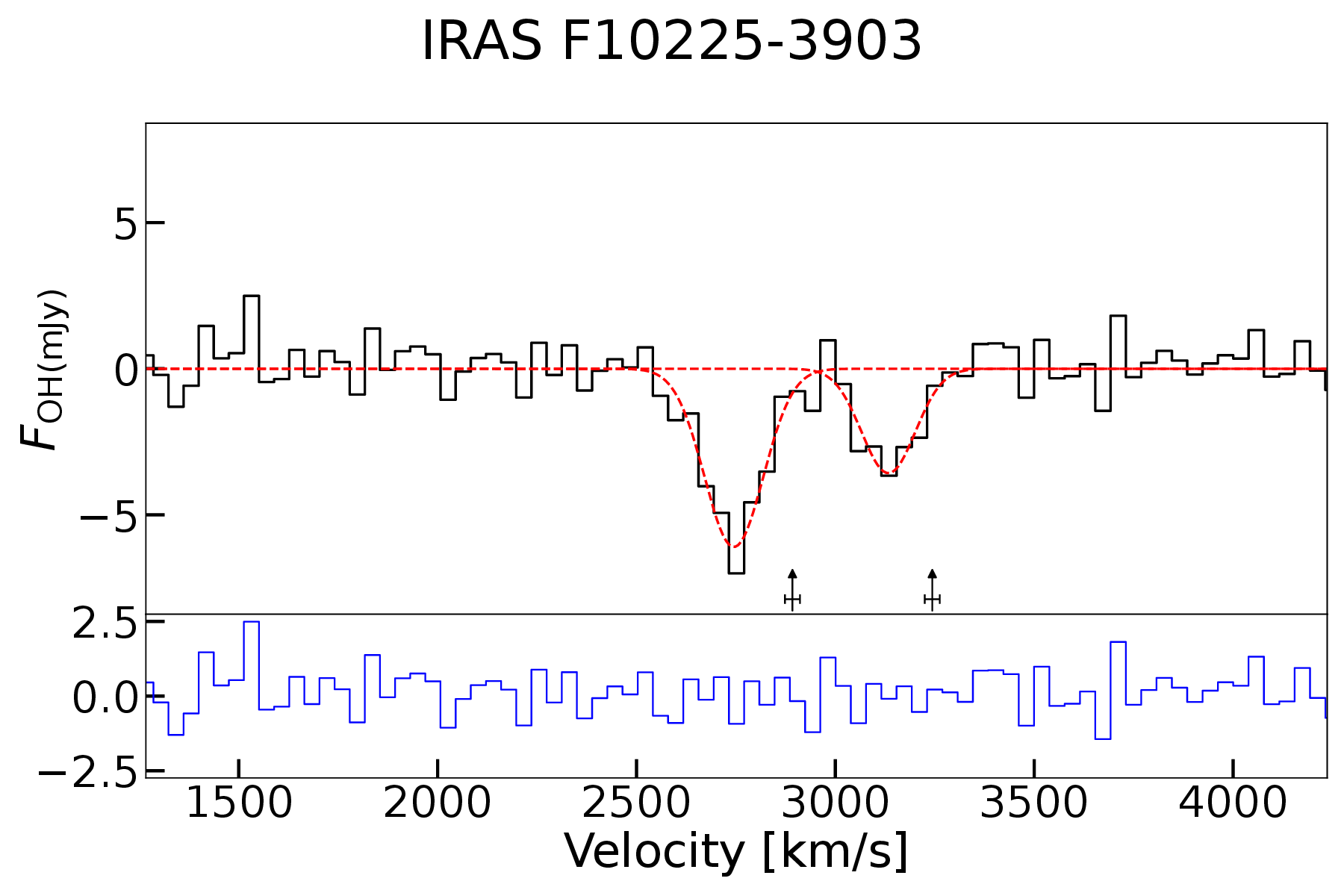}
  \end{minipage}\hfill
  \begin{minipage}[t]{0.31\textwidth}
    \centering
    \includegraphics[width=\textwidth]{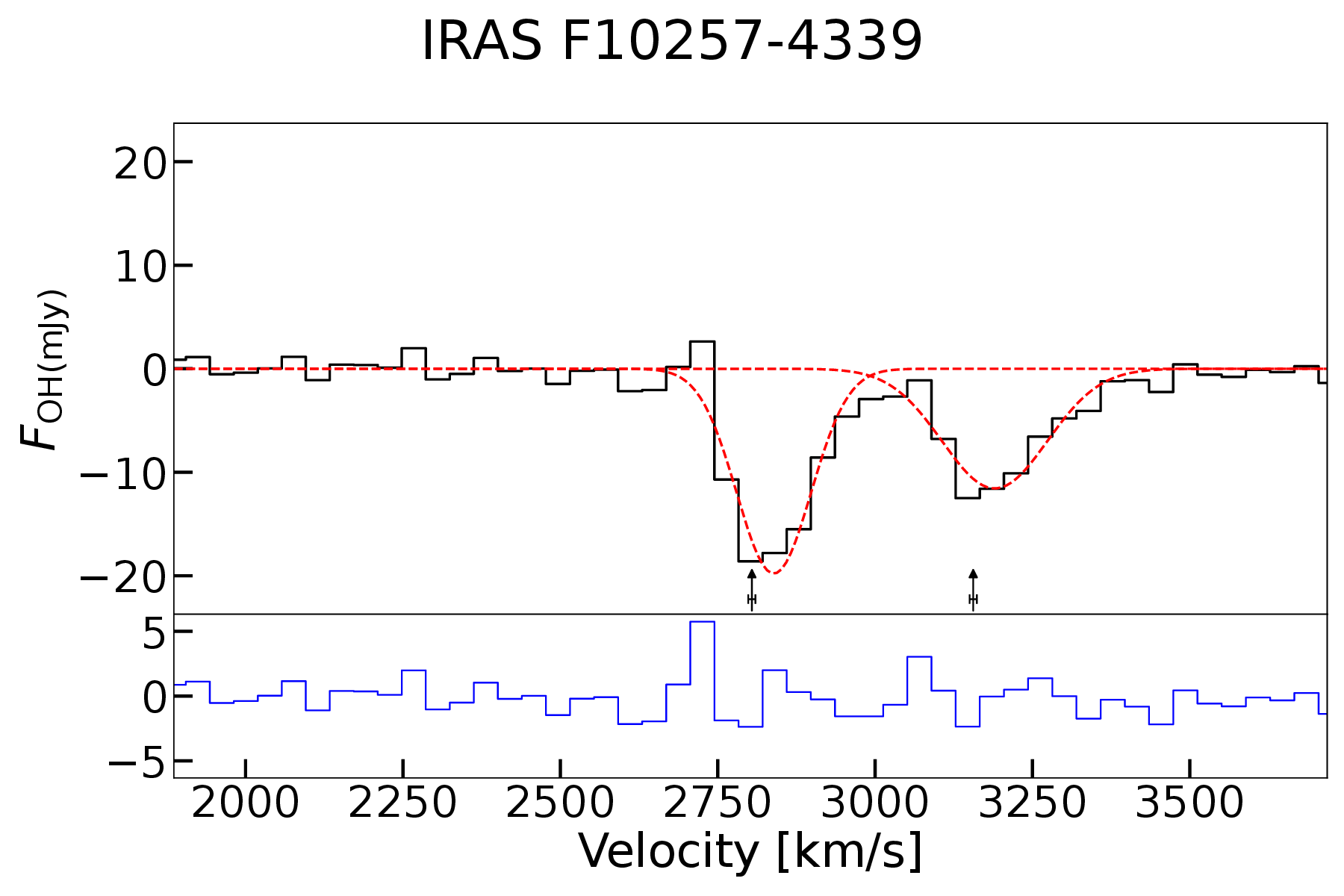}
  \end{minipage}\hfill
  \begin{minipage}[t]{0.31\textwidth}
    \centering
    \includegraphics[width=\textwidth]{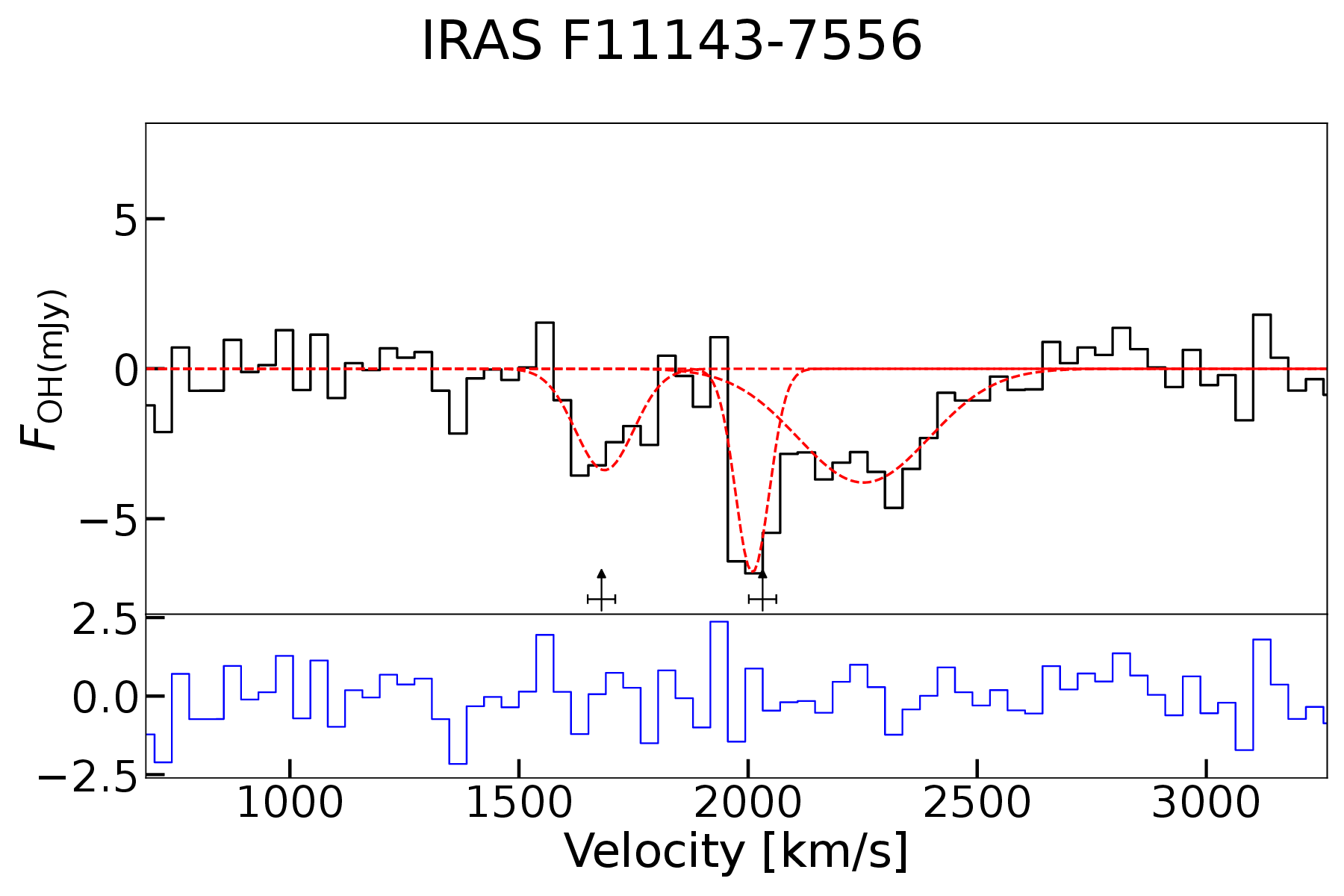}
  \end{minipage}\par
  \begin{minipage}[t]{0.31\textwidth}
    \centering
    \includegraphics[width=\textwidth]{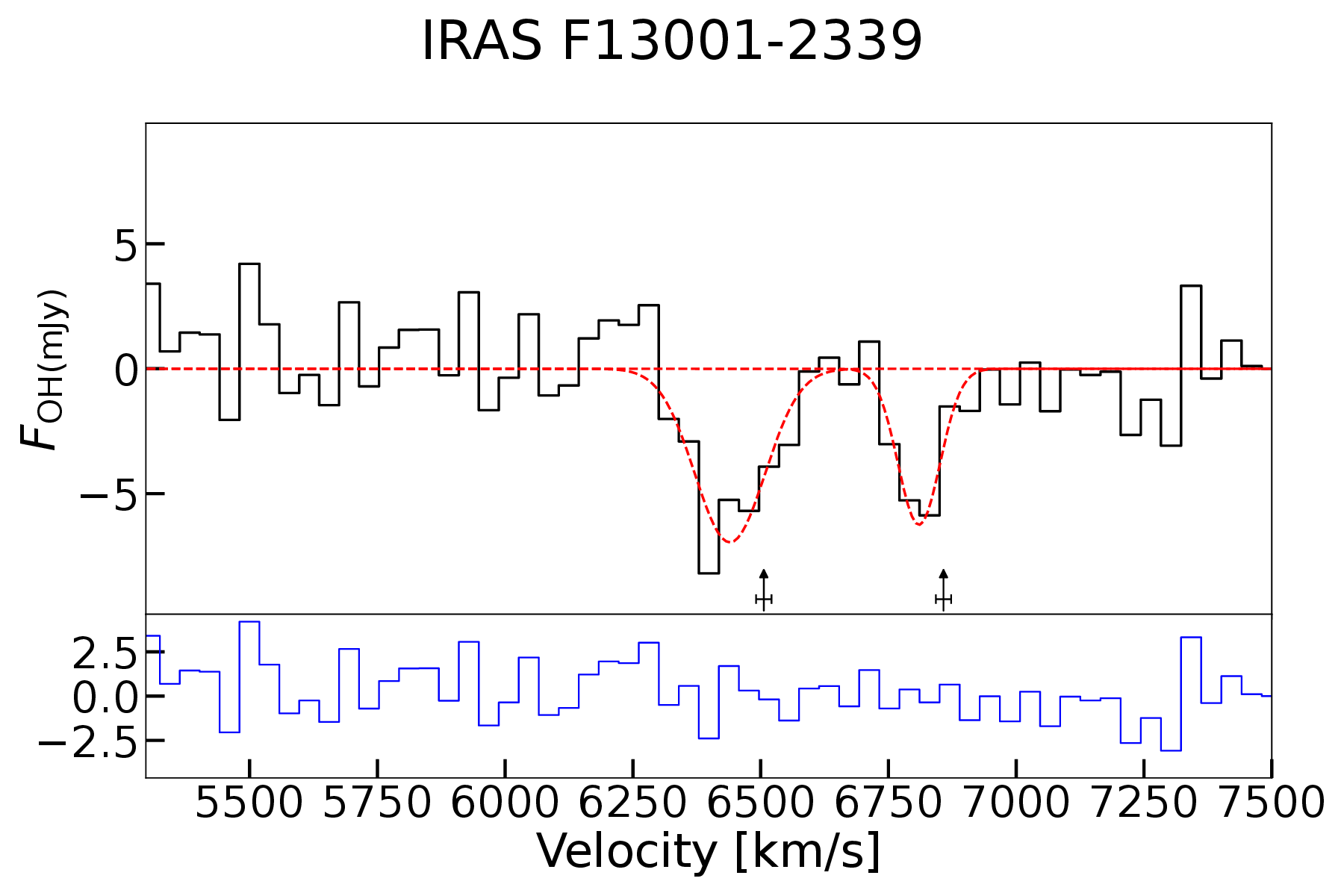}
  \end{minipage}\hfill
  \begin{minipage}[t]{0.31\textwidth}
    \centering
    \includegraphics[width=\textwidth]{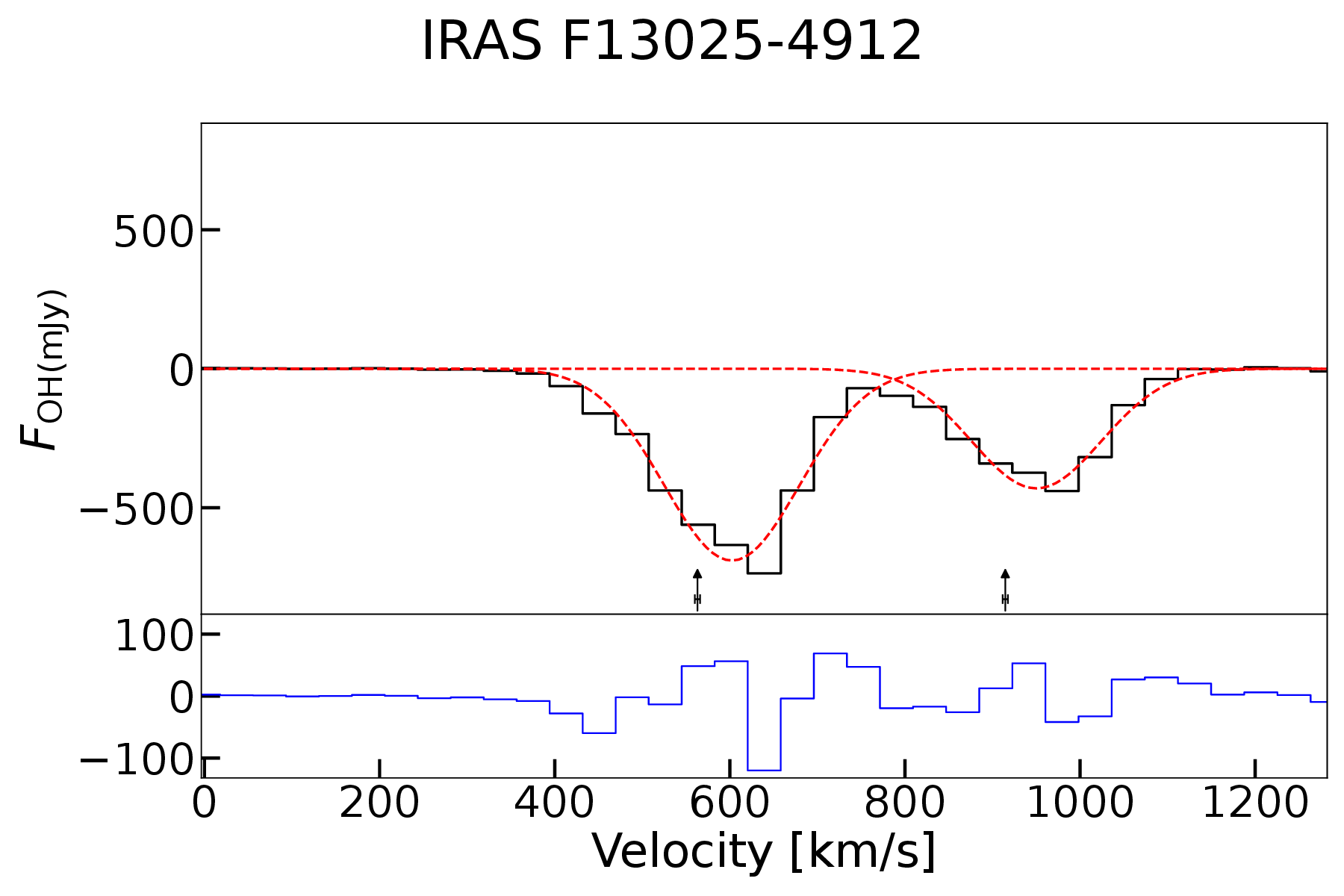}
  \end{minipage}\hfill
  \begin{minipage}[t]{0.31\textwidth}
    \centering
    \includegraphics[width=\textwidth]{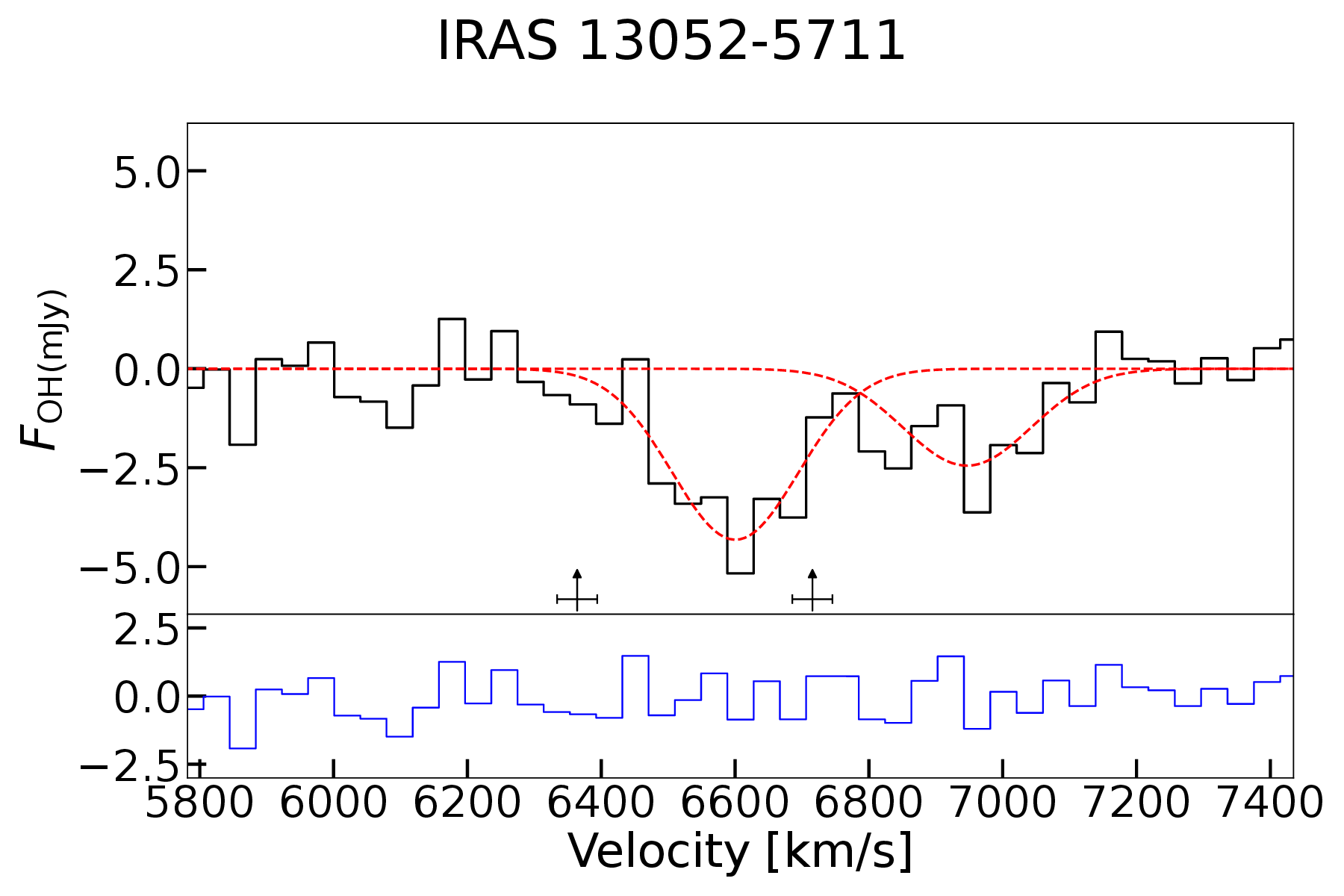}
  \end{minipage}\par
  \begin{minipage}[t]{0.31\textwidth}
    \centering
    \includegraphics[width=\textwidth]{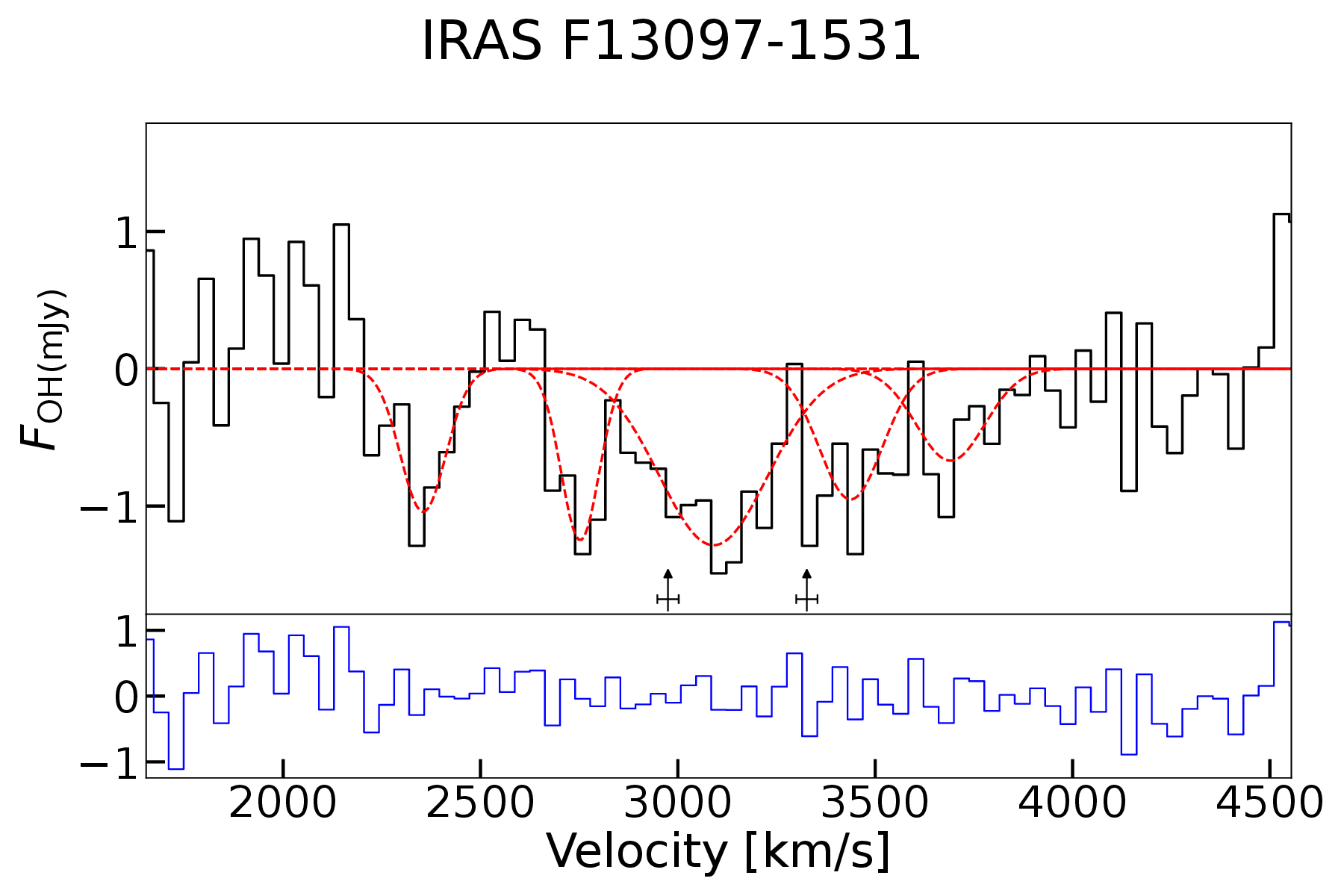}
  \end{minipage}\hfill
  \begin{minipage}[t]{0.31\textwidth}
    \centering
    \includegraphics[width=\textwidth]{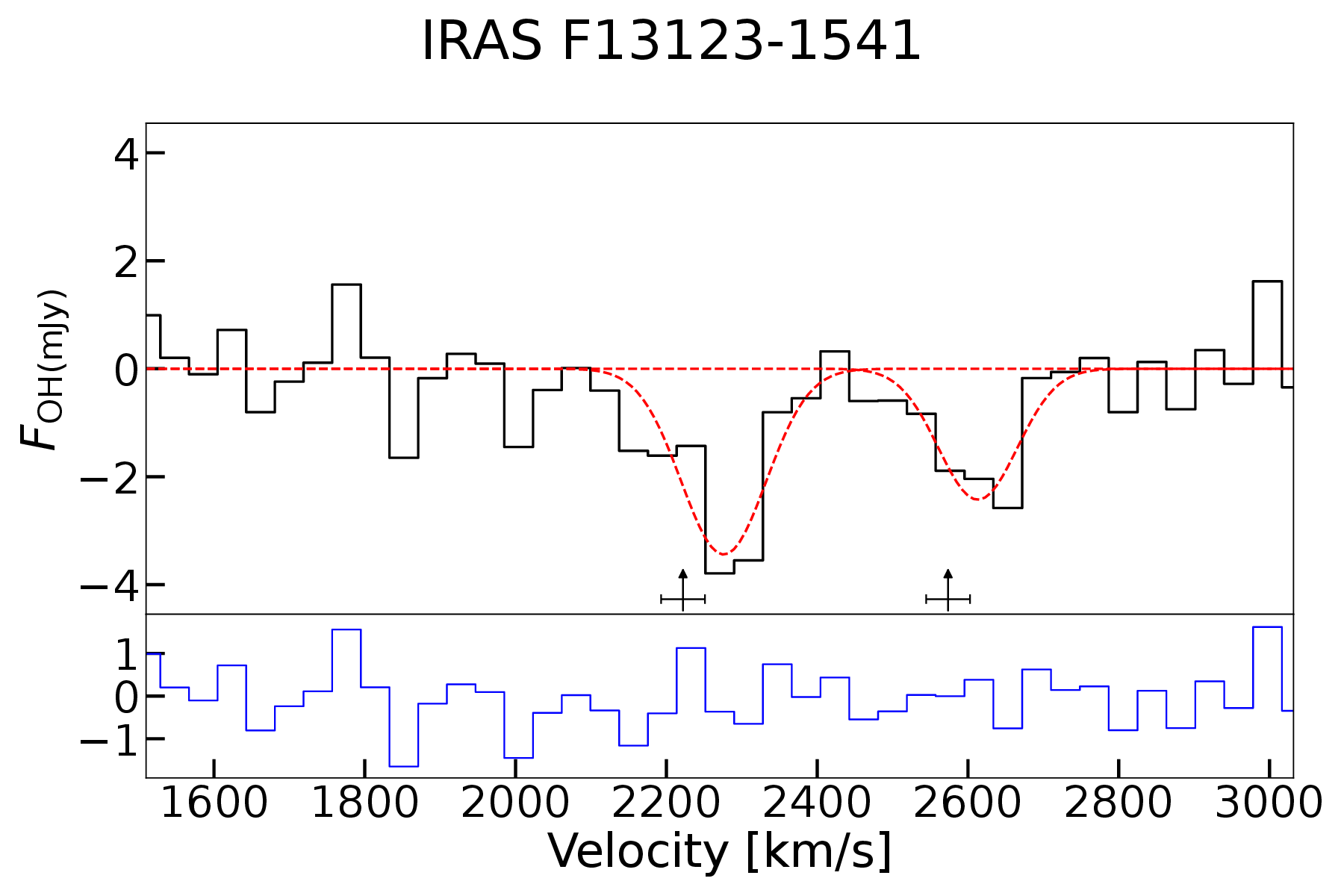}
  \end{minipage}\hfill
  \begin{minipage}[t]{0.31\textwidth}
    \centering
    \includegraphics[width=\textwidth]{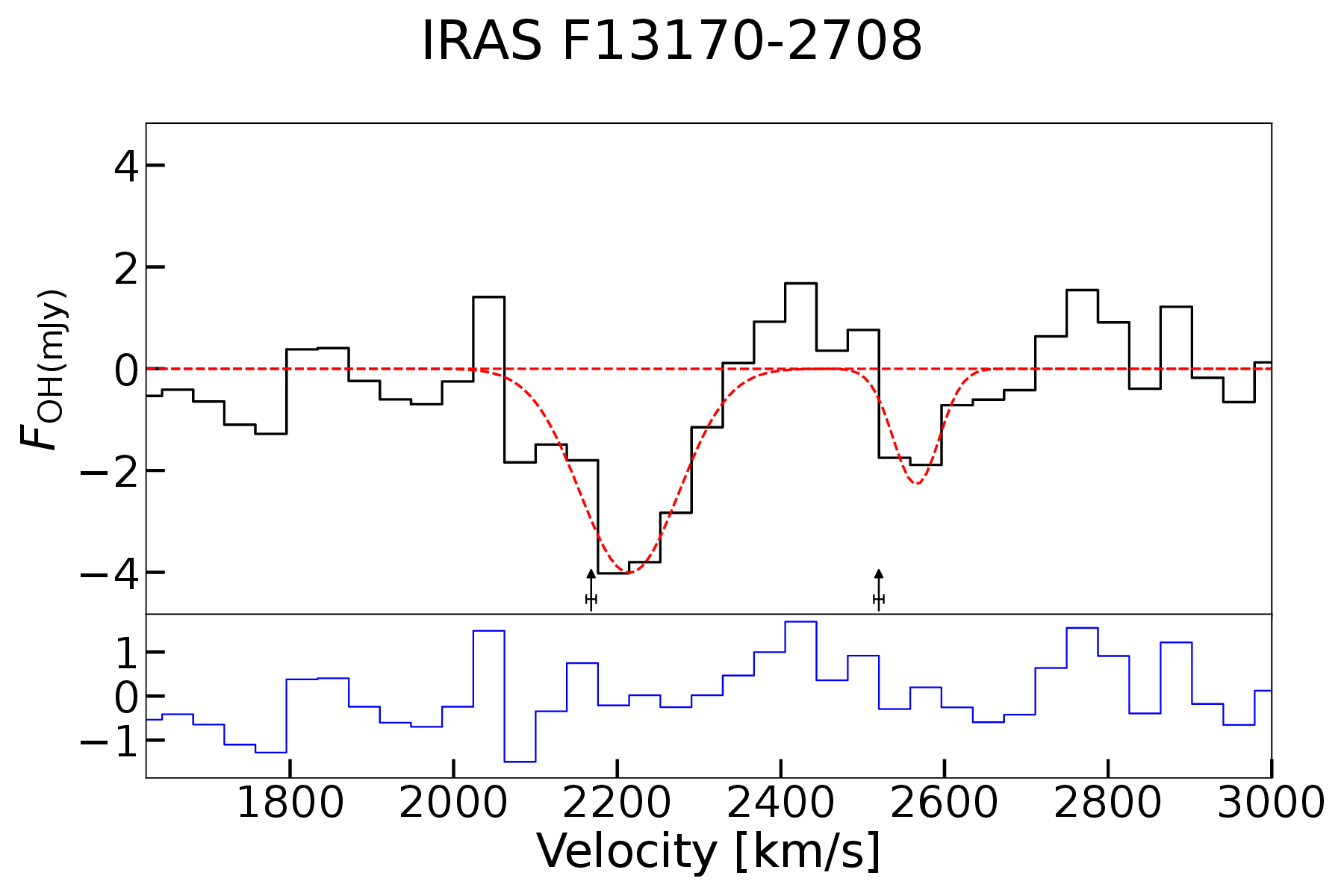}
  \end{minipage}%

  \caption{High S/N OH absorption line profiles. Velocities are given in
the barycentric reference frame using the optical convention. The black dotted lines show the observed OH spectra. The red dotted lines indicate the best-fitting Gaussian absorption components. The blue curves at the bottom of each panel represent the residual spectra (data$-$model). Arrows mark the expected positions of the OH main lines at 1667 MHz (left) and 1665 MHz (right), with error bars indicating the uncertainty in the redshift. The six additional low-S/N OH absorption line profiles are available on Zenodo: \href{https://doi.org/10.5281/zenodo.19904867}{doi:10.5281/zenodo.19904867}. }
  \label{fig:absline1}
\end{figure*}

\begin{figure*}[htbp]\ContinuedFloat
  \centering
  \begin{minipage}[t]{0.31\textwidth}
    \centering
    \includegraphics[width=\textwidth]{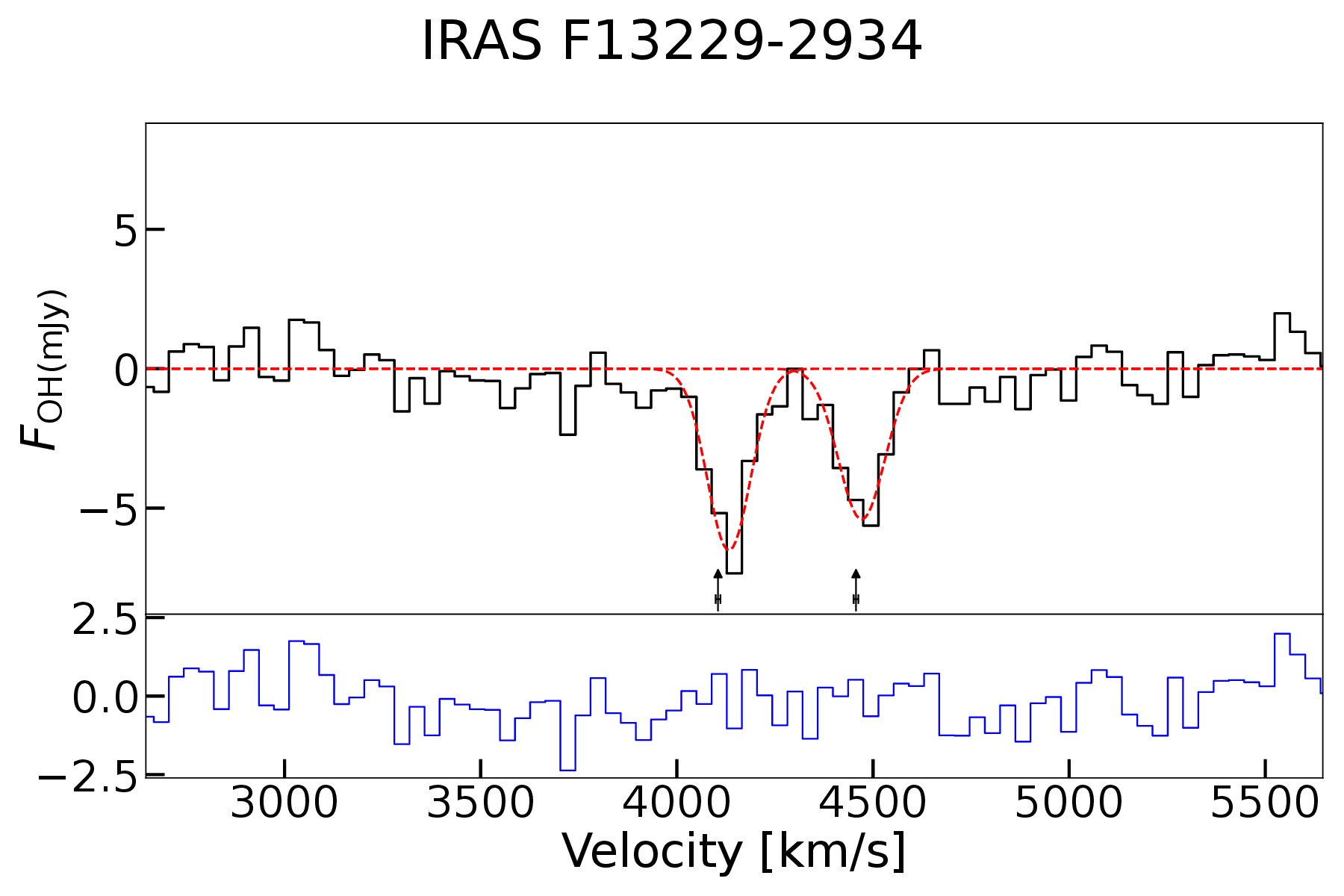}
  \end{minipage}\hfill
  \begin{minipage}[t]{0.31\textwidth}
    \centering
    \includegraphics[width=\textwidth]{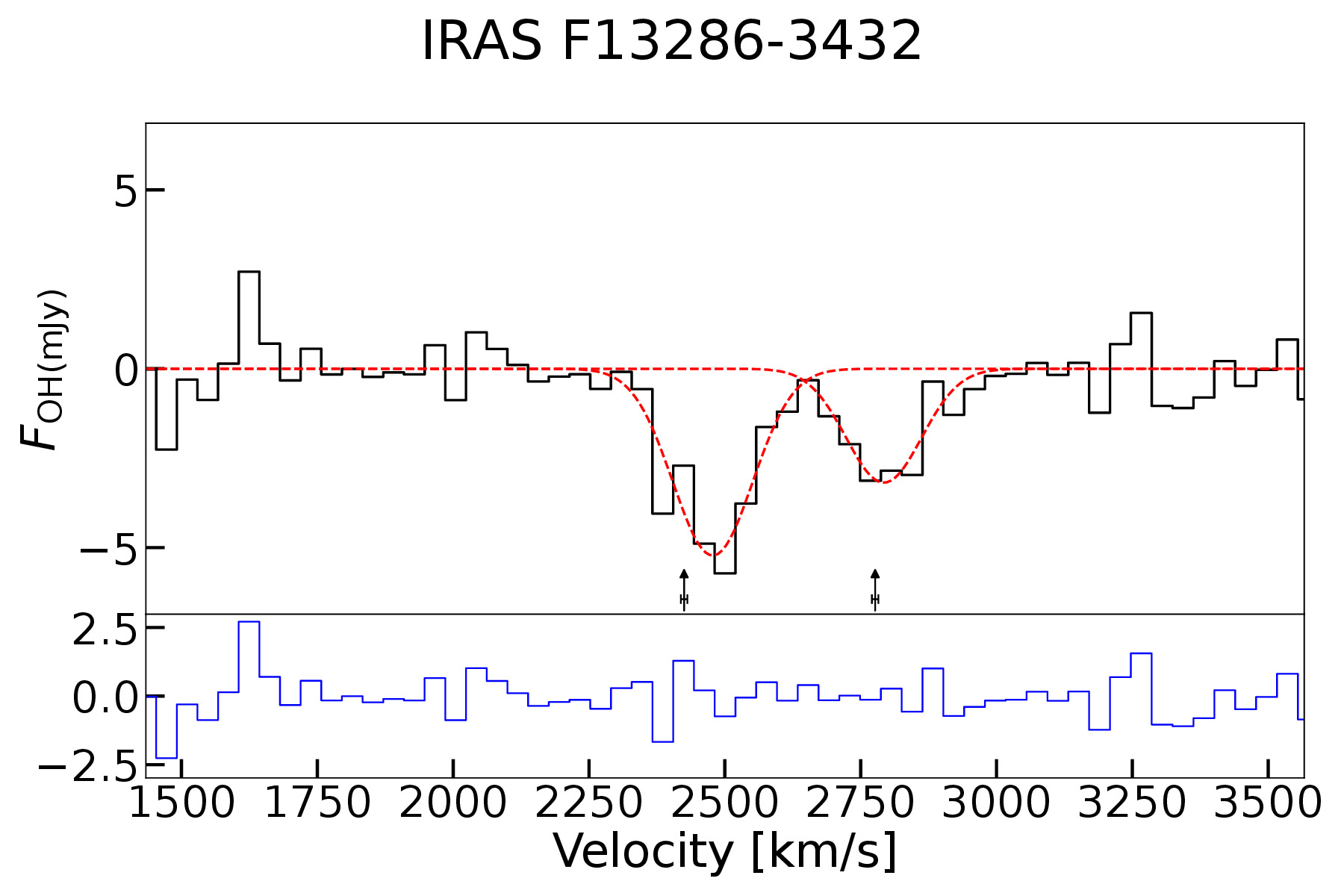}
  \end{minipage}\hfill
  \begin{minipage}[t]{0.31\textwidth}
    \centering
    \includegraphics[width=\textwidth]{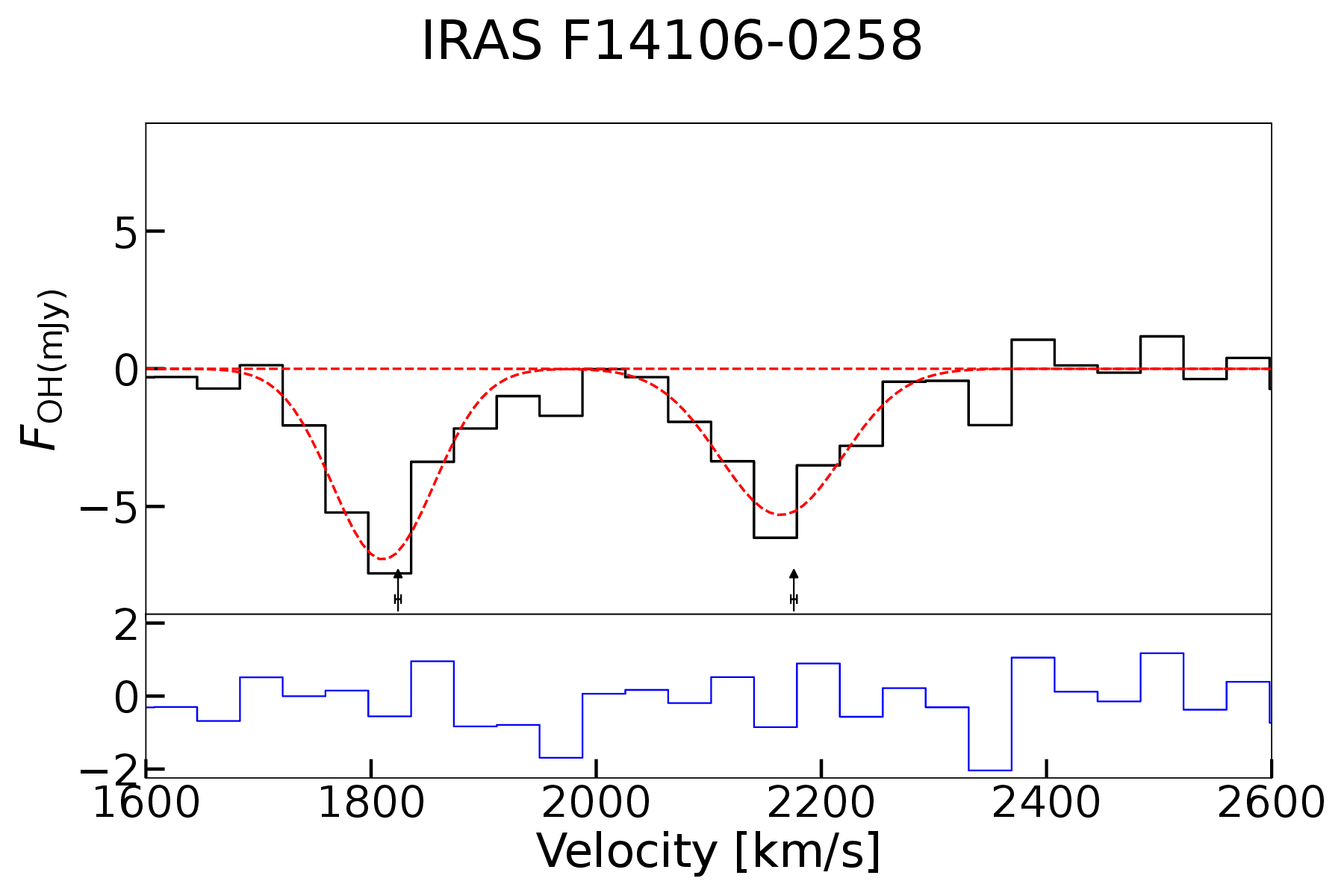}
  \end{minipage}\par
  \begin{minipage}[t]{0.31\textwidth}
    \centering
    \includegraphics[width=\textwidth]{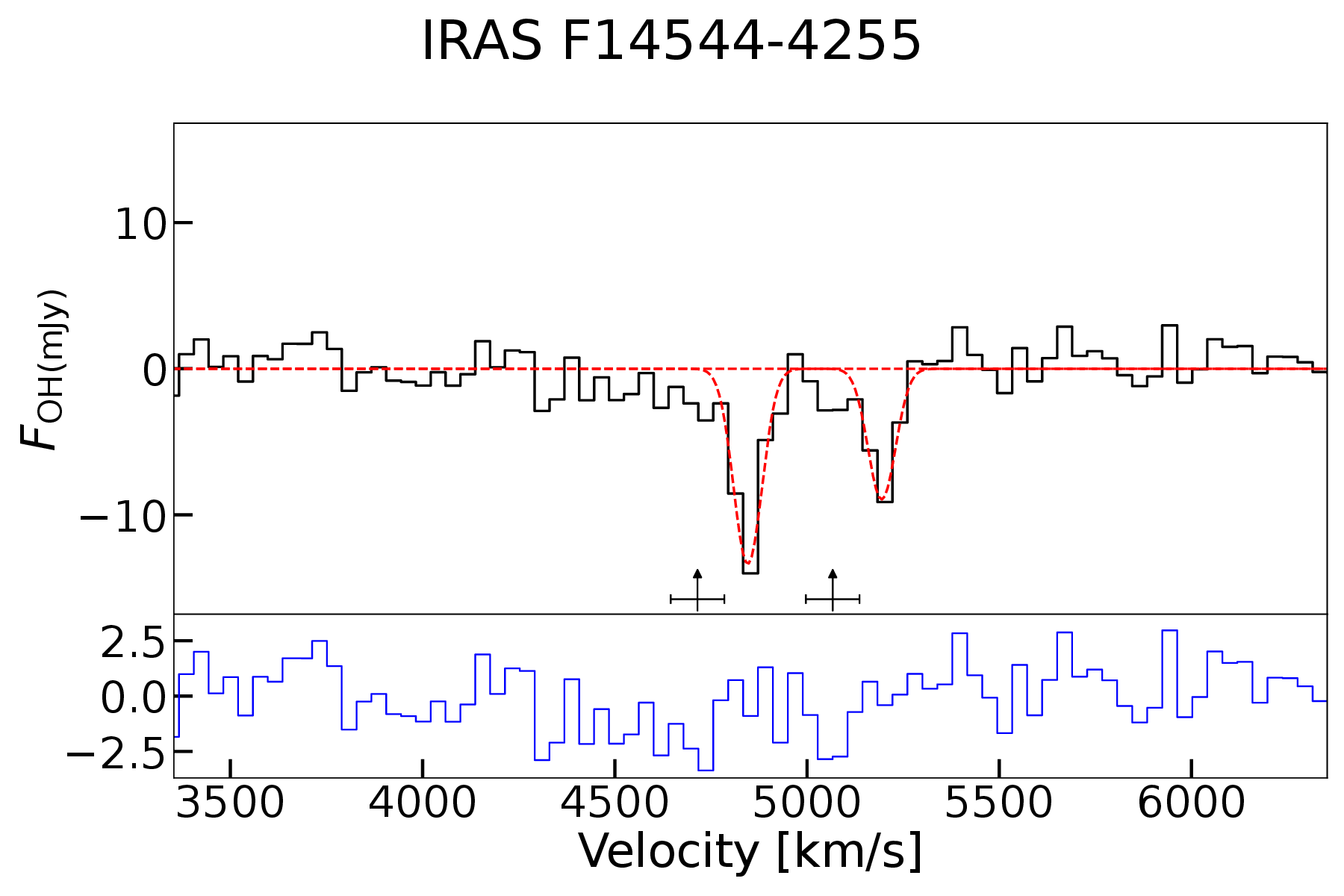}
  \end{minipage}\hfill
  \begin{minipage}[t]{0.31\textwidth}
    \centering
    \includegraphics[width=\textwidth]{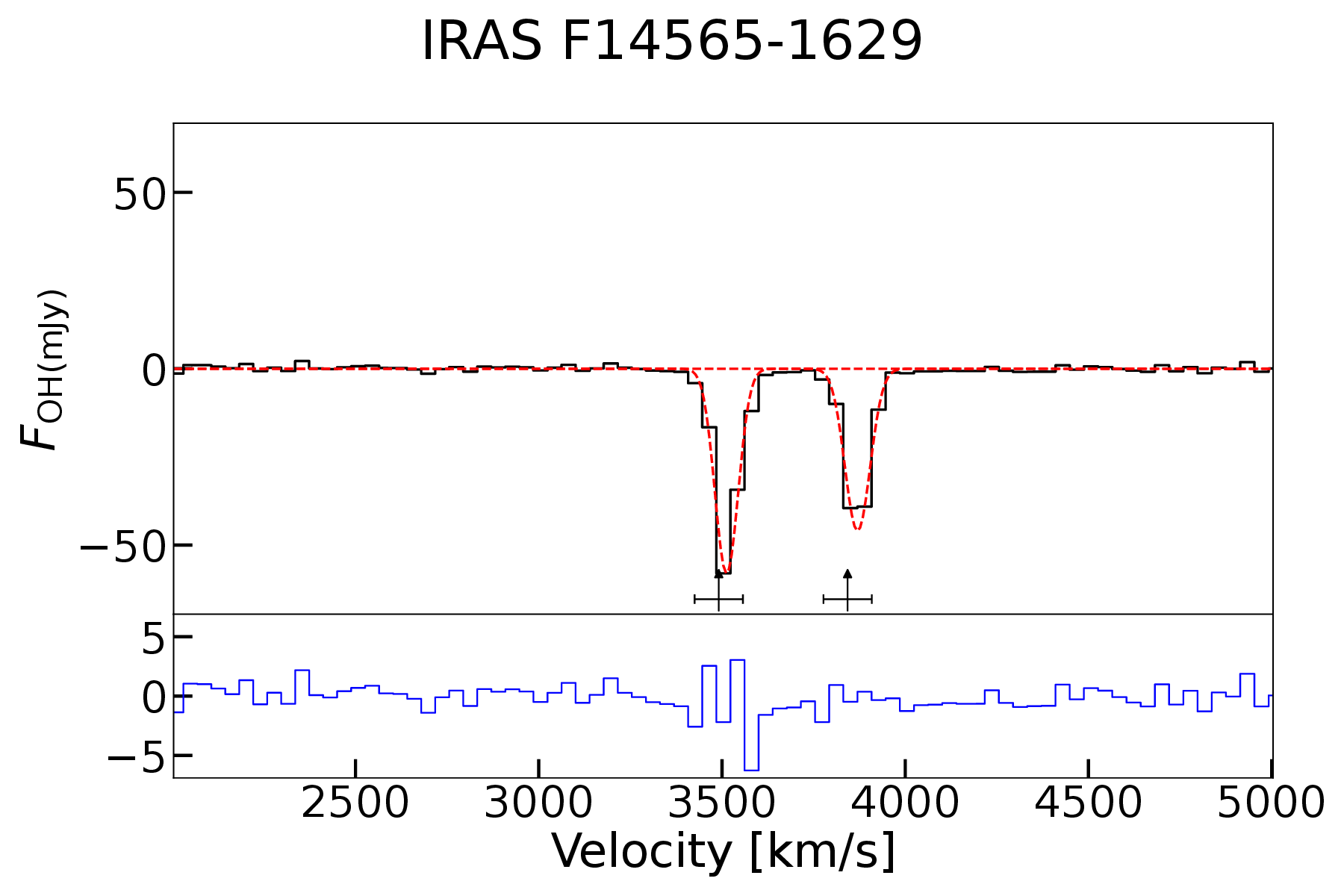}
  \end{minipage}\hfill
  \begin{minipage}[t]{0.31\textwidth}
    \centering
    \includegraphics[width=\textwidth]{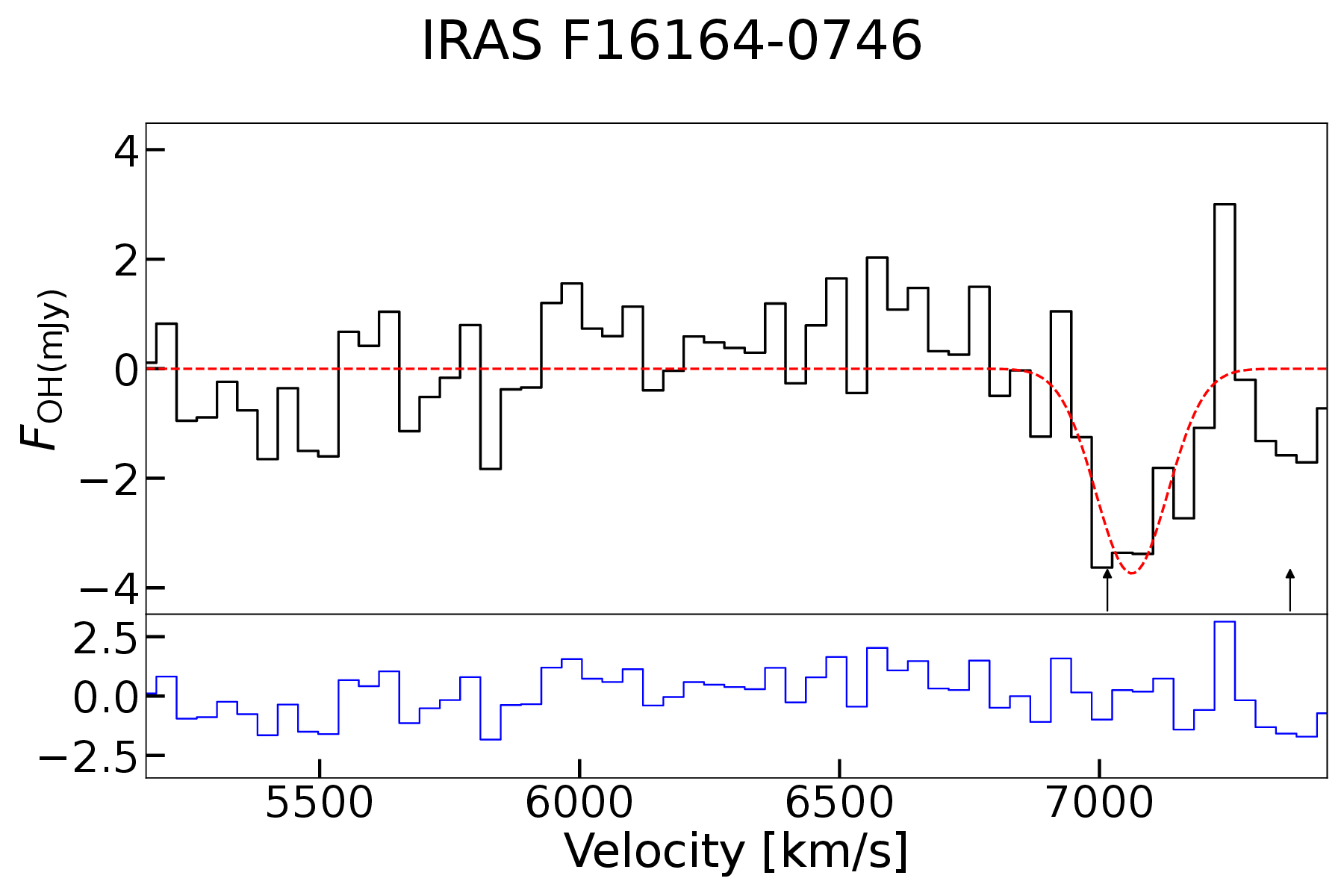}
  \end{minipage}\par
  \begin{minipage}[t]{0.31\textwidth}
    \centering
    \includegraphics[width=\textwidth]{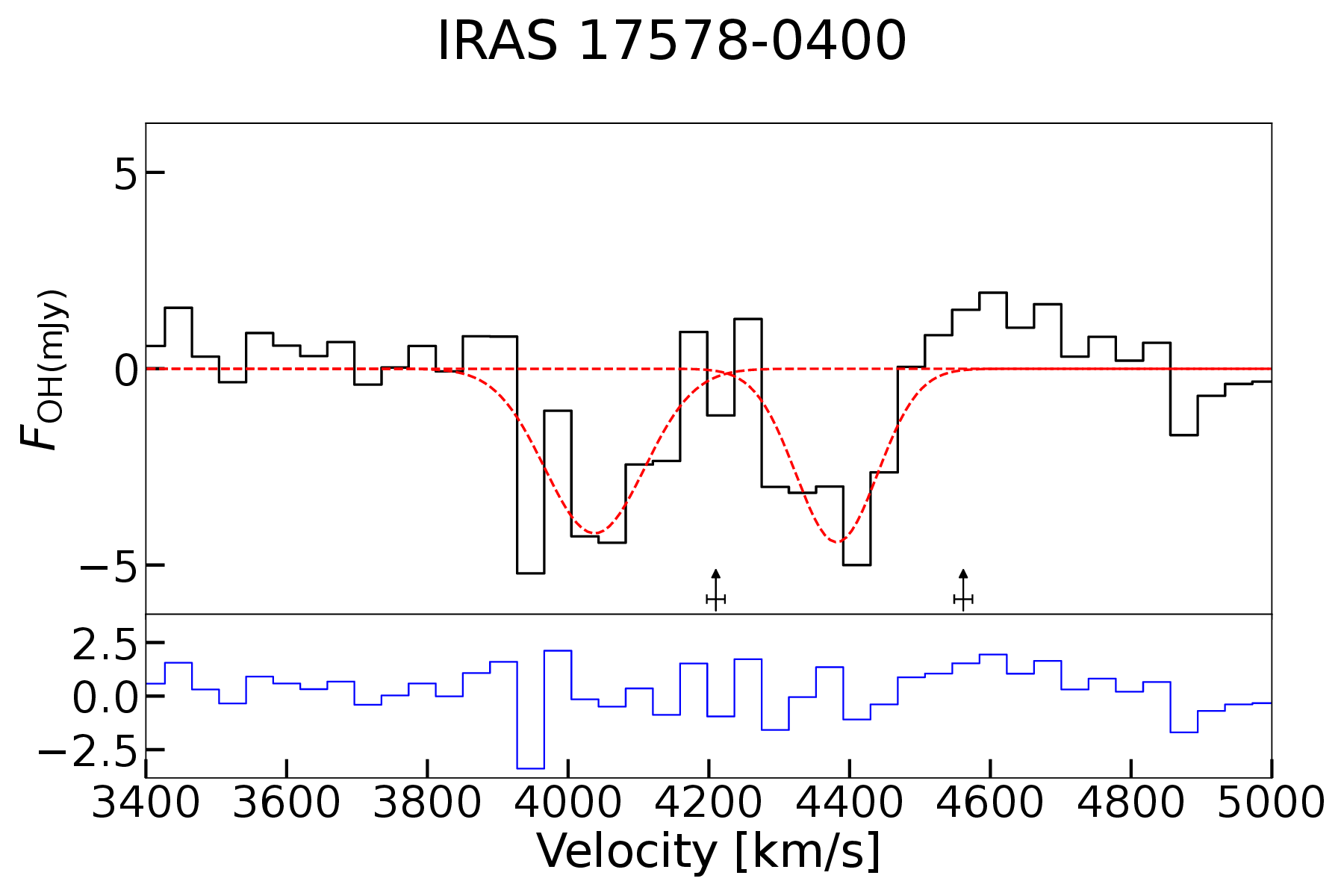}
  \end{minipage}\hfill
  \begin{minipage}[t]{0.31\textwidth}
    \centering
    \includegraphics[width=\textwidth]{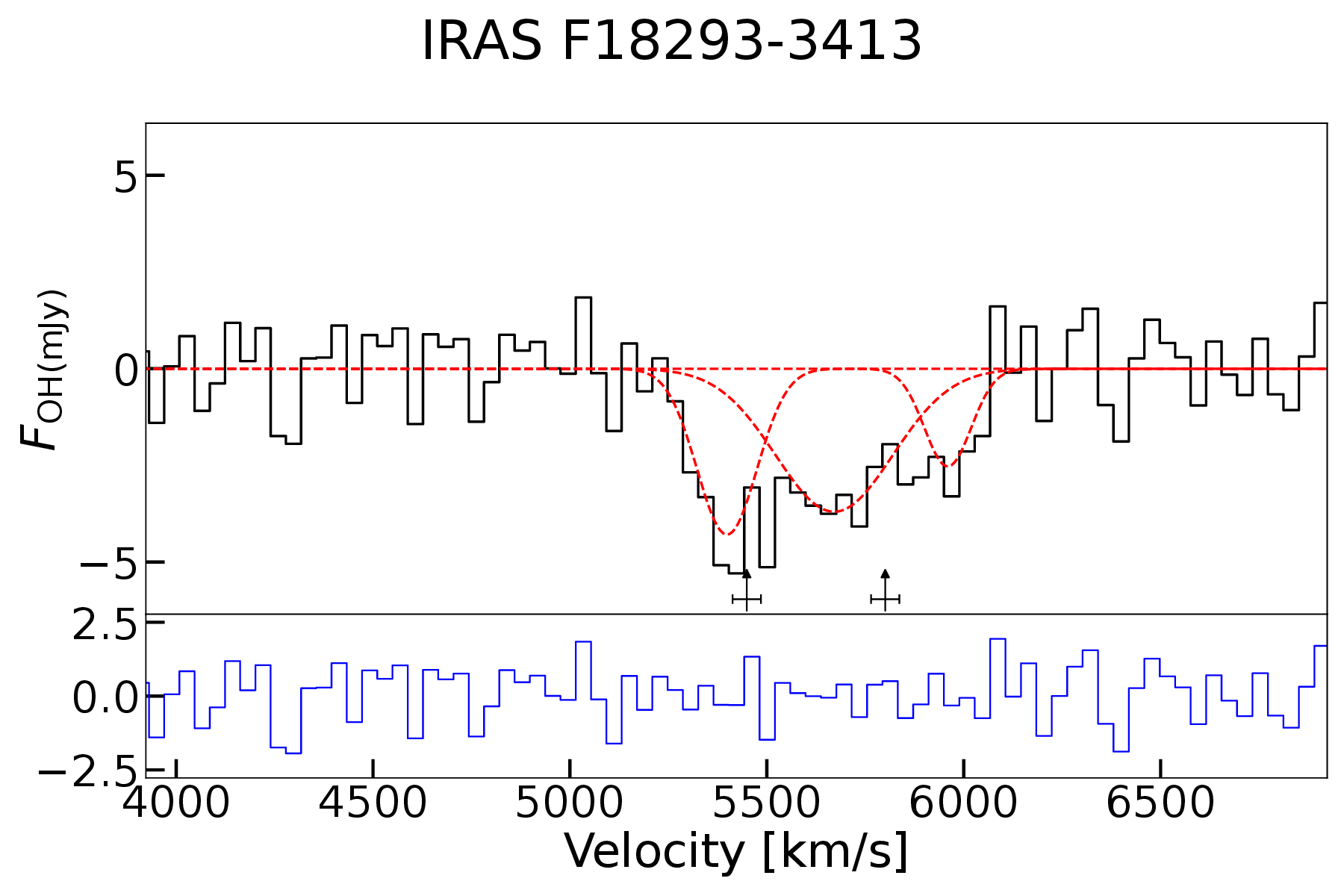}
  \end{minipage}\hfill
  \begin{minipage}[t]{0.31\textwidth}
    \centering
    \includegraphics[width=\textwidth]{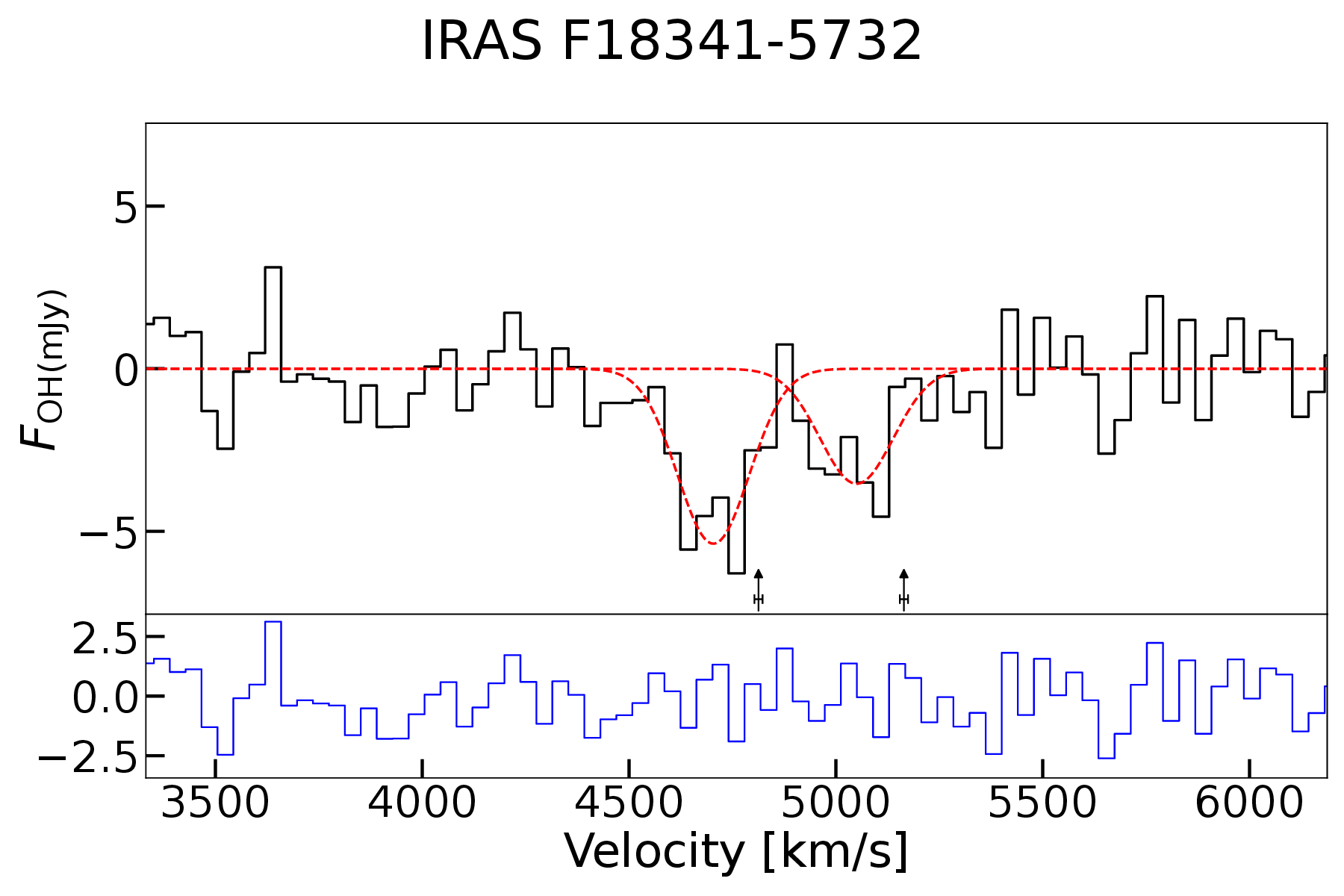}
  \end{minipage}\par
  \begin{minipage}[t]{0.31\textwidth}
    \centering
    \includegraphics[width=\textwidth]{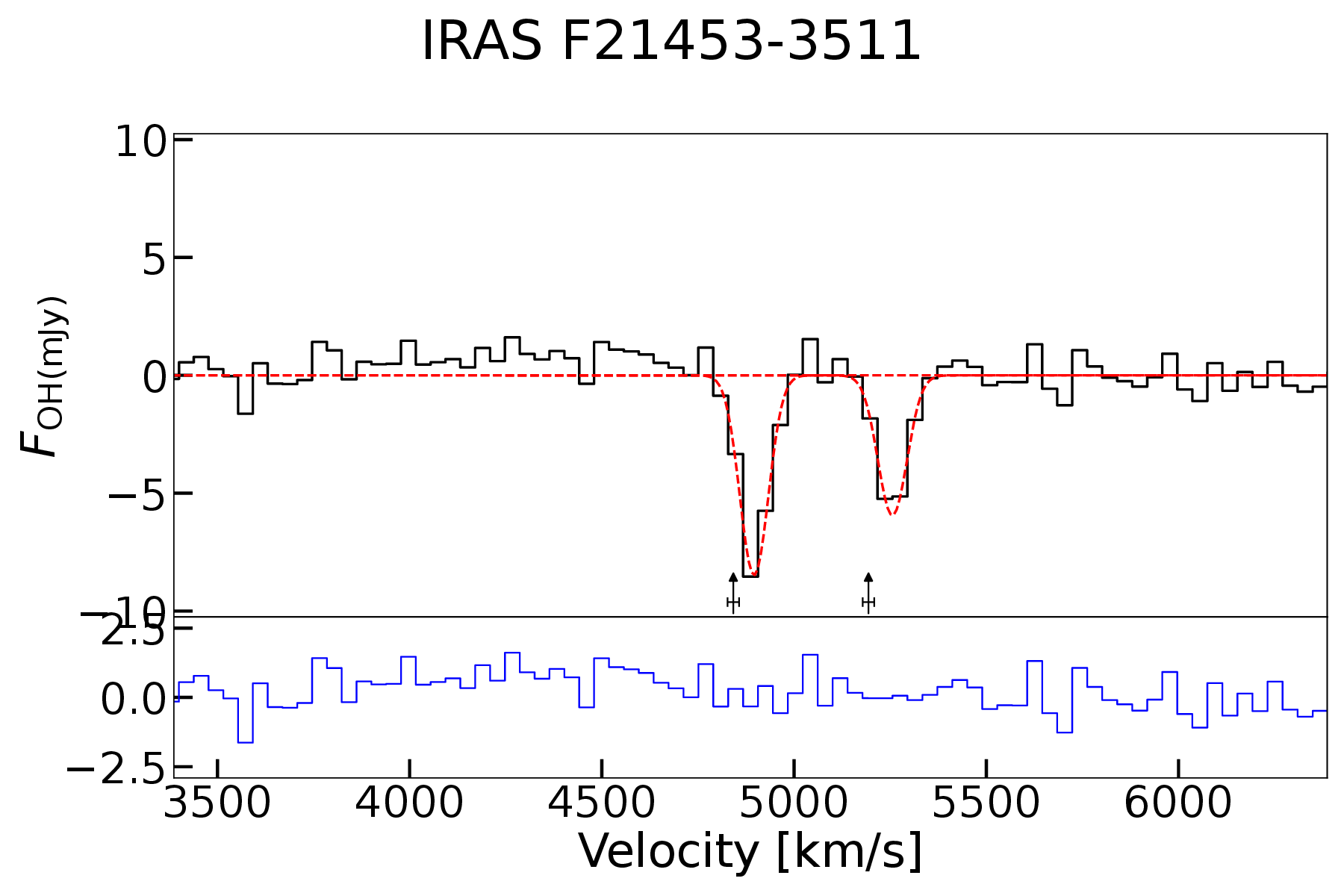}
  \end{minipage}\hfill
  \begin{minipage}[t]{0.31\textwidth}
    \centering
    \includegraphics[width=\textwidth]{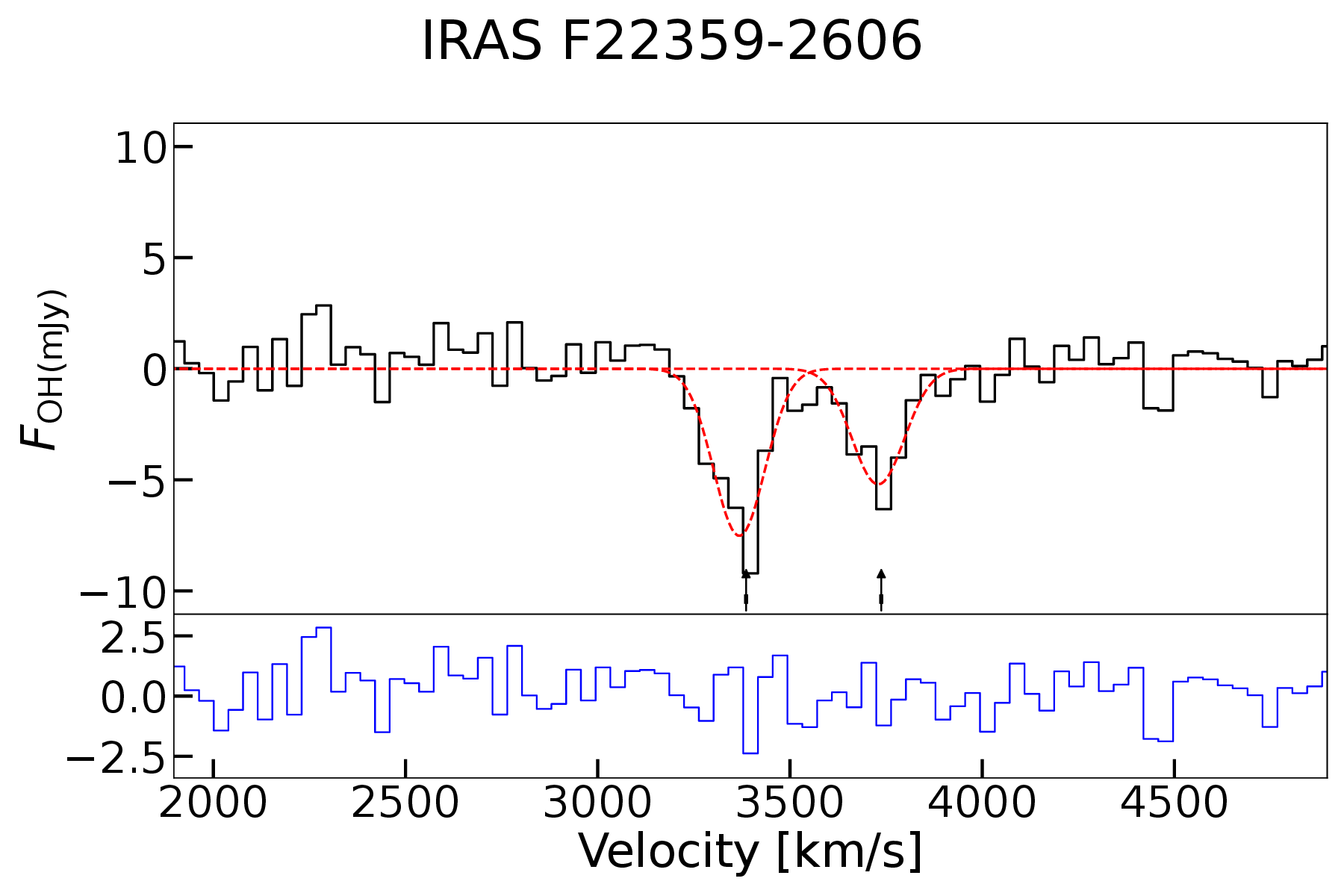}
  \end{minipage}\hfill
  \begin{minipage}[t]{0.31\textwidth}
    \centering
    \includegraphics[width=\textwidth]{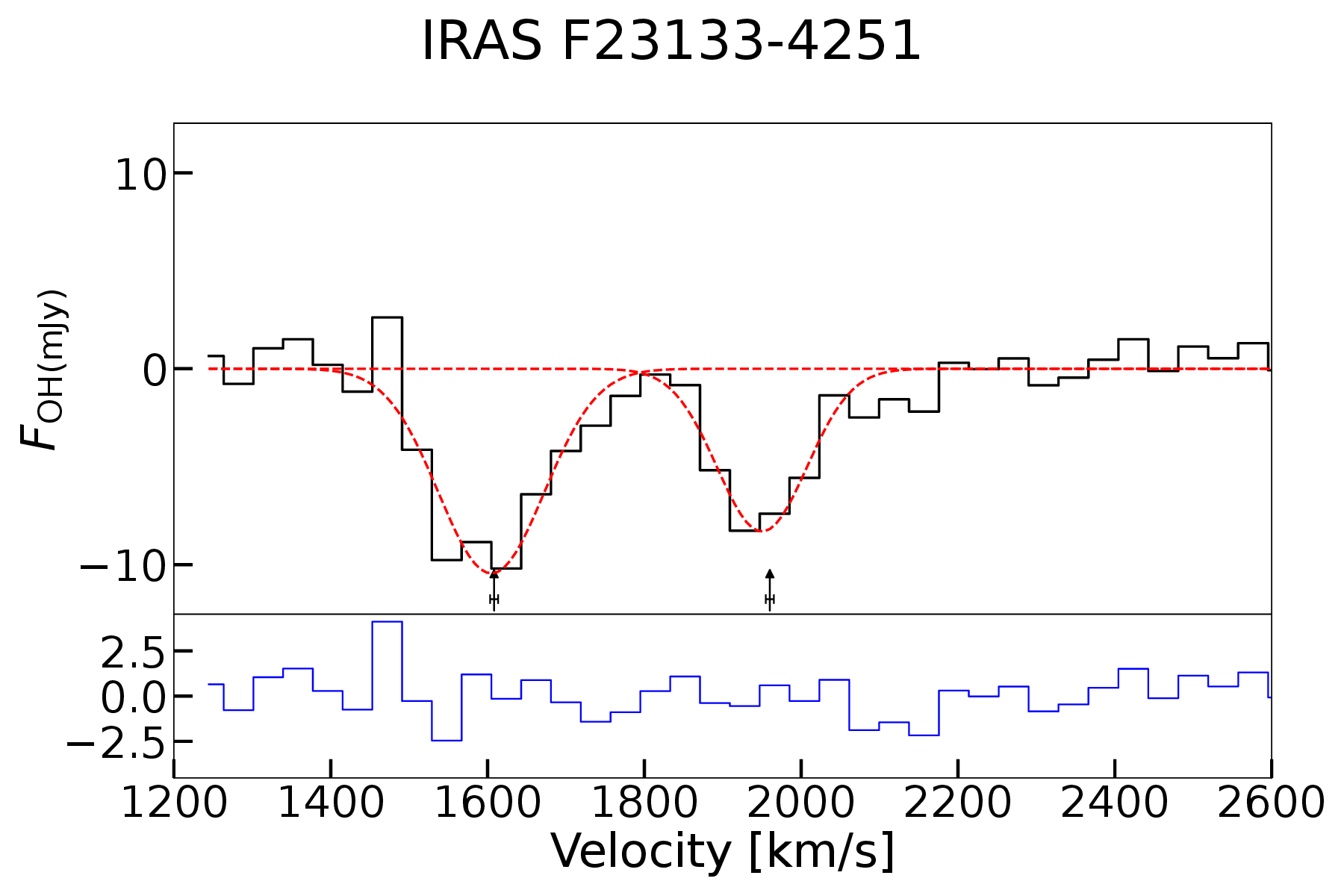}
  \end{minipage}\par
  \par\noindent
  {\centering
  \begin{minipage}[t]{0.31\textwidth}
    \centering
    \includegraphics[width=\textwidth]{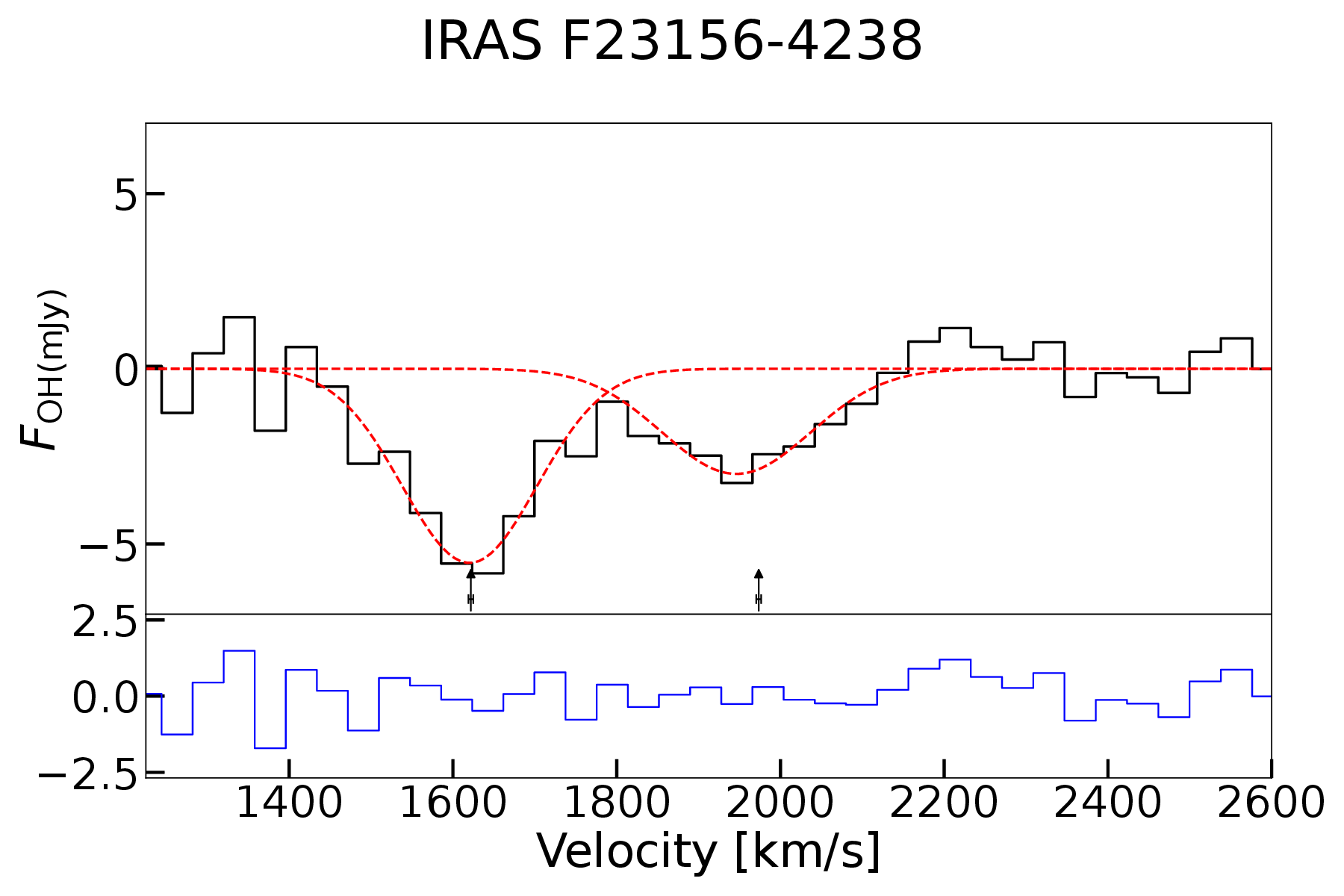}
  \end{minipage}\par}

  \caption{continued.}
\end{figure*}
\begin{figure*}
 \centering
    \subcaptionbox*{}{%
    \begin{minipage}[t]{0.45\textwidth}
      \centering
      \includegraphics[width=\textwidth]{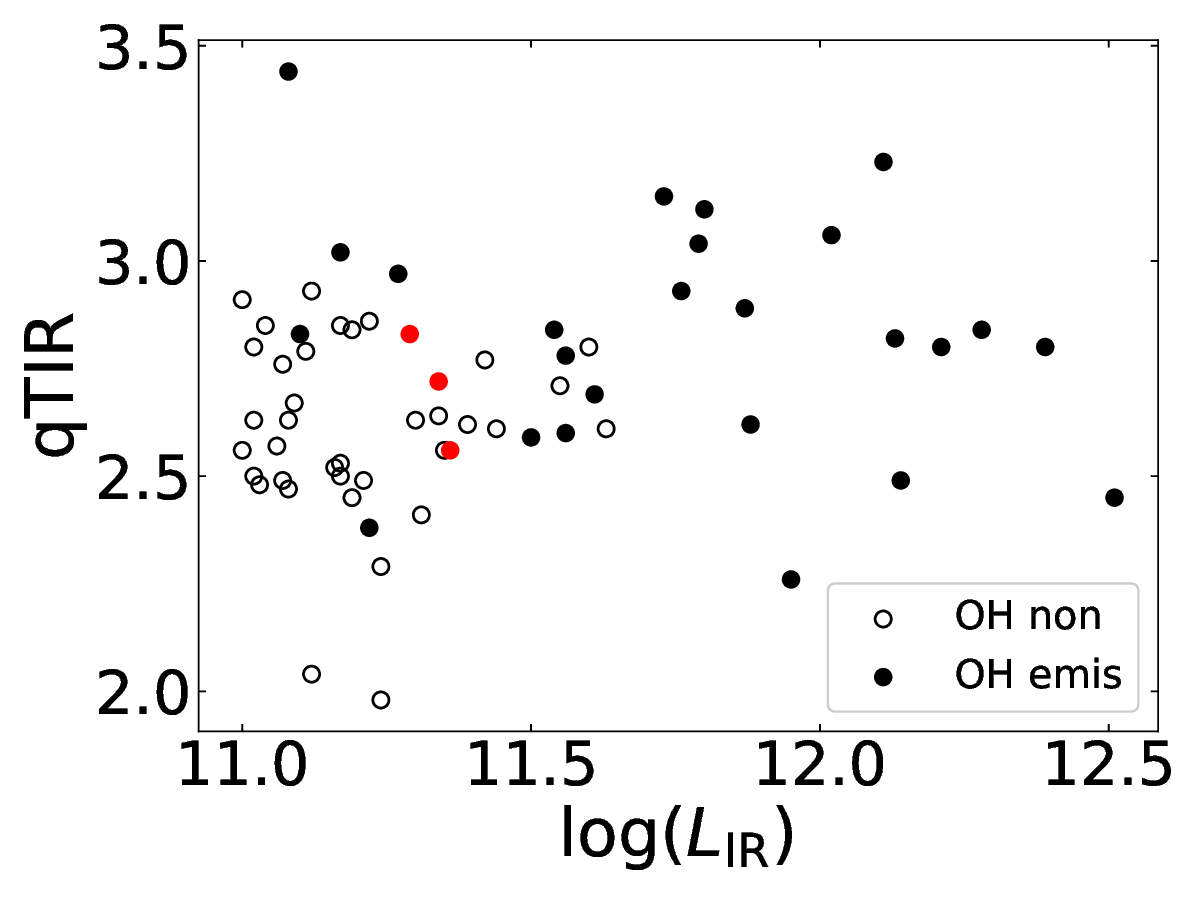}
    \end{minipage}%
    }
    \subcaptionbox*{}{%
    \begin{minipage}[t]{0.45\textwidth}
      \centering
      \includegraphics[width=\textwidth]{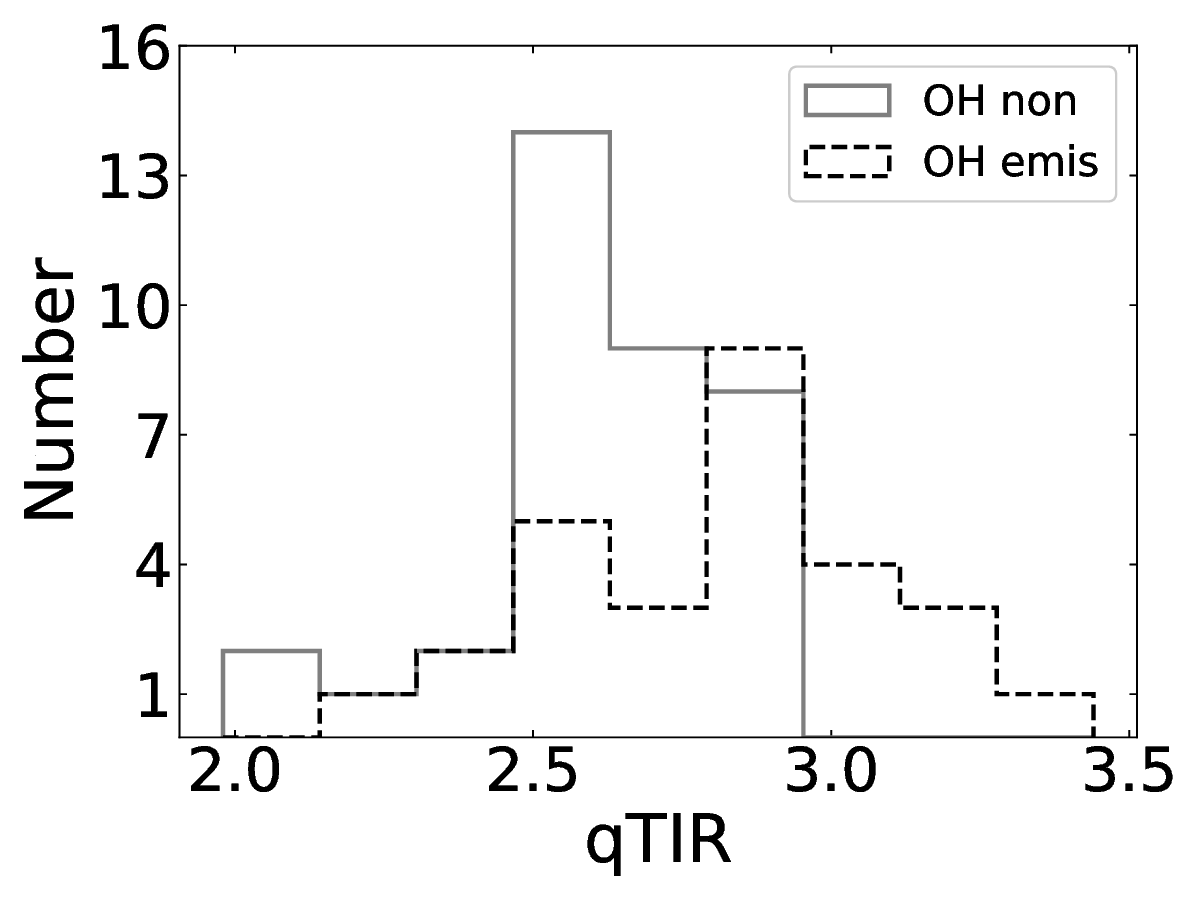}
    \end{minipage}%
   }
   \subcaptionbox*{}{%
    \begin{minipage}[t]{0.45\textwidth}
      \centering
      \includegraphics[width=\textwidth]{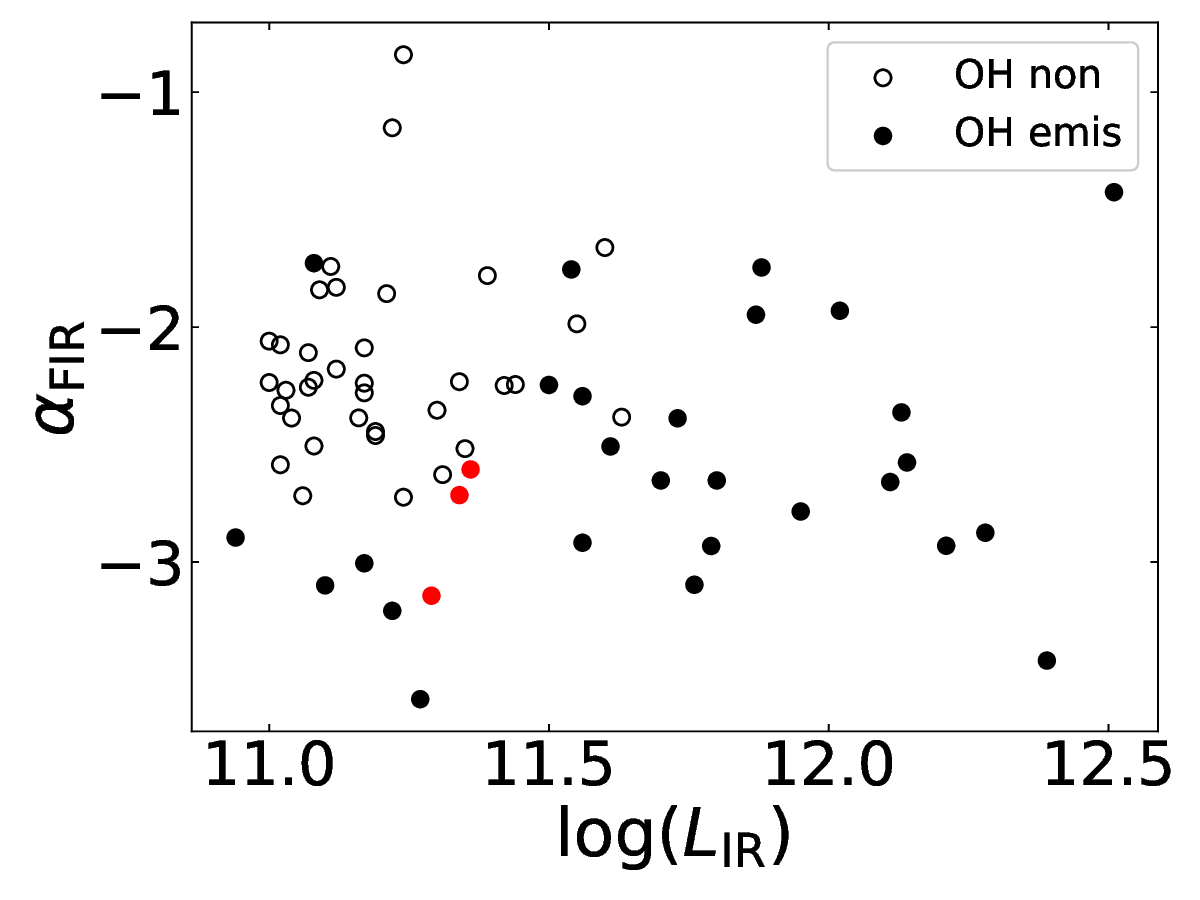}
    \end{minipage}%
   }
   \subcaptionbox*{}{%
    \begin{minipage}[t]{0.45\textwidth}
      \centering
      \includegraphics[width=\textwidth]{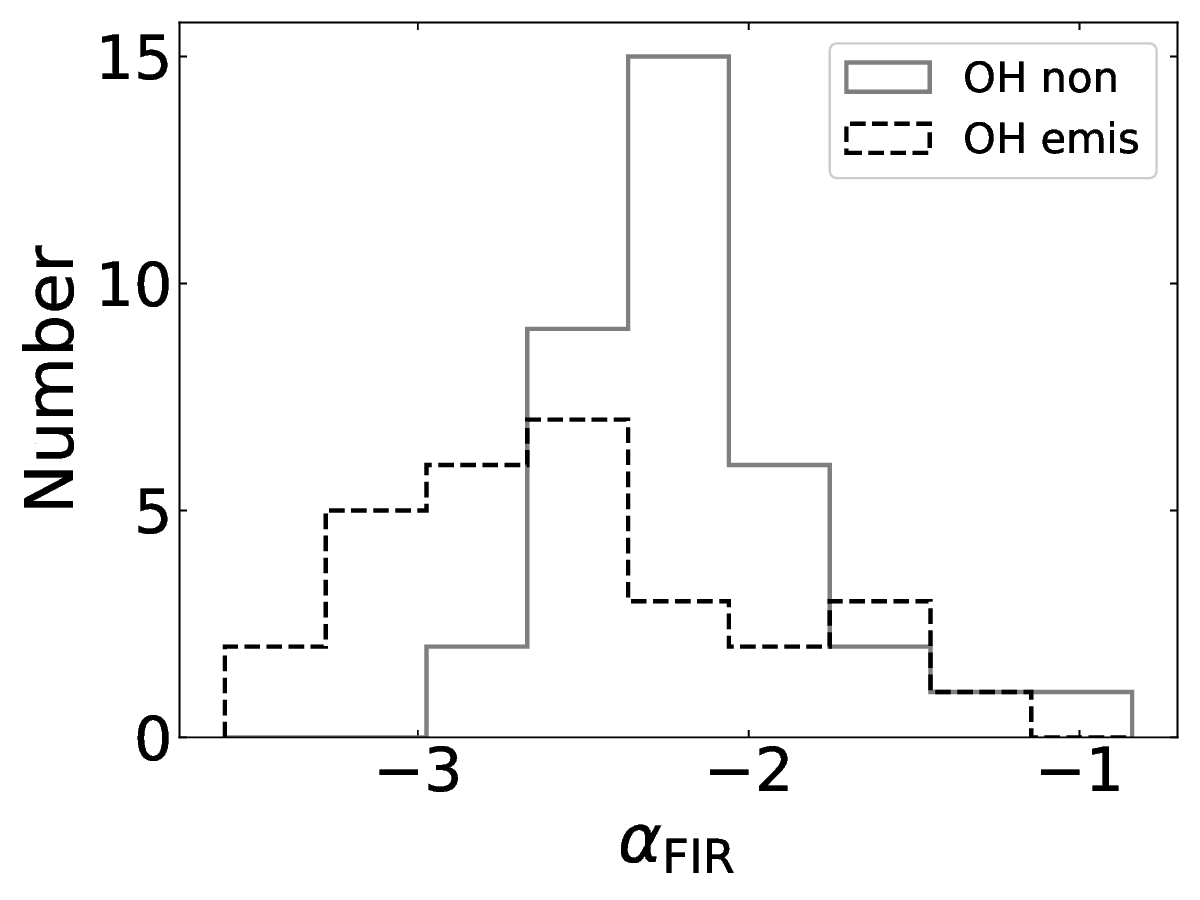}
    \end{minipage}%
   }
\caption{Comparison of the properties of OH non-detected galaxies in this work with OH-emitting galaxies from the RGBS sample. Top left: (qTIR)-ratio versus infrared luminosity. Top right: Distribution of the (qTIR)-ratio for OH non-detected and OH-emitting galaxies. Bottom left: Far-infrared spectral index between \SI{25}{\um} and \SI{60}{\um} ($\alpha_{\mathrm{FIR}}$) versus infrared luminosity. Bottom right: Distribution of $\alpha_{\mathrm{FIR}}$ for the OH non-detected and OH-emitting galaxies. The red filled circles highlight the three newly discovered OH-emitting sources identified in this work. The (qTIR)-ratio values are taken from \cite{2025MNRAS.543.2463M}, and $\alpha_{\mathrm{FIR}}$ is calculated using
$\alpha_{\mathrm{FIR}} = \log\left(S_{25,\mu\mathrm{m}} / S_{60,\mu\mathrm{m}}\right) / \log\left(60 / 25\right)$.}
 
\label{fig:ohnon-prop}
\end{figure*}
\begin{figure*}[htbp]
  \centering
  \subcaptionbox*{}{%
    \begin{minipage}[t]{0.45\textwidth}
      \centering
      \includegraphics[width=\textwidth]{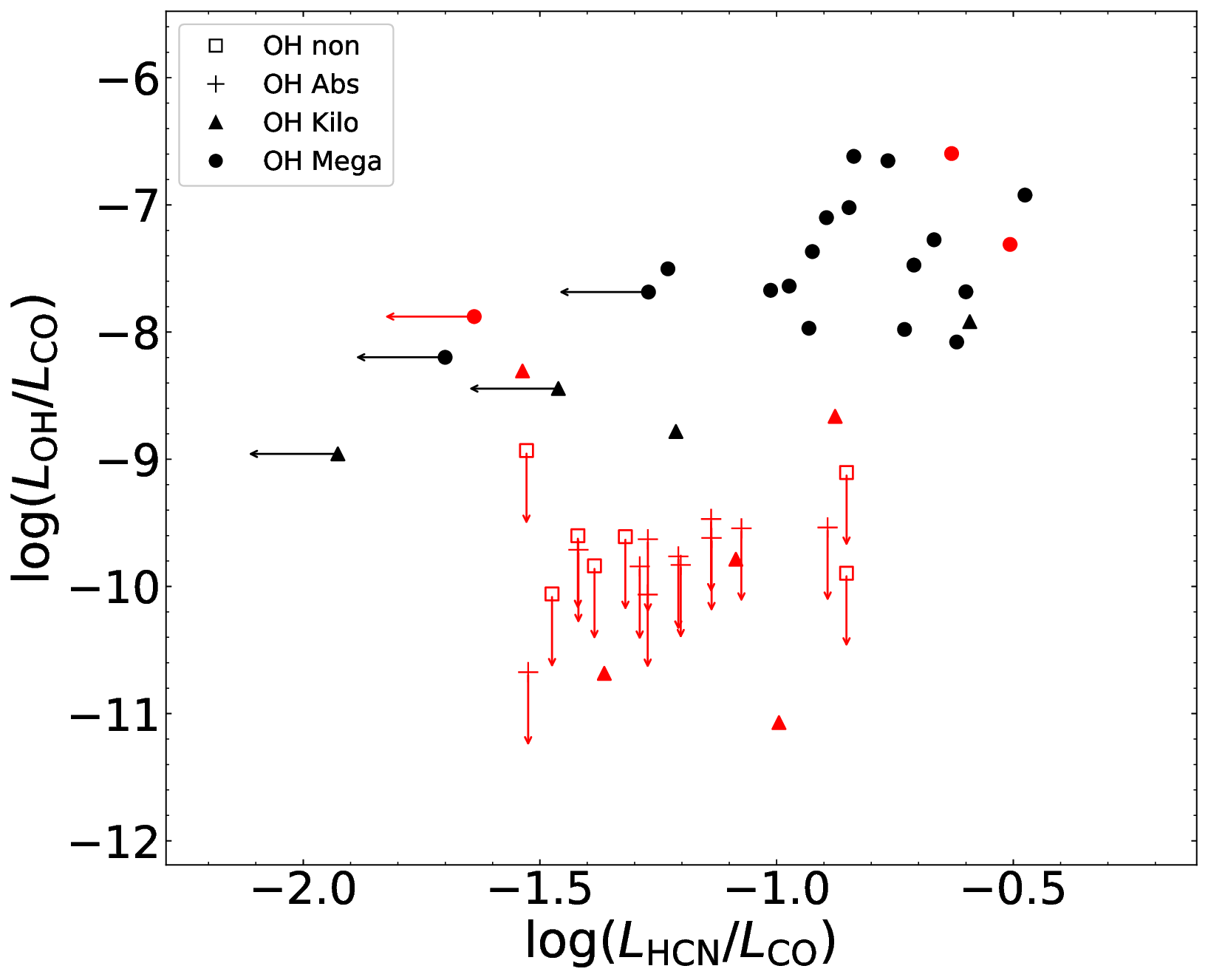}
    \end{minipage}%
    }
\caption{Normalized OH maser luminosity versus HCN luminosity for OH maser galaxies. This figure is a replot of a subpanel from Figure~3 of \citet{2018JApA...39...34H}, updated to include our new results for a subsample of these galaxies. The symbols are adopted from the original figure (as indicated in the upper-left corner), while red symbols denote the revised classifications of the galaxies as OH megamasers, kilomasers, absorbers, and non-detections. Red and black arrows indicate upper limits on the OH luminosity and HCN luminosity, respectively. Upper limits on the OH luminosity are derived using
$L_{\mathrm{OH}}^{\mathrm{max}} = 4\pi D_L^2 , 1.5\sigma \left( \frac{\Delta v}{c} \right) \left( \frac{\nu_0}{1+z} \right)$,
assuming a boxcar line profile with a rest-frame width of $\Delta v$ = 150~$\mathrm{km\,s}^{-1}$ and a peak height of 1.5$\sigma$ \citep[see][for details]{2002AJ....124..100D}.}

\label{fig:ohnon-dense}
\end{figure*}

\begin{figure*}
  \centering
  \begin{tabular}{cc}
    \includegraphics[width=0.48\textwidth]{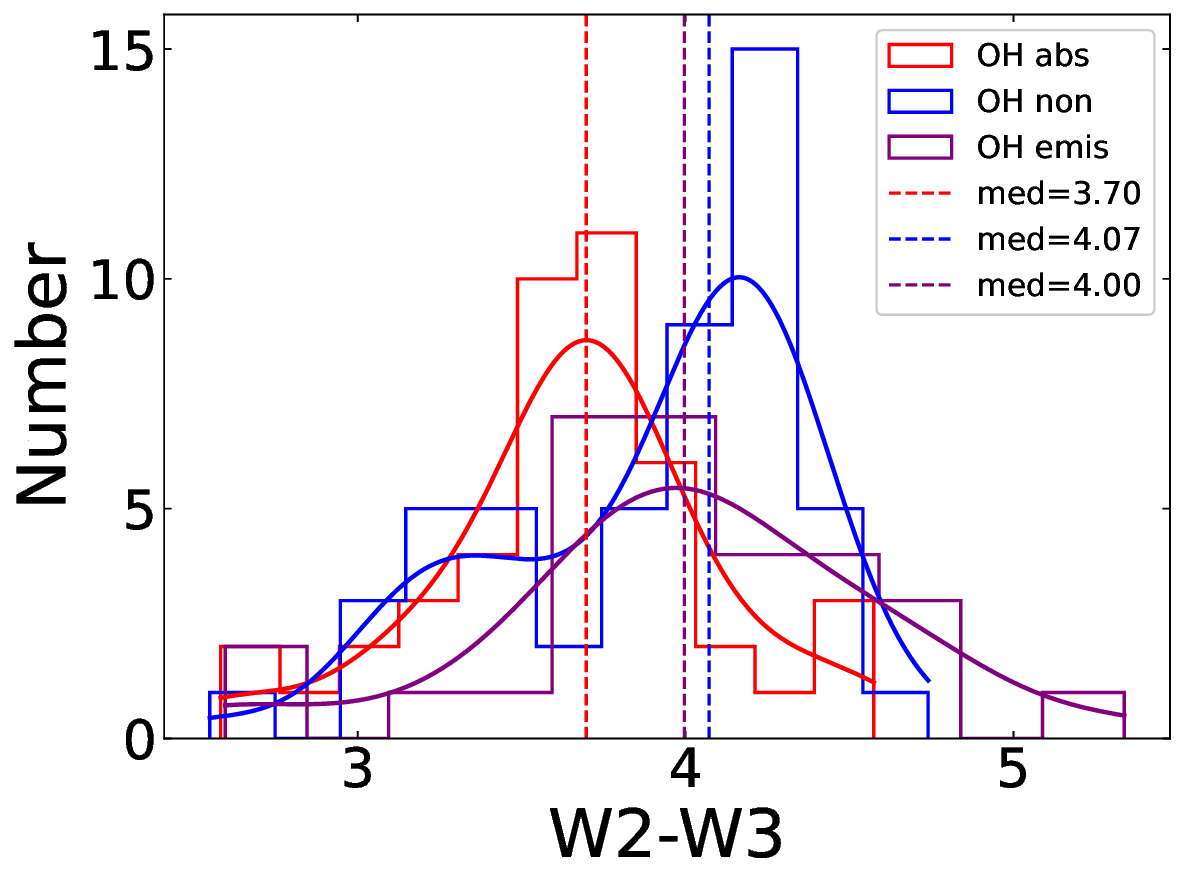} & 
    \includegraphics[width=0.48\textwidth]{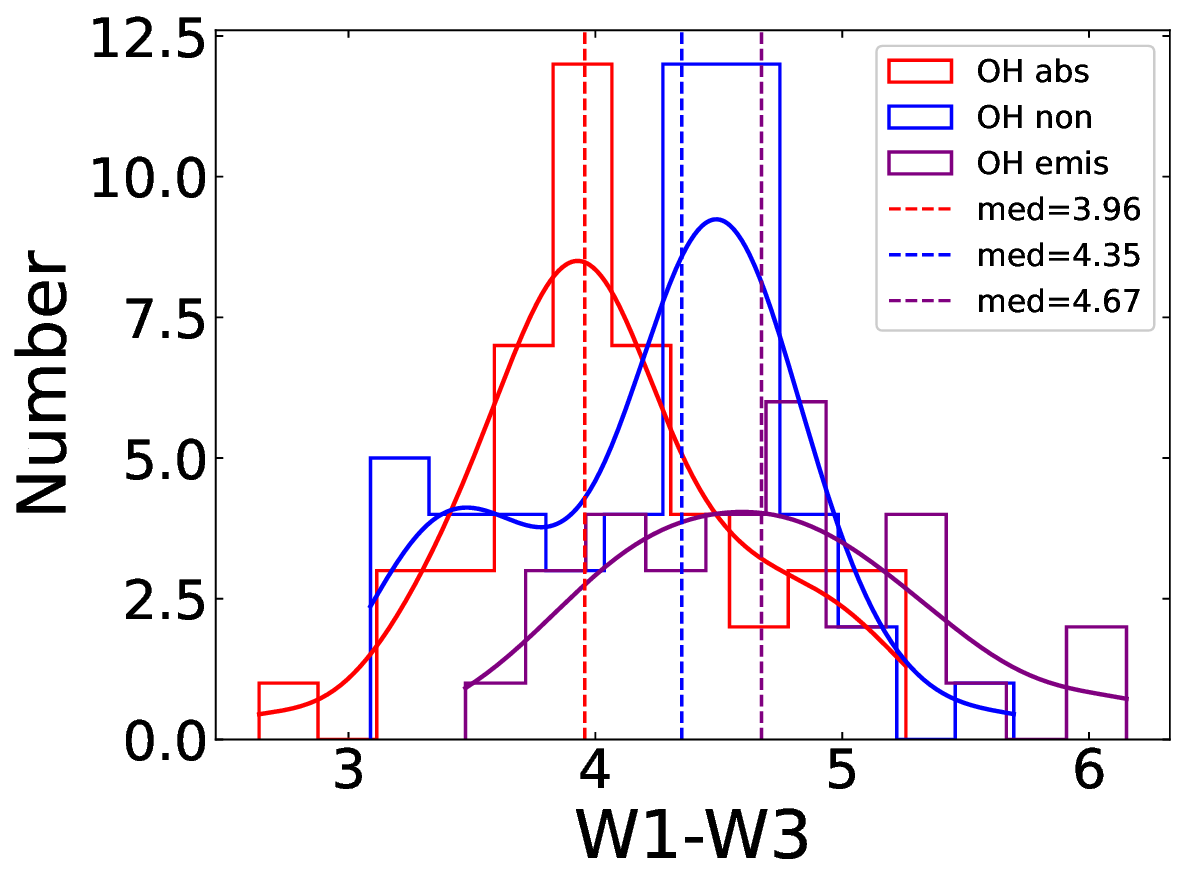} \\
    \\[-5pt] 
    \includegraphics[width=0.48\textwidth]{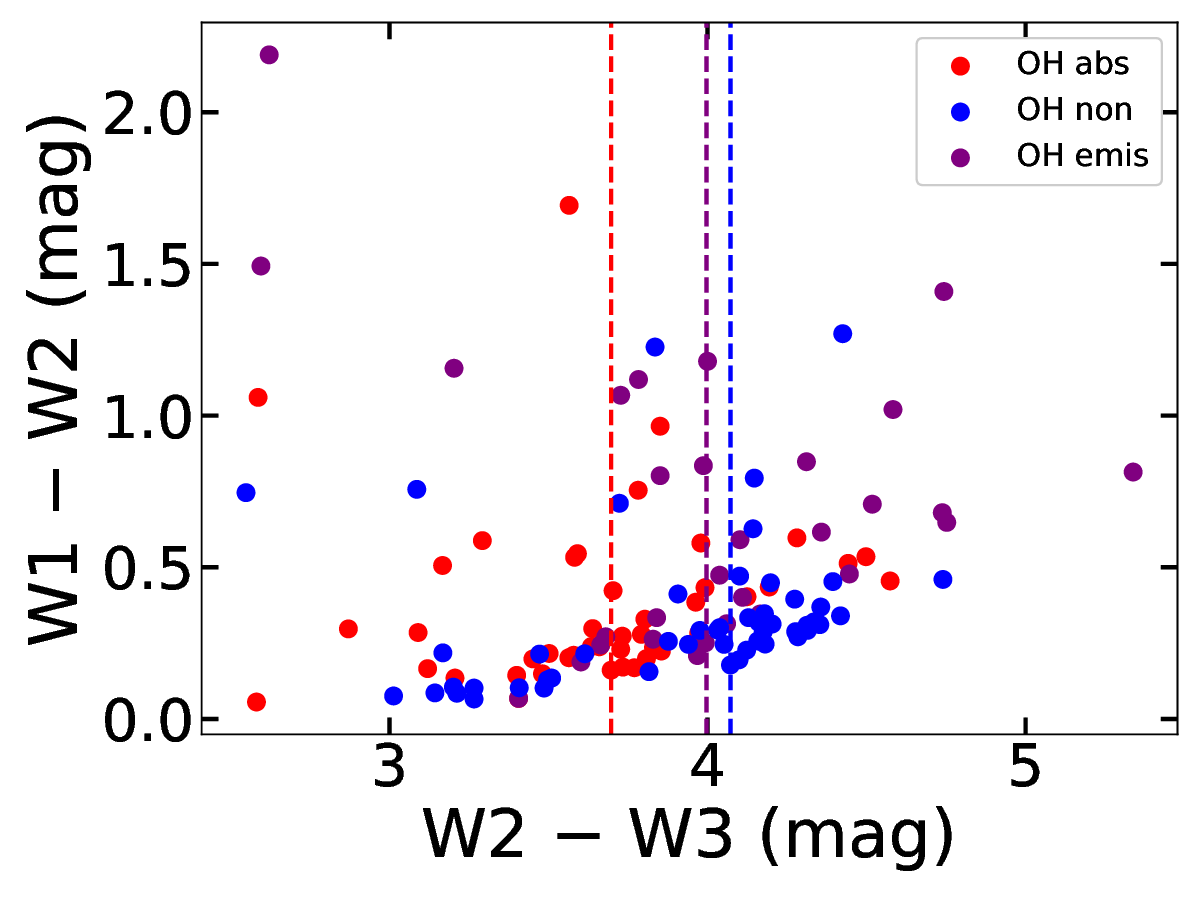} & 
    \includegraphics[width=0.48\textwidth]{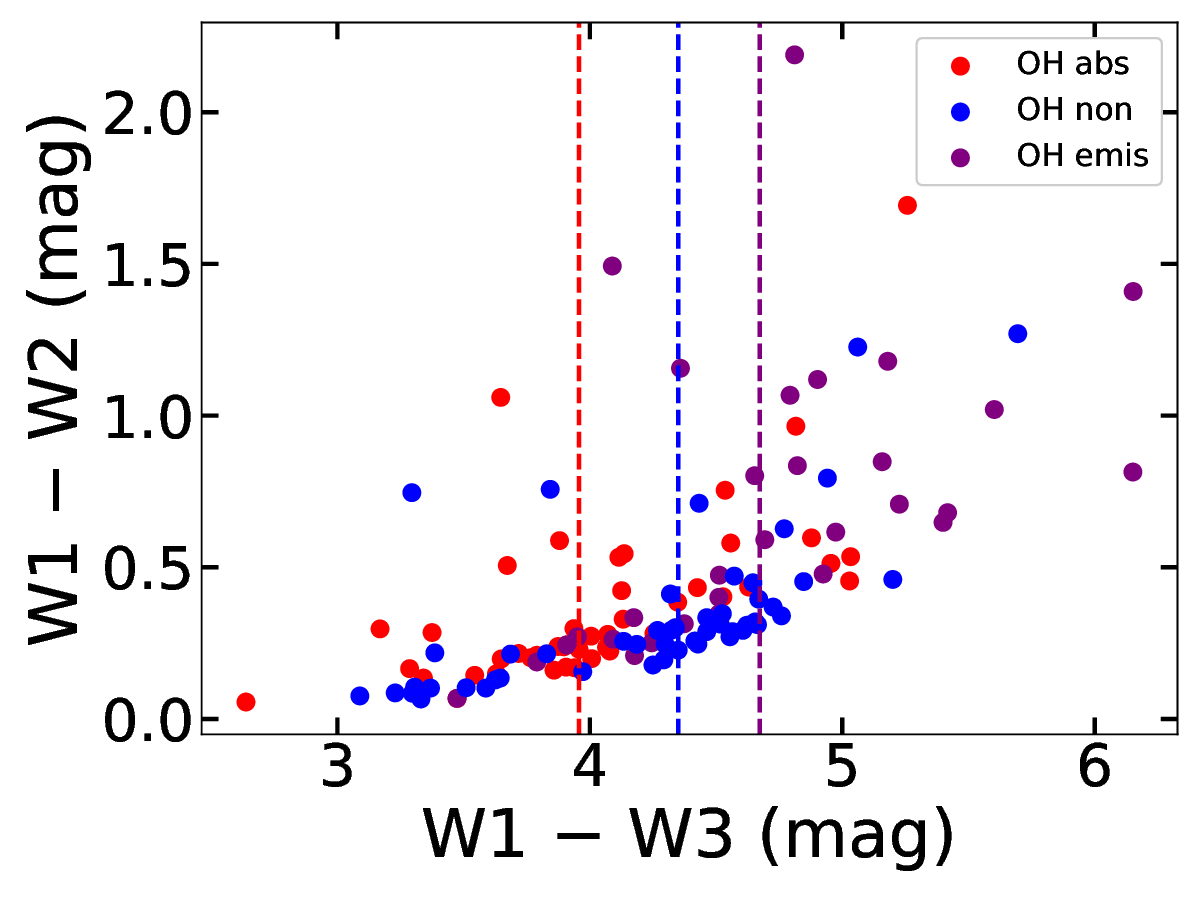} \\
  \end{tabular}
\caption{
Distributions of WISE colors for different OH categories. 
The top left and top right panels show the distributions of W2$-$W3 and W1$-$W3 colors, respectively. The bottom panels present the WISE color--color diagrams of W1$-$W2 versus W2$-$W3 and W1$-$W2 versus W1$-$W3, respectively. The sample is divided into three groups based on their OH spectral features: OH absorption (red), non-detections (blue), and OH emission (purple). 
The stepped histograms in the top panels show the observed number distributions, while the smooth curves are used to guide the visualization of the color distributions. 
The vertical dashed lines in all panels indicate the median colors of each OH category; the same median values are adopted in both the color distributions and the color--color diagrams, as labeled in the legends.
}
\label{fig:ohabs-prop}
\end{figure*}

\begin{figure}[http]
  \centering
  \includegraphics[width=0.45\textwidth]{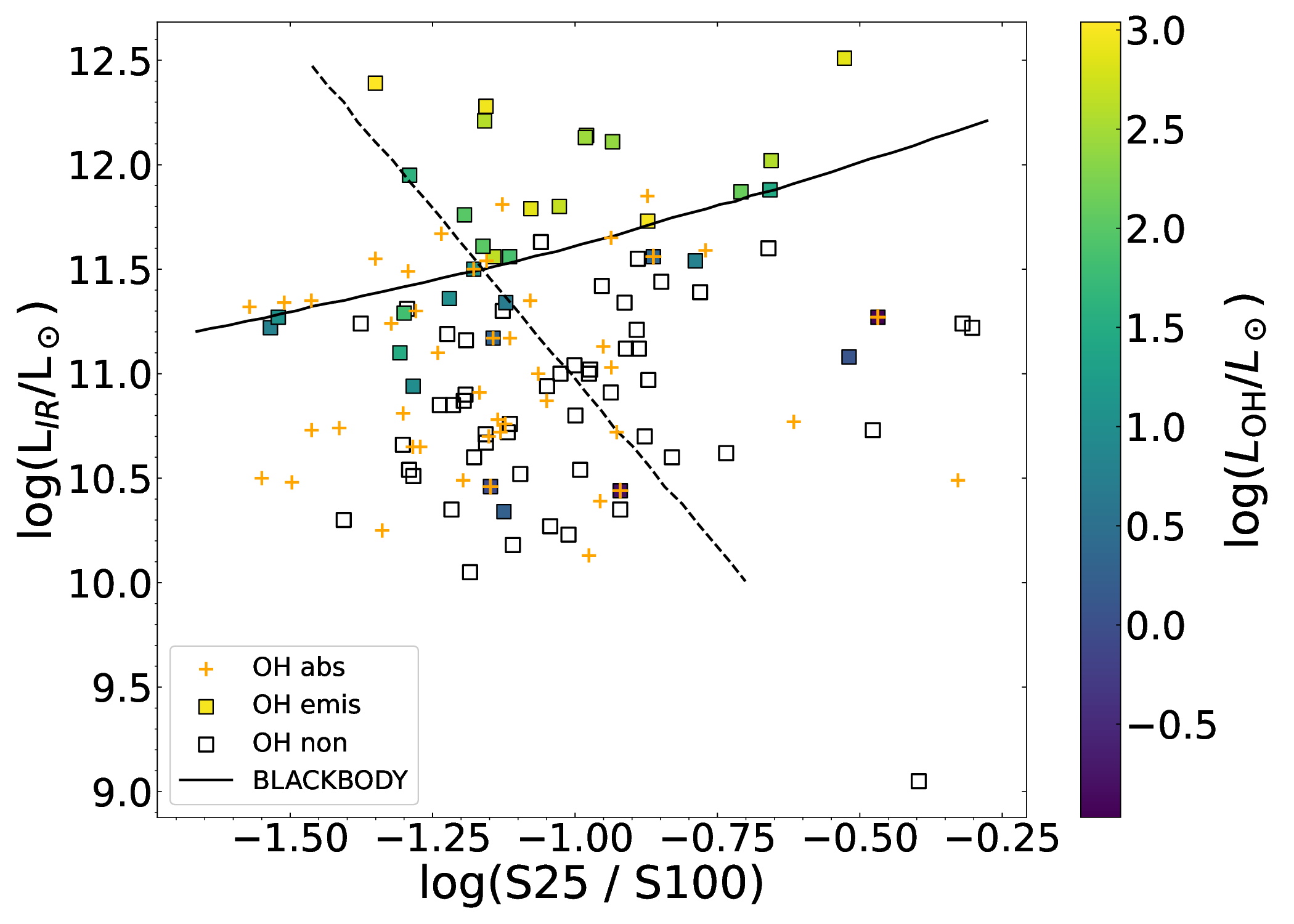}
  \caption{IR luminosity versus 25/100 color diagram \citep{1989ApJ...338..804B} for the OH maser galaxies. The colored squares represent known OH maser galaxies in the IRAS RBGS sample, with the color indicating the OH luminosity as shown by the color bar on the right. The open squares and plus symbols denote OH absorption galaxies and OH non-detections with peak radio continuum emission exceeding 20~mJy~beam$^{-1}$, respectively. The dashed and dotted lines are adopted from \citet{1989ApJ...338..804B}: the dashed line separates OH emission and absorption galaxies, while the dotted line marks the locus of FIR sources with blackbody spectra (see \citealt{1989ApJ...338..804B} for details).}

\label{fig:LFIR-color} 
\end{figure}
\end{document}